\documentclass[11pt]{article}

\usepackage[T1]{fontenc}
\usepackage[cp1251]{inputenc}
\usepackage{textcomp}
\usepackage[centertags]{amsmath}
\usepackage{amsfonts}
\usepackage{amssymb}
\usepackage[hypertex]{hyperref}
\usepackage{graphicx}
\usepackage[numbers,sort&compress]{natbib}

\usepackage{paperinitial}

\paperinitialization{15mm}{15mm}{15mm}{15mm}{2pt}{10pt}

\DeclareMathOperator{\re}{Re}
\DeclareMathOperator{\im}{Im}
\DeclareMathOperator{\sgn}{sgn}

\DeclareMathOperator{\pr}{pr}

\DeclareMathOperator{\cycle}{cycle}
\def\res{\mathop{\mathrm{res}}}
\DeclareMathOperator{\sing}{sing}

\newcommand{\bs}{\boldsymbol}

\newcommand{\ttg}{\stepcounter{equation}\tag{\theequation}}

\newcommand{\e}{\varepsilon}
\newcommand{\vf}{\varphi}
\newcommand{\vk}{\varkappa}
\newcommand{\s}{\sigma}
\newcommand{\Si}{\Sigma}

\newcommand{\be}{\beta}
\newcommand{\ga}{\gamma}
\newcommand{\Ga}{\Gamma}
\newcommand{\de}{\delta}
\newcommand{\De}{\Delta}

\newcommand{\la}{\lambda}
\newcommand{\La}{\Lambda}
\newcommand{\ups}{\upsilon}
\newcommand{\vrho}{\varrho}

\newcommand{\spx}{\mathbf{x}}
\newcommand{\spy}{\mathbf{y}}
\newcommand{\spz}{\mathbf{z}}
\newcommand{\spb}{\mathbf{b}}

\newcommand{\spk}{\mathbf{k}}

\newcommand{\spt}{\bs\tau}
\newcommand{\spbe}{\bs\be}
\newcommand{\spbep}{\bs\be_\perp}
\newcommand{\btbe}{(\mathbf{b}\bs\tau\bs\be)}
\newcommand{\btbep}{(\mathbf{b}\bs\tau\bs\be_\perp)}

\begin{document}
\allowdisplaybreaks[4]
\frenchspacing
\setlength{\unitlength}{1pt}

\title{{\Large\textbf{Self-interaction of an arbitrary moving dislocation}}}

\date{}

\author{P.O. Kazinski\thanks{E-mail: \texttt{kpo@phys.tsu.ru}}\;, V.A. Ryakin\thanks{E-mail: \texttt{vlad-r@sibmail.com}}\;, and A.A. Sokolov\thanks{E-mail: \texttt{alexei.sokolov.a@gmail.com}}\\[0.5em]
{\normalsize Physics Faculty, Tomsk State University, Tomsk, 634050, Russia}
}

\maketitle

\begin{abstract}

The action functional for a linear elastic medium with dislocations is given. The equations of motion following from this action reproduce the Peach-K\"{o}hler and Lorentzian forces experienced by dislocations. The explicit expressions for singular and finite parts of the self-force acting on a curved dislocation are derived in the framework of linear theory of elasticity of an isotropic medium. The velocity of dislocation is assumed to be arbitrary but less than the shear wave velocity. The slow speed and near-sonic motions of a dislocation are investigated. In the near-sonic limit, the explicit expression for the leading contribution to the self-force is obtained. In the case of slowly moving dislocations, the effective equations of motion derived in the present paper reproduce the known results.


\end{abstract}

\section{Introduction}

The description of dynamics of dislocations in crystals is a classical problem of theory of elasticity. The dislocations are used to describe plasticity and strength of crystals, the formation of cracks, the crystal growth, the properties of propagation of sound and heat in deformed solid bodies, and so on (see for review \cite{OrlInden,Friedel64,Kos65,Nabar67,WeertWeert71,KagKravNat74,AlshInden,Mura87,HirthLothe}). A certain type of dislocations also appears in partially ordered structures such as liquid crystals \cite{deGennProst} and even in amorphous solids \cite{BKSZ21}. In order to describe such a plethora of phenomena involving dislocations, the dynamics of dislocations have to be known. In linear elastodynamics, neglecting the influence of atomic structure of the crystal, the forces experienced by a dislocation are the Peach-K\"{o}hler and Lorentzian ones \cite{LLElast,Lund98,Lubarda19,GVSD21,Nabar67,Friedel64,Kos65,HirthLothe,Mura87}. These forces include both the action of external stresses and the action of stresses produced by the dislocation. Thus one encounters the issue of description of self-interacting dynamics of dislocations where the self-force is divergent in the limit of an ideally thin dislocation. This problem is analogous to the self-interaction problem in classical electrodynamics of point and extended singular sources of electromagnetic fields \cite{Lor,Abraham,Dir,BatCarMen,KazShar05,BachMcAll13,GalMelSp16}. In the case of dislocations, the kinetic part of the effective equations of motion describing the inertial properties of a dislocation is of a pure field origin and is contained in the self-force exerted by elastic forces.

For slowly moving dislocations, the integro-differential effective equations of motion of a smeared self-interacting dislocation were obtained in \cite{Kos63,KosNat65,GavBar76,Kos65} keeping only the terms at most linear in velocity. As for arbitrary velocities, the general expression for the Peach-K\"{o}hler contribution to the self-force was described only recently in \cite{Pellegr20}. To our knowledge, the explicit expressions for the divergent and finite parts of the self-force including the Peach-K\"{o}hler and Lorentzian contributions are unknown in the literature. Our aim is to fill this gap. We will obtain the explicit expressions for the singular and finite parts of the total self-force acting on an arbitrary moving dislocation in the framework of linear theory of elasticity of isotropic media. It will become clear from the derivation procedure that there are three regimes of dislocation motion and so there are three types of the effective equations of motion depending on the velocity of dislocation: the subsonic regime, where the dislocation velocity is less than the shear wave velocity; the transsonic (intermediate) regime, where the dislocation velocity is greater than the shear wave velocity but less than the dilatational wave velocity; and the supersonic one, where the dislocation velocity is greater than both sound speeds. We will derive the explicit expressions for the self-force only in the first case, although the procedure we will develop is applicable for all the three cases.

As is well known, in addition to the elastic forces the dynamics of dislocations are affected by many kinds of other forces. The dislocations interact directly with other crystal defects, they experience the effective drag force due to interaction with electrons, phonons, and other crystal excitations, they are subject to the osmotic Bardeen-Herring force facilitating a dislocation climb and to the Peierls-Nabarro force resulting, in particular, in the existence of the minimum stress needed to compel the dislocation to move \cite{Nabarro51,Weert65,LothHirth65,HirthLothe,Mura87,AlshInden,KagKravNat74,Nabar67,Friedel64,SwinDud15}. In the present paper, we do not consider these forces and study only the contribution of elastic forces to the effective equations of motion of dislocations. The only information about the microscopic structure of the medium that we include in our considerations is the gliding condition. Namely, in analyzing the physical consequences of the equations obtained, we introduce the constraint that the dislocation moves along a slip plane, i.e., it performs the so-called conservative motion. Such an approximation works rather well for ductile crystals at low temperatures, which have a low Peierls barrier. Note, however, that the very expression for the self-force will be deduced without appealing to the gliding constraint.

The expression for the self-force we will derive is rather cumbersome. It includes both local and nonlocal terms, and even the local contribution contains about eighty different structures. Nevertheless, such an expression can be employed in computer codes for numerical simulation of dislocation dynamics \cite{GumGao99,MVZWG01}. It can also be used for the study of certain particular cases where it is amenable to  analytical investigations. For example, the use of the gliding condition greatly reduces the number of terms in the expression for the self-force. On the other hand, if one keeps only the terms that are divergent on removing the dislocation core regularization, then the self-force will become local. In particular, employing such approximations, we will reproduce the string model for a slowly moving dislocation \cite{GranLuck66,Pegel66,LaubEsh66,Ninomiya68,HirthLothe,Nabar67,Friedel64}. The expressions for the effective mass density and the string tension coincide with the expressions known in the literature for straight dislocations \cite{Pegel66,LaubEsh66,Ninomiya68,HirthLothe,Kos65}. The stable static configurations of dislocations turn out to be the helices with the axes directed along the Burgers vector. The degenerate cases of such helices are a straight screw dislocation and an edge dislocation in the form of a circle lying in the plane normal to the Burgers vector. In the near-sonic regime, we will find a simple expression for the leading contribution to the self-force. It follows from this expression that the self-force precludes the acceleration of dislocations up to the shear wave velocity \cite{Eshelby53,NiMarken08,Blasch21,Pellegr20,GVSD21,HirthLothe,Mura87,Nabar67,Friedel64,HirZbLoth98}.

The paper is organized as follows. In Sec. \ref{Disl_Dynam}, we introduce the notation and elaborate the action principle for a linear elastic medium with dislocations. In Sec. \ref{Disl_Forces}, using the action functional introduced in the previous section, we deduce the elastic forces acting on dislocations. Here we also discuss the imposition of the gliding motion constraint and the corresponding constraint force. Section \ref{Self_Interaction} is devoted to derivation of the explicit expression for the dislocation self-force. Here we particularize our considerations to the case of an isotropic medium. In Sec. \ref{Lor_Force_Sec}, we start with the contribution of the Lorentzian force. Then, in Sec. \ref{Peach_Kohler_f}, we consider the contribution of the Peach-K\"{o}hler force to the self-interaction. The expression for the total elastic self-force is given in Sec. \ref{Total_Self_Force}. In Sec. \ref{Some_Phys_Cons}, we investigate several simple physical consequences of the obtained effective equations of motion. In Conclusion, we summarize the results. Some bulky formulas and notations are removed to Appendices \ref{Useful_Forms_Deriv}, \ref{Useful_Forms_Ints}, \ref{Useful_Forms_Ints_Coeff},  \ref{App_Conctract} and \ref{Useful_Forms_Local}.

\section{Dislocation dynamics}\label{Disl_Dynam}

The equations describing dynamics of small deformations of an elastic medium read as \cite{LLElast}
\begin{equation}\label{eq_equation}
    \rho\ddot{u}_i=\lambda_{ikls}\partial_kw_{ls}+f_i,\qquad f_i=\partial_k\s^e_{ki},
\end{equation}
where $\rho$ is the matter density, $u_i(t,\spx)$ is the displacement vector, the dots represent time derivatives, $w_{ls}(t,\spx)$ is the distortion tensor, which in absence of dislocations is defined as
\begin{equation}\label{wdu}
    w_{ls}:=\partial_lu_s,
\end{equation}
and $f_i(t,\spx)$ is the force density, $\s^e_{ki}=\s^e_{ik}$ is the tensor of external stresses.

The elasticity tensor possesses the following symmetries
\begin{equation}\label{sym_tensor}
    \lambda_{ikls}=\lambda_{kils}=\lambda_{iksl}=\lambda_{lsik}.
\end{equation}
For example, the elasticity tensor of an isotropic medium takes the form
\begin{equation}\label{lambda_isotropic}
    \lambda_{ikls}=\frac{E}{1+\nu} \Big(\delta_{i(l}\delta_{s)k}+\frac{\nu}{1-2\nu}\delta_{ik}\delta_{ls}\Big) =\rho(c_l^2-2c_t^2)\de_{ik}\de_{ls}+2\rho c_t^2\de_{i(l}\de_{s)k},
\end{equation}
where $\nu$ is the Poisson coefficient, $E$ is the Young's modulus, $c_l$ is the velocity of longitudinal sound waves, $c_t$ is the velocity of transverse sound waves,
\begin{equation}
    c_t^2=\frac{E}{2\rho(1+\nu)},\qquad c_l^2=\frac{E(1-\nu)}{\rho(1-2\nu)(1+\nu)},\qquad q^2:=\frac{c_t^2}{c_l^2}=\frac{1-2\nu}{2(1-\nu)}\in(0,3/4),
\end{equation}
the round brackets denote symmetrization with respect to a pair of indices, i.e., $A_{(ij)}:=(A_{ij}+A_{ji})/2$. We suppose that $\rho$ and $\lambda_{ikls}$ are constant. Moreover, we assume that the elastic medium is infinite and does not contain cavities, i.e., it is $\mathbb{R}^3$. The equations of motion \eqref{eq_equation} are obtained by varying the action functional
\begin{equation}\label{action_0}
    S[u]=\int dtd\spx\big[\frac12\rho\dot{u}^2-\frac12\lambda_{ikls}\partial_iu_k\partial_lu_s-\partial_iu_k\s^e_{ik}\big],
\end{equation}
that possesses a transparent physical interpretation: it is the difference between a kinetic energy of motion of an elastic medium and a potential energy (free energy) of the elastic deformations. It should be noted that this action has a global shift symmetry \cite{Fletcher76}
\begin{equation}\label{symm_trans0}
    \de_\epsilon u_i(t,\spx)=\epsilon_i.
\end{equation}
The conservation law of the corresponding Noether current is the momentum conservation law \eqref{eq_equation} (see also the relation of this symmetry to the $J$-integral in \cite{Marken10}).

The relation \eqref{wdu} will not hold, if there are dislocations in an elastic medium. In this case, the system of equations describing the dynamics of an elastic medium is written as \cite{Kos62}
\begin{equation}\label{Kos_systm}
\begin{split}
    \rho\dot{\ups}_i&=\lambda_{ikls}\partial_kw_{ls}+f_i,\\
    \dot{w}_{kn}&=\partial_k\ups_n+I_{kn},\\
    \e_{ijk}\partial_j w_{kn}&=-D_{in}.
\end{split}
\end{equation}
We have introduced the notation
\begin{equation}\label{D_I}
\begin{split}
    D_{in}(t,\spx)&:=b_n\int_C d\la z'_i(t,\la)\de(\spx-\spz(t,\la)),\\
    I_{in}(t,\spx)&:=b_n\e_{ikl} \int_C d\la z'_k(t,\la) \dot{z}_l(t,\la) \de(\spx-\spz(t,\la)),
\end{split}
\end{equation}
where $b_n=const$ is the Burgers vector, $z_i(t,\la)$ defines a dislocation curve $C$, and the prime denotes the derivative with respect to the parameter $\la$. Equations \eqref{Kos_systm} are self-consistent only if
\begin{equation}\label{divD}
    \partial_i D_{in}(t,\spx)=-b_n [\de(\spx-\spz(t,\la_1))-\de(\spx-\spz(t,\la_0))]=0,
\end{equation}
where $\la_{0,1}$ correspond to the ends of the curve $C$. It is clear that the right-hand side of Eq. \eqref{divD} is equal to zero in the case of a closed curve $C$. Henceforward, we suppose that the dislocation curve $C$ is closed. In this case, the conservation law of a Burgers vector,
\begin{equation}\label{b_cons_law}
    \dot{D}_{in}+\e_{ijk}\partial_jI_{kn}=0,
\end{equation}
is valid. If there are several dislocations in the crystal, the expressions \eqref{D_I} will be given by the sum of contributions of each individual dislocation with its own contour $z^i_a(t,\la_a)$ and Burgers vector $b_{an}$.

The factor multiplying the Burgers vector $b_n$ in $D_{in}$ is called the characteristic current of the manifold $C$ \cite{deRham}. If the contour $C$ can be continuously contracted to a point, then it is a boundary of the surface $\Si$ and
\begin{equation}\label{D_Sigma}
    D_{in}(t,\spx)=-b_n\partial_j\int_\Si d\s d\la \partial_\s h_{[j}\partial_\la h_{i]}\de(\spx-\mathbf{h}(t,\s,\la)),
\end{equation}
where the square brackets denote antisymmetrization of indices without the factor $1/2$ and the two-dimen\-sional surface $\Si$ is specified by the embedding map $h^i(t,\s,\la)$. In certain cases, it is more convenient to employ the representation \eqref{D_Sigma} to solve Eqs. \eqref{Kos_systm}. Let
\begin{equation}
    \chi_{in}:=b_n\e_{ijk}\int_\Si d\s d\la \partial_\s h_{j}\partial_\la h_{k}\de(\spx-\mathbf{h}(t,\s,\la)).
\end{equation}
Then
\begin{equation}\label{chi_pot}
    \e_{ijk}\partial_j\chi_{kn}=D_{in}.
\end{equation}
Substituting this relation into \eqref{b_cons_law}, we obtain
\begin{equation}\label{kappa_pot}
    \dot{\chi}_{in}+I_{in}+\partial_i\vk_n=0,
\end{equation}
where
\begin{equation}\label{kappa_pot_expl}
    \vk_n=-\De^{-1}\partial_n(\dot{\chi}_{in}+I_{in}).
\end{equation}
We shall employ this formula below in order to find the variation of the action.

Let us find the distortion tensor and the velocity vector field that are generated exclusively by a dislocation. Combining the equations in the system \eqref{Kos_systm} and assuming $f_i=0$, we obtain
\begin{equation}\label{disl_eqn}
    L_{ij}\ups_j=\la_{ikls}\partial_k I_{ls},\qquad L_{ij}w_{nj}=\rho\dot{I}_{ni}-\la_{ikls}\e_{nlm}\partial_k D_{ms},
\end{equation}
where
\begin{equation}
    L_{ij}:=\rho\de_{ij}\partial_t^2-\la_{iklj}\partial_{kl}.
\end{equation}
The first equation in \eqref{disl_eqn} is derived by differentiating the first equation of the system \eqref{Kos_systm} with respect to $t$ and the second one is obtained by differentiating the first equation of the system \eqref{Kos_systm} with respect to $x_n$. After that we have used the second and the third equations of the system \eqref{Kos_systm}.

Let $G_{ij}$ be a retarded Green's function of the operator $L_{ij}$ in an infinite medium. Then
\begin{equation}\label{Green_func}
    L_{ij}G_{jk}(t,\spx;t',\spy)=\de_{ik}\de(t-t')\de(\spx-\spy).
\end{equation}
Due to the fact that $\la_{iklj}\partial_{kl}$ is real and self-adjoint in the Hilbert space of square integrable vector functions $\ups_i(\spx)$, the following symmetry,
\begin{equation}
    G_{ij}(t,\spx;t',\spy)=G_{ji}(t,\spy;t',\spx),
\end{equation}
is valid. We regularize \eqref{D_I} by replacing the delta function with a certain infinitely differentiable delta sequence
\begin{equation}\label{regul_disl}
    \de_\e(\spx-\spz(t,\la))\underset{\e\rightarrow+0}{\rightarrow} \de(\spx-\spz(t,\la)) ,\qquad \de_\e(\spx-\spz)=\de_\e(\spz-\spx),
\end{equation}
which is a rapidly decreasing function at spatial infinity, for example, a function with a compact support. The conservation laws \eqref{divD}, \eqref{b_cons_law} are preserved by this regularization. The regularization is necessary for a self-consistent description of self-interaction of dislocations. The degree of smearing of the delta function, $\e$, is of the order of $5|\spb|$ (see, e.g., \cite{WeertWeert71}).

We suppose that for $t<t_0$ the tensors $D_{in}(t,\spx)$ and $I_{in}(t,\spx)$ vanish. Furthermore, we suppose that they are infinitely differentiable functions of $t$. Then a particular solution of \eqref{disl_eqn} is written as \cite{Mura87,Pellegr20}
\begin{equation}\label{disl_sol}
\begin{split}
    \ups_i[\spz(\tau,\lambda)]&\equiv\xi_i=G_{ij}\la_{jkls}\partial_k I_{ls}=\partial_kG_{ij}\la_{jkls} I_{ls},\\ w_{ni}[\spz(\tau,\lambda)]&\equiv\eta_{ni}=G_{ij}(\rho\dot{I}_{nj}-\la_{jkls}\e_{nlm}\partial_k D_{ms})=\rho\dot{G}_{ij} I_{nj}-\partial_k G_{ij}\la_{jkls}\e_{nlm} D_{ms},
\end{split}
\end{equation}
where it is implied that $G_{ij}$ acts as a distribution on the corresponding expressions. In the absence of dislocations, this particular solution becomes zero. It is easy to check that this solution satisfies the second and third equations of the system \eqref{Kos_systm}. Then it follows from the method of derivation of Eqs. \eqref{disl_eqn} that
\begin{equation}
    \rho\dot{\xi}_i-\lambda_{ikls}\partial_k\eta_{ls}=const.
\end{equation}
Since $\xi_i(t,\spx)$ and $\eta_{ls}(t,\spx)$ vanish for $t<t_0$, the constant on the right-hand side of the last equation is zero. Notice that in virtue of the assumptions specified above about the behaviour of dislocations, the fields $\xi_i(t,\spx)$ and $\eta_{ni}(t,\spx)$ vanish at the points $(t,\spx)$ lying outside the influence domain of the point $(t_0,\spx_0)$ where the dislocation was created.

We seek for a general solution of the system \eqref{Kos_systm} in the following form
\begin{equation}\label{Kos_systm_sol}
    \ups_i=\xi_i+\ups^0_i,\qquad w_{in}=\eta_{in}+w^0_{in}.
\end{equation}
Then Eqs. \eqref{Kos_systm} are written as
\begin{equation}\label{Kos_systm_1}
    \rho\dot{\ups}^0_i=\lambda_{ikls}\partial_kw^0_{ls}+f_i,\qquad
    \dot{w}^0_{kn}=\partial_k\ups^0_n,\qquad
    \e_{ijk}\partial_j w^0_{kn}=0.
\end{equation}
Since we consider the case where any closed contour can be continuously contracted to a point through the elastic medium, we deduce from the last equation that
\begin{equation}
    w^0_{kn}=\partial_k u_n.
\end{equation}
Supposing $\ups^0_n(t,\spx)\rightarrow0$ for $|\spx|\rightarrow\infty$, it follows from the second equation of the system \eqref{Kos_systm_1} that
\begin{equation}
    \ups^0_{n}=\dot{u}_n,
\end{equation}
whereas the first equation in \eqref{Kos_systm_1} leads to Eqs. \eqref{eq_equation}, \eqref{wdu}. Thus we have shown that \eqref{Kos_systm_sol} is indeed the general solution of the system of equations \eqref{Kos_systm}.

The equations given above do not allow one to find the self-consistent dynamics of dislocations. To describe such dynamics it is necessary to introduce an additional principle. It seems that the most natural principle in the considered case is the variation principle that is a generalization of \eqref{action_0}. To be specific, we suppose that the true trajectories $u_i(t,\spx)$, $z^i_a(t,\la)$ provide the extremum to the functional
\begin{equation}\label{action}
    S[u,z_a]=\int dtd\spx  \big[\frac12\rho(\dot{u}_i+\xi_i)^2-\frac12\lambda_{ikls}(\partial_iu_k+\eta_{ik})(\partial_lu_s+\eta_{ls})-(\partial_iu_k+\eta_{ik})\s^e_{ik}\big],
\end{equation}
where the nonlocal functionals,
\begin{equation}\label{xi_eta}
    \xi_i=G_{ij}\la_{jkls}\sum_a\partial_k I_{ls}[\spy_a(t,\la)],\qquad \eta_{ni}=G_{ij}\sum_a(\rho\dot{I}_{nj}[\spy_a(t,\la)]-\la_{jkls}\e_{nlm}\partial_k D_{ms}[\spy_a(t,\la)]),
\end{equation}
are constructed by the displaced dislocations loops
\begin{equation}\label{shift}
    z_a^i(t,\la)\rightarrow y^i_a(t,\la):= z_a^i(t,\la)-u^i(t,\spy_a(t,\la)).
\end{equation}
The tensors \eqref{xi_eta} satisfy the system \eqref{Kos_systm} with $f_i=0$, where $D_{in}$ and $I_{in}$ are given by the sum of expressions \eqref{D_I} over all the dislocations displaced as in \eqref{shift}. If the deformation of the crystal is static, then the action principle \eqref{action} is just the requirement that the free energy of the system should be minimized.

The expression \eqref{shift} for the arguments of the nonlocal functionals \eqref{xi_eta} is chosen such that the global transformation,
\begin{equation}\label{shift_symm}
    \de_\epsilon u^i(t,\spx)=\epsilon^i,\qquad \de_\epsilon z^i_a(t,\la)=\epsilon^i,
\end{equation}
be a symmetry of the action \eqref{action}, i.e., the system remains unchanged when all the atoms and dislocations are shifted simultaneously by the same vector. Then the symmetry transformation \eqref{shift_symm} gives
\begin{equation}
    \de_\epsilon y^i_a(t,\la)=0.
\end{equation}
The expression \eqref{shift} specifies implicitly the dependence of $y^i_a(t,\la)$ on $z_a^i(t,\la)$ and $u^i(t,\spx)$. For small $\partial_j u^i$, this equation is solvable since in this case
\begin{equation}
    \det\Big(\frac{\partial z^i_a}{\partial y^j_a}\Big)=\det\big(\de^i_j+\partial_j u^i\big)\neq0.
\end{equation}
In the leading order, we have
\begin{equation}
    y^i_a(t,\la)\approx z_a^i(t,\la)-u^i(t,\spz_a(t,\la)).
\end{equation}

The action functional \eqref{action} can be simplified. Let us consider the contributions to \eqref{action} of the form
\begin{equation}
    I:=\int dt d\spx(\rho\dot{u}_i\xi_i-\la_{ikls}\partial_i u_k\eta_{ls}).
\end{equation}
We integrate by parts in the last term. As follows from the asymptotic behaviour of the solution $\eta_{ls}(t,\spx)$ for $|\spx|\rightarrow\infty$ discussed above, the integrated term is zero. Taking into account that $\xi_i$ and $\eta_{ls}$ satisfy the system \eqref{Kos_systm} with $f_i=0$ for arbitrary $z^i_a(t,\la)$, we obtain
\begin{equation}
    I=\int dt d\spx\frac{\partial}{\partial t}(\rho u_i\xi_i)=\int d\spx(\rho u_i\xi_i)\Big|_{t=t_1}^{t=t_2},
\end{equation}
where $t_2\rightarrow+\infty$ and $t_1\rightarrow-\infty$. Since $\xi_i(t,\spx)$ vanishes for $t<t_0$, the contribution from the lower integration limit also vanishes. As for the contribution at $t_2\rightarrow\infty$, we suppose that the elastic medium possesses a small absorption. In that case, $\xi_i(t,\spx)$ tends to zero proportionally to $\exp(-\epsilon t)$ for $t\rightarrow\infty$, where the parameter $\epsilon$ characterizes the absorption. Formally, such a behaviour of the solutions will be achieved, if $\epsilon$ in the $i\epsilon$-prescription for the Fourier transform of the retarded Green's function $G_{ij}$ is a small positive parameter that is to be put to zero in the final answer. As a result, $I=0$ and
\begin{equation}\label{action1}
    S[u,z_a]=\int dtd\spx  \big[\frac12\rho(\dot{u}^2_i+\xi^2_i)-\frac12\lambda_{ikls}(\partial_iu_k \partial_lu_s+\eta_{ik} \eta_{ls})-(\partial_iu_k+\eta_{ik})\s^e_{ik}\big].
\end{equation}

Let us find the equations of motion following from the action \eqref{action}, \eqref{action1}. On varying the action with respect to $z^i_a(t,\la)$, we have
\begin{equation}
    \frac{\de S[u,z_a]}{\de z^i_a(t,\la)}=\frac{\partial y^k_a}{\partial z^i_a}(t,\la)\frac{\de S[u,z]}{\de y^k_a(t,\la)}=0.
\end{equation}
The matrix $\partial y^k_a/\partial z^i_a$ is invertible. Therefore, the equation of motion of the dislocation can be cast into the form
\begin{equation}\label{disl_eq_mot}
    \frac{\de S[u,z_a]}{\de y^i_a(\tau,\la)}=\int dtd\spx  \Big[\rho\xi_k\frac{\de\xi_k}{\de y^i_a(\tau,\la)}-\lambda_{nkls}\eta_{nk}\frac{\de\eta_{ls}}{\de y^i_a(\tau,\la)}-\s^e_{ls}\frac{\de\eta_{ls}}{\de y^i_a(\tau,\la)}\Big]=0.
\end{equation}
Varying the action \eqref{action1} with respect to $u_i$, we deduce
\begin{equation}\label{eq_mot_u}
    \frac{\de S[u,z_a]}{\de u_i(t,\spx)}=\frac{\de S[u,z_a]}{\de u_i(t,\spx)}\Big|_{\spy_a=const} +\int d\tau d\la \sum_a\frac{\de S[u,z_a]}{\de y^j_a(\tau,\la)}\Big|_{\mathbf{u}=const} \frac{\de y^j_a(\tau,\la)}{\de u_i(t,\spx)}=0.
\end{equation}
Using the equations of motion \eqref{disl_eq_mot}, we finally obtain
\begin{equation}
    \frac{\de S[u,z_a]}{\de u_i(t,\spx)}\Big|_{\spy_a=const}=0.
\end{equation}
This equation has the form \eqref{eq_equation} with the substitution \eqref{wdu}. It is clear that
\begin{equation}
    \ups_i:=\xi_i+\dot{u}_i,\qquad w_{in}:=\eta_{in}+\partial_i u_n
\end{equation}
satisfy the system of equations \eqref{Kos_systm}, where $D_{in}$ and $I_{in}$ are given by the sums of the expressions \eqref{D_I} over all the dislocations displaced in accordance with \eqref{shift}.

\section{Forces experienced by a dislocation}\label{Disl_Forces}

The expression \eqref{disl_eq_mot} for the equations of motion of a dislocation can be substantially simplified. Let us consider
\begin{equation}
    I_1:=\int dtd\spx\s^e_{ni}\de\eta_{ni}= \int dtd\spx\s^e_{ni}G_{ij}(\rho\de\dot{I}_{nj}-\la_{jkls}\e_{nlm}\partial_k \de D_{ms}).
\end{equation}
Hereinafter, $D_{in}$ and $I_{in}$ are given by the sums of expressions \eqref{D_I} over all the dislocations displaced as in \eqref{shift} and $\de$ denotes the variation of the expression with respect to $y^i_a(t,\la)$. Substituting \eqref{chi_pot} into $I_1$, we arrive at
\begin{equation}
\begin{split}
    I_1&=\int dtd\spx\s^e_{ni}G_{ij}(\rho\de\dot{I}_{nj}-\la_{jkls}\partial_{kl}\de\chi_{ns} -\la_{jkls}\partial_{kn}\de\chi_{ls})=\\
    &=\int dtd\spx\s^e_{ni}G_{ij}(\rho\de\dot{I}_{nj} +\rho\de\ddot{\chi}_{nj} -L_{js}\de\chi_{ns} -\la_{jkls}\partial_{kn}\de\chi_{ls}).
\end{split}
\end{equation}
Then we employ the relations \eqref{kappa_pot}, \eqref{Green_func} and obtain
\begin{equation}
    I_1=-\int dtd\spx\s^e_{ni}\big[\de\chi_{ni} +G_{ij}(\rho\partial_n\de\dot{\vk}_j+\la_{jkls}\partial_{nk}\de\chi_{ls})\big].
\end{equation}
Accounting for the equations of motion \eqref{eq_mot_u},
\begin{equation}
    \partial_n\s^e_{ni}=L_{ij}u_j,
\end{equation}
and integrating by parts, we come to
\begin{equation}
    I_1=-\int dtd\spx\big[(\s^e_{ni}+\la_{nikl}\partial_ku_l)\de\chi_{ni} +\rho\dot{u}_i\de\vk_i\big].
\end{equation}

Now we consider
\begin{equation}
    I_2:=\int dt d\spx\lambda_{lsni}\eta_{ls}\de\eta_{ni}.
\end{equation}
Repeating the steps as above and taking into account that
\begin{equation}
    \lambda_{lsni}\partial_n\eta_{ls}=\rho\dot{\xi}_i,
\end{equation}
we have
\begin{equation}
    I_2=-\int dtd\spx \big[\lambda_{lsni}\eta_{ls}\de\chi_{ni} -\rho\dot{\xi}_iG_{ij}(\rho\de\dot{\vk}_i +\la_{jkls}\partial_k\de\chi_{ls})\big].
\end{equation}
Let
\begin{equation}
    I_3:=\int dtd\spx\big[\rho\xi_i\de\xi_i -\rho\dot{\xi}_iG_{ij}(\rho\de\dot{\vk}_i +\la_{jkls}\partial_k\de\chi_{ls})\big],
\end{equation}
where the last term in the square brackets is the last contribution in $I_2$. Then, using \eqref{kappa_pot}, we deduce
\begin{equation}
    I_3=-\int dtd\spx\big[\rho\xi_i G_{ij}\lambda_{jkls}\partial_{kl}\de\vk_s +\rho\dot{\xi}_iG_{ij}\de\dot{\vk}_j\big]= \int dtd\spx \rho\xi_i\de\vk_i.
\end{equation}
As a result, the variation of the action functional appearing in the equations of motion \eqref{disl_eq_mot} becomes
\begin{equation}
    \de S[u,z_a]=\int dtd\spx\big[(\s^e_{ni}+\la_{nikl} w_{kl})\de\chi_{ni} +\rho\ups_i\de\vk_i\big].
\end{equation}

Let us represent the total stress tensor $\s_{ni}$ as the sum of longitudinal and transversal parts with respect to the index $n$. To this end, we introduce the auxiliary potentials $A_{si}$ and $\vf_{ni}$ such that
\begin{equation}
    \s_{ni}=\s^e_{ni}+\la_{nikl}w_{kl}=:-\e_{nks}\partial_kA_{si}+\dot{\vf}_{ni},\qquad \vf_{ni}:=\De^{-1}\partial_n(\rho\ups_i).
\end{equation}
Taking the divergence of the right- and left-hand sides of this equation with respect to the index $n$, we reproduce the first equation in the system \eqref{Kos_systm}. Then, employing the expression \eqref{kappa_pot_expl} for $\vk_i$, we obtain
\begin{equation}
    \de S[u,z_a]=\int dtd\spx(\de I_{ni}\vf_{ni}-\de\chi_{ni}\e_{nks}\partial_kA_{si})=\int dtd\spx(\vf_{si}\de I_{si} -A_{si} \de D_{si} ).
\end{equation}
Substituting the explicit expressions for $I_{si}$ and $D_{si}$, it is not difficult to cast the equations of motion of a dislocation \eqref{disl_eq_mot} into the form
\begin{equation}\label{PK_force_reg}
    \mathbf{f}^{\text{PK}}+\mathbf{f}^{\text{L}}=0,
\end{equation}
where
\begin{equation}\label{fPK}
    f^{\text{PK}}_i:=\int d\spx\de_\e(\spy_a(\tau,\la)-\spx) \e_{ikn}\s_{ns}(\tau,\spx)b^s_a y'^k_a(\tau,\la),
\end{equation}
and
\begin{equation}\label{fL}
    f^{\text{L}}_i=-\int d\spx\de_\e(\spy_a(\tau,\la)-\spx) \e_{ikn} \dot{y}_a^k(\tau,\la) y'^n_a(\tau,\la) \rho \xi_s(\tau,\spx)b^s_a.
\end{equation}
The first contribution \eqref{fPK} is the Peach-K\"{o}hler force with the time dependent stress tensor. The second contribution \eqref{fL} is the analog of the Lorentz force with the intensity of the ``magnetic field''
\begin{equation}
    \spy'_a(\tau,\la) \rho \xi_s(\tau,\spx)b^s_a.
\end{equation}
We will refer to this force as the Lorentzian one \cite{Lund98}. Due to invariance of the action with respect to the change of the parameter $\la$ on dislocation curves, both forces are orthogonal to $y'^i_a(\tau,\la)$.

In Eq. \eqref{PK_force_reg}, we single out the contribution of the dislocation $a$ to the stress tensor and the medium velocity
\begin{equation}
    \s_{ns}=:\la_{nskl}\eta^a_{kl}+\tilde{\s}^a_{ns},\qquad \xi_s=:\xi^a_{s}+\tilde{\xi}^a_{s}.
\end{equation}
Then, neglecting the contributions vanishing in the limit $\e\rightarrow+0$, from \eqref{PK_force_reg} we have
\begin{equation}\label{PK_force_fin}
    \mathbf{f}^{\text{sPK}} +\mathbf{f}^{\text{sL}}=-\mathbf{f}^{\text{ext}},
\end{equation}
where
\begin{equation}
\begin{split}
    f^{\text{sPK}}_{i}&:=\int d\spx\de_\e(\spy_a(\tau,\la)-\spx) \e_{ikn}\la_{nsmr}\eta^a_{mr}(\tau,\spx)b^s_a y'^k_a(\tau,\la),\\
    f^{\text{sL}}_{i}&:=-\int d\spx\de_\e(\spy_a(\tau,\la)-\spx) \e_{ikn} \dot{y}_a^k(\tau,\la) y'^n_a(\tau,\la) \rho \xi^a_{s}(\tau,\spx)b^s_a,\\
    f^{\text{ext}}_{i}&:=\big[\e_{ikn}\tilde{\s}^a_{ns}(\tau,\spy_a(\tau,\la)) b^s_a y'^k_a(\tau,\la) -\e_{ikn} \dot{y}_a^k(\tau,\la) y'^n_a(\tau,\la) \rho \tilde{\xi}^a_s(\tau,\spy_a(\tau,\la))b^s_a\big]_{\e=0}.
\end{split}
\end{equation}
The contributions on the left-hand side of Eq. \eqref{PK_force_fin} describe the self-force of a dislocation and diverge on removing the regularization. In particular, these contributions contain the kinetic terms describing the inertial properties of a dislocation. The contribution on the right-hand side of the equation does not depend on $\e$ and describes the external forces acting on the dislocation.

It should be noted that the action functional \eqref{action}, \eqref{action1} and, consequently, Eqs. \eqref{PK_force_fin} are obtained solely within the framework of linear theory of elasticity and do not take into account the microscopic crystal structure. For example, it is known \cite{LLElast,Friedel64,HirthLothe,WeertWeert71} that the dislocation climb is virtually absent at low temperatures and dislocations move only along the slip plane that is spanned by the vectors $b^i_a$ and $y'^i_a(t,\la)$, i.e., the following relation holds true
\begin{equation}\label{constrants}
    \e_{ijk}b^i_a y'^j_a \dot{y}^k_a\equiv0.
\end{equation}
To account for these constraints, it is necessary to add the extra term,
\begin{equation}\label{action_constr}
    S_c[u,z_a,\vk_a]=\int dt d\la\sum_a\vk_a\e_{ijk}b^i_a y'^j_a\dot{y}^k_a,
\end{equation}
to the action \eqref{action}, \eqref{action1}, where $\vk_a(t,\la)$ are the Lagrange multipliers ensuring the fulfilment of the constraints \eqref{constrants} (do not confuse them with $\vk_n$ introduced in \eqref{kappa_pot_expl}). The action \eqref{action_constr} is invariant under a change of the parameter $\la$ on dislocation curves provided $\vk_a(t,\la)$ are transformed as scalars with respect to such transformations. In addition to \eqref{constrants}, the variation of the action \eqref{action_constr} results in the appearance of the constraint reaction force density,
\begin{equation}\label{constr_reac_force}
    f^c_i:=\frac{\de S_c[u,z_a,\vk_a]}{\de y^i_a(\tau,\la)}=\e_{ijk}(\dot{\vk}_ay'^j_a-\vk'_a\dot{y}^j_a)b^k_a,
\end{equation}
on the left-hand side of the equations of motion \eqref{PK_force_fin}. The gauge invariance discussed above implies that the constraint reaction force is orthogonal to $y'^i_a$ on the constraints surface \eqref{constrants}. As is known, the dislocations are affected by the lattice friction force (the Peierls-Nabarro force) and, in the presence of vacancies and interstitial atoms, by the osmotic Bardeen-Herring force \cite{Weert65,LothHirth65,HirthLothe,Mura87,Nabar67,Friedel64}. Besides, the dislocations are subject to the effective drag force due to their interaction with photons, electrons, and other crystal excitations (see, e.g., \cite{Mason66,Nabarro51,AlshInden,KagKravNat74,SwinDud15}). These forces can also be added to Eqs. \eqref{PK_force_fin} when the corresponding effects impact substantially the dislocation dynamics. In that case, the condition \eqref{constrants} is not met and the Bardeen-Herring force replaces the constraint reaction force \eqref{constr_reac_force}.

In the present paper, we consider the simplest case when Eqs. \eqref{PK_force_fin} are fulfilled with the constraints \eqref{constrants} and the constraint reaction force \eqref{constr_reac_force}. Nevertheless, in deriving the explicit expression for the self-force standing on the left-hand side of \eqref{PK_force_fin}, we will not take into account the constraints \eqref{constrants}. Then Eqs. \eqref{PK_force_fin} are readily generalized to the case when the dislocation climb is possible.

\section{Self-force}\label{Self_Interaction}

Let us rewrite the left-hand side of the dislocation equations of motion \eqref{PK_force_fin} in a more convenient form. To this aim, we introduce the regularized Green's function
\begin{equation}
    G^{2\e}_{ij}(t,\spz-\spz'):=\int d\spx d\spy\de_\e(\spz-\spx)G_{ij}(t,\spx-\spy)\de_\e(\spy-\spz').
\end{equation}
Then the left-hand side of \eqref{PK_force_fin} becomes
\begin{equation}\label{self_force}
   f^{\text{self}}_i=f^{\text{sPK}}_i+f^{\text{sL}}_i,
\end{equation}
where
\begin{equation}\label{sPK_sL}
\begin{split}
    f^{\text{sPK}}_i&=\e_{ipn}b^s_a y'^p_a(\tau,\la) \la_{nsmr}\int d t d\spx G^{2\e}_{rj}\big(\tau-t,\spy_a(\tau,\la)-\spx\big) [\rho\dot{I}^a_{mj}(t,\spx)-\la_{jklq}\e_{mll'}\partial_k D^a_{l'q}(t,\spx)],\\
    f^{\text{sL}}_i&=\e_{ipn} y'^p_a(\tau,\la) \dot{y}_a^n(\tau,\la) \rho b^s_a \int d t d\spx G^{2\e}_{sj}\big(\tau-t,\spy_a(\tau,\la)-\spx\big) \la_{jklr}\partial_k I^a_{lr}(t,\spx).
\end{split}
\end{equation}
The expressions for $I^a_{lr}$ and $D^a_{mr}$ have the form \eqref{D_I} with the replacement $\spz(t,\la)\rightarrow \spy_a(t,\la)$, i.e., they do not contain the regularization parameter $\e$ and they are strictly localized on the curve $C_a: \spx=\spy_a(t,\la)$.

Our goal in this section is to derive the asymptotic expansion of the integrals \eqref{sPK_sL} for $\e\rightarrow+0$ and to single out the divergent terms and the finite part of the self-force. In the general case of anisotropic medium, the expression for the Green's function $G_{ij}$ is rather huge. Therefore, in this paper we restrict our considerations to the case of a homogenous isotropic medium \eqref{lambda_isotropic}. Then
\begin{equation}
\begin{split}
    G_{ij}(t,\spx)&=G^\perp_{ij}(t,\spx)+G^\parallel_{ij}(t,\spx)=\\
    &=\frac{\theta(t)}{\rho}\Big[\frac{\de_{ij}-n_in_j}{2\pi c_t}\de(r^2-c_t^2t^2) +\frac{n_i n_j}{2\pi c_l}\de(r^2-c_l^2t^2) +\frac{\de_{ij}-3 n_i n_j}{4\pi r^3}t\big(\theta(r-c_l t) -\theta(r-c_t t) \big) \Big],\\
    G^\perp_{ij}(t,\spx)&=\frac{\theta(t)}{\rho}\int\frac{d\spk}{(2\pi)^3}\frac{\pr_{ij}}{c_t k}\sin(c_tkt) e^{i\spk\spx},\\
    G^\parallel_{ij}(t,\spx)&=\frac{\theta(t)}{\rho}\int\frac{d\spk}{(2\pi)^3}\frac{k_ik_j}{c_lk^3}\sin(c_lkt) e^{i\spk\spx},
\end{split}
\end{equation}
where $k:=|\spk|$ and $\pr_{ij}:=\de_{ij}-k_ik_j/k^2$.

Let us choose the following regularization of the delta function \cite{Pellegr20}
\begin{equation}\label{delta_reg}
    \de_\e(\spx)=\int\frac{d\spk}{(2\pi)^3}e^{i\spk\spx-\e k}.
\end{equation}
Notice that any other spherically-symmetric regularization, $\tilde{\de}_\s(\spx)$, of the delta function can be obtained as
\begin{equation}\label{delta_reg_gen}
    \tilde{\de}_\s(\spx)=\int_0^\infty \frac{d\e}{\s} g\Big(\frac{\e}{\s}\Big)\de_\e(\spx)=\int\frac{d\spk}{(2\pi)^3}e^{i\spk\spx}\int_0^\infty \frac{d\e}{\s} g\Big(\frac{\e}{\s}\Big)e^{-\e k}=:\int\frac{d\spk}{(2\pi)^3}e^{i\spk\spx} g_L(\s\spk),
\end{equation}
where the spherically-symmetric form-factor $g_L(\s\spk)$ is the Laplace image of $g(\e)$. It satisfies the normalization condition
\begin{equation}\label{norm_cond}
    \int_0^\infty d\e g(\e)=1.
\end{equation}
With such a regularization, the Green's function $G^{2\e}_{ij}(t,\spx)$ entering into \eqref{sPK_sL} is to be replaced by
\begin{equation}\label{regul_convol}
    \int_0^\infty\frac{d\e d\e'}{\s^2}g\Big(\frac{\e}{\s}\Big)g\Big(\frac{\e'}{\s}\Big)G^{\e+\e'}_{ij}(t,\spx).
\end{equation}
This formula allows one to generalize easily the expression for the self-force \eqref{self_force} obtained with the use of the regularization \eqref{delta_reg} to the case of an arbitrary regularization of the form \eqref{delta_reg_gen}.

Notice that the use of the spherically-symmetric regularization of the delta function \eqref{delta_reg} does not mean that the dislocation core is assumed to be spherically-symmetric. This regularized delta function is integrated along the dislocation line as in Eq. \eqref{D_I} that leads to a tube-like smearing of an infinitely thin dislocation. The radius of this tube is characterized by the regularization parameter $\varepsilon$. This is the simplest regularization of the dislocation core that preserves all the symmetries of the non-regularized theory and does not introduce additional structures apart from the scalar parameter $\varepsilon$. In particular, Eqs. \eqref{divD} and \eqref{b_cons_law} are fulfilled in the regularized theory. Of course, for \textit{ab initio} modeling of the dislocation core, one needs to know its microscopic structure which depends on the crystal structure of the material, the dislocation type, and the external stress. Nevertheless, as is known (see, e.g., \cite{Friedel64}), the energy of the dislocation core is about $1/5$ or less of the total energy of the dislocation. Therefore, the simplest model of the dislocation core given by the regularization described above is enough to catch the main peculiarities of the dislocation dynamics with self-interaction. The fine structure of the core gives only small corrections to this dynamics. Moreover, as long as the Green's function is actually regularized in evaluating the self-force, the expression for the self-force that will be derived in the next subsections is applicable to the case where the dislocation curves are not smeared but the Green's function is modified in the ultraviolet by a cutoff at the Debye momentum \cite{Ninomiya68}.

For the regularization \eqref{delta_reg}, we have
\begin{equation}\label{Green_func_reg}
    G^\e_{ij}(t,\spx)=\frac{\theta(t)}{\rho}\int\frac{d\spk}{(2\pi)^3}\Big[\frac{\pr_{ij}}{c_t k}\sin(c_tkt) +\frac{k_ik_j}{c_lk^3}\sin(c_lkt) \Big] e^{i\spk\spx-\e k}.
\end{equation}
Let us employ the Mellin representation
\begin{equation}
    e^{-\e k}=\int_C\frac{ds}{2\pi i}\Ga(s)(\e k)^{-s},
\end{equation}
where the integration contour $C$ runs from below upwards parallel to the imaginary axis and it is convenient to choose $\re s\in(0,1)$. Integrating over $\spk$, we obtain
\begin{equation}\label{Green_func_reg_split}
\begin{split}
    G^\e_{ij}(t,\spx)&=G^{\e\perp}_{ij}(t,\spx)+G^{\e \parallel}_{ij}(t,\spx),\\
    G^{\e\perp}_{ij}(t,\spx)&=-\frac{\theta(t)}{8\pi \rho}\int_C\frac{ds}{2\pi i}\frac{\e^{-s}}{\cos(\pi s/2)}\Big[\de_{ij} \frac{d(c_t;s)}{c_t r} -\partial_{ij}\frac{d(c_t;s+2)}{s(s+1)c_tr}\Big],\\
    G^{\e\parallel}_{ij}(t,\spx)&=-\frac{\theta(t)}{8\pi \rho}\int_C\frac{ds}{2\pi i}\frac{\e^{-s}}{\cos(\pi s/2)}\partial_{ij}\frac{d(c_l;s+2)}{s(s+1)c_lr},
\end{split}
\end{equation}
where $r:=|\spx|$ and
\begin{equation}
    d(c;s):=|r+ct|^{s-1}-|r-ct|^{s-1}.
\end{equation}

\subsection{Lorentzian force}\label{Lor_Force_Sec}

The expression for the contribution of the Lorentzian force, $f^{\text{sL}}_i$, to the dislocations self-force can be cast into the form
\begin{equation}\label{sL_1}
    f^{\text{sL}}_i=\e_{ipn} y'^p_a(\tau,\la) \dot{y}_a^n(\tau,\la) \rho b^s_a \int d t d\s \partial_kG^{2\e}_{sj}\big(-t,\spy_a(\tau,\la)-\spy_a(\tau+t,\la+\s)\big) \vf_{jk}(t,\s),
\end{equation}
where
\begin{equation}
\begin{split}
     \vf_{jk}(t,\s):&=\la_{jklr} b^r_a \e_{lmn} y'^m_a(\tau+t,\la+\s) \dot{y}^n_a(\tau+t,\la+\s)=\\
     &=\rho\big[(c_l^2-2c_t^2)\de_{jk} b^r_a\e_{rmn}y'^m_a \dot{y}^n_a +2c_t^2 b_{a(j}\e_{k)mn}y'^m_a \dot{y}^n_a\big](\tau+t,\la+\s).
\end{split}
\end{equation}
In the limit $\e\rightarrow+0$, the divergencies of the contribution \eqref{sL_1} come solely from integration over the neighborhood of the point $(t,\s)=0$. Let us develop the vector-function specifying the dislocation position as a series in the vicinity of this point
\begin{equation}
    \spy_a(\tau+t,\la+\s)-\spy_a(\tau,\la)=\overset{(1)}{\spy}_a+\overset{(2)}{\spy}_a+\cdots=\dot{\spy}_at+\spy'_a\s +\frac12\big(\ddot{\spy}_at^2+2\dot{\spy}'_at\s +\spy''_a\s^2 \big)+\cdots,
\end{equation}
where, in the last expression, all the derivatives of $\spy_a$ are evaluated at the point $(\tau,\la)$. To simplify the further calculations, we suppose that $\la$ is the natural parameter on the curve, viz.,
\begin{equation}
    \spy'_a\spy'_a=1.
\end{equation}
Notice that this condition does not fix completely the gauge freedom of the action functional \eqref{action1}. There are still the gauge transformations of the form
\begin{equation}\label{gauge_trans}
    y^i_a(\tau,\la_a)\rightarrow y^i(\tau,\la_a+\epsilon_a(\tau)),
\end{equation}
where $\epsilon_a(\tau)$ are the gauge transformation parameters. These transformations describe an unphysical (unobservable) flow of a closed dislocation along itself with the velocity whose projection onto the tangent vector of the dislocation curve is constant along this curve. The same gauge freedom arises, for example, in describing the effective dynamics of a closed uniformly charged string \cite{el_str}. Further, we shall use the remaining gauge freedom to simplify the calculations.

Let
\begin{equation}
\begin{split}
    J_i:&=J^\perp_i+J^\parallel_i,\\
    J^{\perp,\parallel}_i:&=\rho\int d t d\s \partial_kG^{2\e\perp,\parallel}_{ij}\big(-t,\spy_a(\tau,\la)-\spy_a(\tau+t,\la+\s)\big) \vf_{jk}(t,\s).
\end{split}
\end{equation}
Denote as $D_{\perp,\parallel}$ the small neighborhoods of the point $(t,\s)=0$ and as $\bar{D}_{\perp,\parallel}$ the remaining integration regions. The precise definitions of the vicinities $D_{\perp,\parallel}$ will be given below. In that case,
\begin{equation}
    J^{\perp,\parallel}_i=J^{\perp,\parallel}_{iD_{\perp,\parallel}}+J^{\perp,\parallel}_{i\bar{D}_{\perp,\parallel}} =\rho\Big(\int_{D_{\perp,\parallel}} + \int_{\bar{D}_{\perp,\parallel}} \Big)d t d\s \partial_kG^{2\e\perp,\parallel}_{ij}\big(-t,\spy_a(\tau,\la)-\spy_a(\tau+t,\la+\s)\big) \vf_{jk}(t,\s).
\end{equation}
It is clear that $J^{\perp,\parallel}_{i\bar{D}_{\perp,\parallel}}$ is finite on removing the regularization, $\e\rightarrow+0$. From now on, we suppose that this limit is evaluated in $J^{\perp,\parallel}_{i\bar{D}_{\perp,\parallel}}$.

Substituting the explicit expression \eqref{Green_func_reg} for the Green's function into $J^{\perp,\parallel}_{iD_{\perp,\parallel}}$, we see by power counting that all the divergences of $J^{\perp,\parallel}_{iD_{\perp,\parallel}}$ are contained in the contribution
\begin{equation}\label{sing_int}
\begin{split}
    \bar{J}^{\perp,\parallel}_i:=&\,\rho\int_{D_{\perp,\parallel}} d t d\s \Big\{\partial_kG^{2\e\perp,\parallel}_{ij}\big(-t,-\overset{(1)}{\spy}_a) \big[\vf_{jk}(0,0) +\dot{\vf}_{jk}(0,0)t +\vf'_{jk}(0,0)\s\big]-\\
    &-\partial_{kl}G^{2\e\perp,\parallel}_{ij}\big(-t,-\overset{(1)}{\spy}_a\big) \vf_{jk}(0,0) \overset{(2)}{y}{}^l_a \Big\}.
\end{split}
\end{equation}
Denote as
\begin{equation}
    \tilde{J}^{\perp,\parallel}_i:=\lim_{\e\rightarrow+0} (J^{\perp,\parallel}_{iD_{\perp,\parallel}}-\bar{J}^{\perp,\parallel}_i).
\end{equation}
By construction, this limit is finite. Therefore, up to the terms vanishing in the limit $\e\rightarrow+0$, we have
\begin{equation}
    J_i=J^{\perp}_{i\bar{D}_{\perp}}+\tilde{J}^{\perp}_i+\bar{J}^{\perp}_i + J^{\parallel}_{i\bar{D}_{\parallel}}+\tilde{J}^{\parallel}_i+\bar{J}^{\parallel}_i.
\end{equation}
Thus we have to find the divergent and finite parts of the integrals $\bar{J}^{\perp,\parallel}_i$.

Introduce
\begin{equation}\label{syst_coords}
    t=-(\vrho/c)\cos\psi,\qquad\s=\vrho\sin\psi,\quad\psi\in[-\pi/2,\pi/2],
\end{equation}
where $c=c_t$ for the domain $D_\perp$ and $c=c_l$ for the domain $D_\parallel$. We also denote as
\begin{equation}\label{notation1}
\begin{gathered}
    \dot{\spy}_a(\tau,\la)\dot{\spy}_a(\tau,\la)/c^2=:\be^2, \qquad\dot{\spy}_a(\tau,\la)\spy'_a(\tau,\la)/c=:\be_\parallel,\\
    \be_k:=\dot{y}^k_a(\tau,\la)/c\qquad\tau_k:=y'^k_a(\tau,\la).
\end{gathered}
\end{equation}
Hereinafter, we assume that $|\spbe|<1$ and use the shorthand notation $\mathbf{b}\equiv\mathbf{b}_a$, $\spy\equiv\spy_a$. It is convenient to choose the domains as
\begin{equation}
    D_{\perp,\parallel}=\{(\varrho,\psi):\varrho\in[0,\La], \psi\in[-\pi/2,\pi/2]\},
\end{equation}
where $\La>0$ is some parameter which has the dimension of length and which is independent of $\vrho$ and $\psi$. It is also supposed that $\La\gg\e$. The domains $D_\perp$ and $D_\parallel$ are different because the different sound velocities $c=c_t$ and $c=c_l$ are used in the respective changes of variables \eqref{syst_coords}. Employing this notation, we can write
\begin{align*}
    \vf_{jk}(0,0)&=c\rho\big[(c_l^2-2c_t^2)\de_{jk}(\mathbf{b}\bs\tau\bs\be)+2c_t^2 b_{(j}\e_{k)mn}\tau_m\be_n \big],\\
    \dot{\vf}_{jk}(0,0)&=c\rho\big[(c_l^2-2c_t^2)\de_{jk}(\mathbf{b}\bs\tau\bs\be)^{\cdot}+2cc_t^2 b_{(j}\e_{k)mn}\be'_m\be_n + 2c_t^2 b_{(j}\e_{k)mn}\tau_m\dot{\be}_n\big],\ttg\label{phi_dphi}\\
    \vf'_{jk}(0,0)&=c\rho\big[(c_l^2-2c_t^2)\de_{jk}(\mathbf{b}\bs\tau\bs\be)'+2c_t^2 b_{(j}\e_{k)mn}\tau'_m\be_n + 2c_t^2 b_{(j}\e_{k)mn}\tau_m\be'_n\big],
\end{align*}
where the triple product is defined as usual
\begin{equation}
    (\mathbf{b},\bs\tau,\bs\be)\equiv(\mathbf{b}\bs\tau\bs\be):=\e_{ijk}b_i\tau_j\be_k.
\end{equation}
Changing the variables as in \eqref{syst_coords}, we obtain
\begin{equation}
    n_k:=-\frac{\overset{(1)}{y}{}_a^k}{r}=\frac{\be_k\cos\psi-\tau_k\sin\psi}{\bar{r}},
\end{equation}
where $\bar{r}$ is defined in \eqref{b_quants_not}.

The gauge transformation \eqref{gauge_trans} alters the vectors in the self-force as follows
\begin{equation}
\begin{gathered}
    \be_i\rightarrow\be_i+\tau_i \dot{\epsilon}/c,\qquad \tau_i\rightarrow\tau_i,\qquad \dot{\be}_i\rightarrow\dot{\be}_i +2\be'_i\dot{\epsilon} +\tau'_i \dot{\epsilon}^2/c +\tau_i\ddot{\epsilon}/c,\\
    \be'_i\rightarrow\be'_i+\tau'_i\dot{\epsilon}/c,\qquad\tau'_i\rightarrow\tau'_i.
\end{gathered}
\end{equation}
Let us fix the coordinates of the point $(\tau,\la)$ where the self-force is calculated and choose the parameter of the gauge transformation in the form
\begin{equation}
    \epsilon(\tau)=0,\qquad\dot{\epsilon}(\tau)=-c\be_\parallel,\qquad \ddot{\epsilon}(\tau)=-c\dot{\be}_\parallel +c^2\be_\parallel\be'_\parallel+c^2(\bs\be_\perp\bs\be'_\perp),
\end{equation}
where the expressions on the right-hand sides are evaluated at the point $(\tau,\la)$. The dot and prime denote the derivatives with respect to the first and the second arguments of the functions, respectively. Furthermore,
\begin{equation}
    \be^i_\perp:=\be^i-\be_\parallel\tau^i.
\end{equation}
The explicit form of the higher order derivatives of $\epsilon(\tau)$ at the point $\tau$ is not important for the further analysis. Such a gauge transformation does not move the point $(\tau,\la)$ while the derivatives of the embedding map specifying the dislocation position change as
\begin{equation}\label{gauge_trans_1}
\begin{gathered}
    \be^i\rightarrow\be^i_\perp,\qquad \tau^i\rightarrow\tau^i,\qquad \dot{\be}^i\rightarrow D\be_\perp^i+c(\bs\be_\perp\bs\be'_\perp)\tau^i,\\
    \be'^i\rightarrow \be'^i_\perp+\be'_\parallel\tau^i,\qquad \tau'_i\rightarrow\tau'_i.
\end{gathered}
\end{equation}
where we have introduced the covariant derivative
\begin{equation}
    D:=\partial_\tau-c\be_\parallel\partial_\la.
\end{equation}
The following relations hold
\begin{equation}\label{gauge_rels}
    \bs\tau^2=1,\qquad\bs\tau\bs\be'=0,\qquad\bs\tau\bs\tau'=0,\qquad (\bs\tau D\bs\be_\perp)=-c(\bs\be_\perp\bs\be'_\perp),\qquad (\bs\tau\bs\be'_\perp)=-(\spt'\spbep)=-\be'_\parallel.
\end{equation}
It is convenient to perform the gauge transformation \eqref{gauge_trans_1} not in the original expression \eqref{sing_int}, \eqref{phi_dphi} but when all the contractions listed in Appendix \ref{App_Conctract} have been calculated. Then
\begin{equation}\label{gauge_trans_cons}
\begin{gathered}
    \btbe^\cdot\rightarrow D\btbep,\qquad \btbe'\rightarrow \btbep',\qquad (\spbe\spt\dot{\spbe})\rightarrow (\spbep\spt D\spbep), \qquad (\spt\spbe'\spbe)\rightarrow (\spt\spbep'\spbep),\\
    (\spb\dot{\spbe})\rightarrow (\spb D\spbep)+c(\spbep\spbep')(\spb\spt), \qquad
    (\spb \spbe')\rightarrow (\spb \spbep') -(\spt\spbep') (\spb\spt),\\
    (\spbe\dot{\spbe})\rightarrow (\spbep D\spbep),\qquad (\spbe\spbe')\rightarrow (\spbep \spbep'),\qquad (\spt\dot{\spbe})\rightarrow 0.
\end{gathered}
\end{equation}

\paragraph{First contribution.}

Now we substitute \eqref{Green_func_reg_split} into \eqref{sing_int}. Then the power counting reveals that the leading contribution at $\e\rightarrow+0$ is given by the integrals (see the notation in Appendices \ref{Useful_Forms_Deriv}, \ref{Useful_Forms_Ints})
\begin{align*}
    V_k(s):=&\,\frac{(2\e)^{-s}}{8\pi c\cos(\pi s/2)}\int_{-\pi/2}^{\pi/2}d\psi\int_0^\La d\vrho \vrho \partial_kf(s)= \frac{(2\e)^{-s}\La^{s-1}}{8\pi c(s-1)\cos(\pi s/2)}\big[\be_k N^{10}(s) -\tau_k N^{01}(s)\big],\ttg\label{V_k_V_kij_0}\\
    V_{kij}(s):=&\,\frac{(2\e)^{-s}}{8\pi cs(s+1)\cos(\pi s/2)}\int_{-\pi/2}^{\pi/2}d\psi\int_0^\La d\vrho \vrho \partial_{ijk}f(s+2)=\\
    =&\,\frac{(2\e)^{-s}\La^{s-1}}{8\pi c(s-1)s(s+1)\cos(\pi s/2)}\big[(\de_{ij}\be_k+\cycle(i,j,k)) K^{10}(s) -(\de_{ij}\tau_k+\cycle(i,j,k)) K^{01}(s)+\\
    &+\be_i\be_j\be_k L^{30}(s) -(\be_i\be_j\tau_k +\cycle(i,j,k)) L^{21}(s) +(\be_i\tau_j\tau_k +\cycle(i,j,k)) L^{12}(s) -\tau_i\tau_j\tau_k L^{03}(s)\big],
\end{align*}
where it is necessary to move the integration contour in the $s$ plane to the right up to the line $\re s=1+0$ before the integration with respect to the variable $\vrho$. The integral $V_k$ originates from the first term in expression \eqref{Green_func_reg_split} for the transversal part of the Green's function and it is contracted with $\vf_{ik}(0,0)$. The integral $V_{kij}$ describes the contribution of the second term in the transversal Green's function and the contribution of the longitudinal part of the Green's function. It is contracted with $\vf_{jk}(0,0)$. In the expressions \eqref{V_k_V_kij_0}, we have introduced the notation \eqref{N_K_L_def} and have employed the relations \eqref{partial_f} and \eqref{K_L_M}. Owing to the fact that the gauge transformation \eqref{gauge_trans_1} eliminates the longitudinal projection the velocity, $\be_\parallel$, at the given point, the integrals $D^{kn}_l(s)$, $H^{kn}_l(s)$ and, consequently, the functions \eqref{N_K_L_def} vanish when $n$ is an odd number. Hence
\begin{equation}\label{V_k_V_kij}
\begin{split}
    V_k(s):=&\,\frac{(2\e)^{-s}\La^{s-1}}{8\pi c(s-1)\cos(\pi s/2)} \be_k N^{10}(s),\\
    V_{kij}(s):=&\,\frac{(2\e)^{-s}\La^{s-1}}{8\pi c(s-1)s(s+1)\cos(\pi s/2)}\big[(\de_{ij}\be_k+\cycle(i,j,k)) K^{10}(s) +\be_i\be_j\be_k L^{30}(s)+\\
    &+(\be_i\tau_j\tau_k +\cycle(i,j,k)) L^{12}(s) \big].
\end{split}
\end{equation}

In order to obtain the asymptotic expansion of the contributions \eqref{V_k_V_kij} to the integral \eqref{sing_int} as $\e\rightarrow+0$, it is necessary to move the integration contour with respect to $s$ to the line $\re s=-0$ and to take into account the singularities of the integrand in the strip swept by the contour \cite{GSh,ParKam01,Wong,KalKaz3}. As follows from the explicit expressions \eqref{DH_analyt}, \eqref{tilde_c} for the analytical continuation of the functions $D^{kn}(s)$ and $H^{kn}(s)$, they tend to infinity not faster than some power of $s$ for $|\im s|\rightarrow\infty$. Therefore, $V_k(s)$ and $V_{kij}(s)$ tend exponentially to zero as $|\im s|\rightarrow\infty$. This allows us to neglect the contributions of the segments of the integration contour at $\im s=\pm\infty$ when we move it along the real axis. On shifting the integration contour to the line $\re s=-0$, we are left with the contributions of the integrand poles located in the strip that is swept by the contour and also with the integral along the shifted contour $\re s=-0$. It is not difficult to see that the last integral tends to zero as $\e\rightarrow+0$.

The singularities of $V_k(s)$ and $V_{kij}(s)$ in the strip $\re s\in[0,1]$ for finite $s$ are located only at the points  $s=0$ and $s=1$. It turns out (see \eqref{rels1}, \eqref{rels2}) that the terms of the series for $K^{kn}(s)$ near the point $s=1$ that are given in \eqref{K_L_expans} vanish for $(k,n)$ equal to $(1,0)$ and $(0,1)$. As for $L^{kn}(s)$, the terms of the series in the vicinity of the point $s=1$ given in \eqref{K_L_expans} are zero for $(k,n)$ equal to $(3,0)$, $(2,1)$, $(1,2)$, and $(0,3)$. It is an expected result since the divergences must be logarithmic and local in the case at hand. The last condition implies, in particular, that there should not be any divergent contributions depending on $\La$. As a result, in the limit  $\e\rightarrow+0$, the nonvanishing contributions of the integrals \eqref{V_k_V_kij} to $\bar{J}_{\perp,\parallel}$ come from the pole at the point $s=0$ and take the form
\begin{equation}\label{V_k_V_kij_fin}
\begin{split}
    r_k:=\res_{s=0}V_k=&\,-\frac{1}{8\pi c\La} \be_k \overset{(-1)}{N}{}^{10}(0),\\
    r_{kij}:=\res_{s=0}V_{kij}=&\,-\frac{1}{8\pi c\La} \big[(\de_{ij}\be_k+\cycle(i,j,k)) K^{10}(0)+\\
    &+\be_i\be_j\be_k L^{30}(0)+(\be_i\tau_j\tau_k +\cycle(i,j,k)) L^{12}(0)\big].
\end{split}
\end{equation}
As we see, the integrals \eqref{V_k_V_kij} contribute only to the finite part of the expression \eqref{sing_int}.

The sum of the transversal and longitudinal contributions is written as
\begin{equation}
    \bar{J}_i^{(1)}=c_t^{-1}[r_k\vf_{ik}-r_{kij}\vf_{jk}]_{c\rightarrow c_t}+c_l^{-1}r_{kij}\vf_{jk}|_{c\rightarrow c_l},
\end{equation}
where $\vf_{jk}\equiv\vf_{jk}(0,0)$. The self-force contains the contraction
\begin{equation}\label{bJ1}
    b_i\bar{J}_i^{(1)}=-\frac{\rho}{8\pi\La}\Big\{(\spb\spbep)\btbep \frac{a_1}{c}\Big|_{c\rightarrow c_t} - (\spb\spbep)\btbep \frac{\bar{a}_1}{c}\Big|_{c\rightarrow c_l}\Big\},
\end{equation}
where, for brevity, we have introduced the coefficients $a_1$ and $\bar{a}_1$ (see \eqref{a16}, \eqref{a1ba1}). The coefficients $a_1$ and $\bar{a}_1$ depend only on $\spbep^2$ and on the sound velocities $c_{t,l}$. In deriving \eqref{bJ1}, we have used the relations \eqref{recurr_rels} and $L^{kn}_{-2}$ denotes the expression $L^{kn}$ defined in \eqref{N_K_L_def} where all the functions $D^{kn}_l$ and $H^{kn}_l$ are taken with the shifted index $l\rightarrow l-2$.

\paragraph{Second contribution.}

Now we consider the contributions of higher order expansion terms in \eqref{sing_int}. We begin with the terms linear in $t$ and $\s$ that are multiplied by $\dot{\vf}_{ij}(0,0)$ and $\vf'_{ij}(0,0)$, respectively. Discarding the contributions of the integrals $D^{kn}_l$ and $H^{kn}_l$ with odd $n$ since they are equal to zero, we obtain (see \eqref{V_k_V_kij_0})
\begin{align*}
    V^t_k(s):=&\,\frac{-(2\e)^{-s}}{8\pi c^2\cos(\pi s/2)}\int_{-\pi/2}^{\pi/2}d\psi\int_0^\La d\vrho \vrho^2\cos\psi \partial_kf(s)= \frac{-(2\e)^{-s}\La^{s}}{8\pi c^2s\cos(\pi s/2)} \be_k N^{20}(s),\\
    V^\s_k(s):=&\,\frac{(2\e)^{-s}}{8\pi c\cos(\pi s/2)}\int_{-\pi/2}^{\pi/2}d\psi\int_0^\La d\vrho \vrho^2\sin\psi \partial_kf(s)= \frac{-(2\e)^{-s}\La^{s}}{8\pi cs\cos(\pi s/2)}\tau_k N^{02}(s),\\
    V^t_{kij}(s):=&\,\frac{-(2\e)^{-s}}{8\pi c^2s(s+1)\cos(\pi s/2)}\int_{-\pi/2}^{\pi/2}d\psi\int_0^\La d\vrho \vrho^2\cos\psi \partial_{ijk}f(s+2)=\\
    =&\,\frac{-(2\e)^{-s}\La^s}{8\pi c^2s^2(s+1)\cos(\pi s/2)}\big[(\de_{ij}\be_k+\cycle(i,j,k)) K^{20}(s) +\ttg\label{Vt_k_Vt_kij}\\
    &+\be_i\be_j\be_k L^{40}(s)  +(\be_i\tau_j\tau_k +\cycle(i,j,k)) L^{22}(s) \big],\\
    V^\s_{kij}(s):=&\,\frac{(2\e)^{-s}}{8\pi cs(s+1)\cos(\pi s/2)}\int_{-\pi/2}^{\pi/2}d\psi\int_0^\La d\vrho \vrho^2\sin\psi \partial_{ijk}f(s+2)=\\
    =&\,\frac{-(2\e)^{-s}\La^s}{8\pi cs^2(s+1)\cos(\pi s/2)}\big[ (\de_{ij}\tau_k+\cycle(i,j,k)) K^{02}(s)+\\
    & +(\be_i\be_j\tau_k +\cycle(i,j,k)) L^{22}(s)  +\tau_i\tau_j\tau_k L^{04}(s)\big],
\end{align*}
where the integration contour of the variable $s$ runs along the line $\re s=+0$. Using the expansions \eqref{eps_La_exp_0}, in the limit $\e\rightarrow+0$, the nonvanishing contributions to the self-force coming from the integrals \eqref{Vt_k_Vt_kij} are written as
\begin{equation}\label{Vt_k_Vt_kij_fin}
\begin{split}
    r_k^t:=\res_{s=0}V^t_k=&\,\frac{1}{8\pi c^2} \be_k n^{20},\\
    r_k^\s:=\res_{s=0}V^\s_k=&\,\frac{1}{8\pi c} \tau_k n^{02},\\
    r_{kij}^t:=\res_{s=0}V^t_{kij}=&\,\frac{1}{8\pi c^2} \big[(\de_{ij}\be_k+\cycle(i,j,k)) k^{20}
     +\be_i\be_j\be_k l^{40}+
    (\be_i\tau_j\tau_k +\cycle(i,j,k)) l^{22}\big],\\
    r_{kij}^\s:=\res_{s=0}V^\s_{kij}=&\,\frac{1}{8\pi c} \big[(\de_{ij}\tau_k+\cycle(i,j,k)) k^{02}+(\be_i\be_j\tau_k +\cycle(i,j,k)) l^{22} +\tau_i\tau_j\tau_k l^{04}\big],
\end{split}
\end{equation}
where we have introduced the notation \eqref{n_k_l}. As we see, these contributions contain logarithmic divergences.

The sum of the transversal and longitudinal contributions to \eqref{sing_int} is given by
\begin{equation}
    \bar{J}_i^{(2)}=c_t^{-1}[r^t_k\dot{\vf}_{ik}-r^t_{kij}\dot{\vf}_{jk} +r^\s_k \vf'_{ik}-r^\s_{kij} \vf'_{jk}]_{c\rightarrow c_t}
    +c_l^{-1}[r^t_{kij}\dot{\vf}_{jk} +r^\s_{kij} \vf'_{jk}]_{c\rightarrow c_l},
\end{equation}
where $\dot{\vf}_{jk}$ and $\vf'_{jk}$ are evaluated at the origin of the system of coordinates. The self-force depends on the contraction
\begin{align*}
    b_i\bar{J}_i^{(2)}=&\,\frac{\rho}{8\pi}\Big\{D\btbep\frac{(\spb\spbep)}{c^2} a_2+\btbep'\frac{(\spb\spt)}{c} a_3-\\
    &-2c_t^2\frac{(\spbep\spt D\spbep)}{c^2}\big[\spb^2(k^{20}-n^{20}/2) +(\spb\spt)^2 l^{22} +(\spb\spbep)^2 l^{40}\big]-\ttg\label{bJ2}\\
    &-2c_t^2\frac{(\spt\spt' \spbep)}{c}\big[\spb^2(k^{02}-n^{02}/2) +(\spb\spbep)^2 l^{22} +(\spb\spt)^2 l^{04}\big] -8c_t^2 (\spt\spbep'\spbep) \frac{(\spb\spbep)(\spb\spt)}{c}l^{22} \Big\}_{c\rightarrow c_t}-\\
    &-\frac{\rho}{8\pi}\Big\{\cdots\Big\}_{c\rightarrow c_l,n^{kn}\rightarrow0}.
\end{align*}
The coefficients $a_{2,3}$ are defined in \eqref{a16}. The omission marks in the curly brackets in \eqref{bJ2} describe the contribution of the longitudinal part of the Green's function and denotes the same expression as in the braces above but with the prescribed replacements. This contribution contains the coefficients $\bar{a}_{2,3}$. The coefficients $a_{2,3}$ and $\bar{a}_{2,3}$ depend on $\spbep^2$ and the sound velocities $c_{t,l}$. The explicit expressions for the divergences containing in the coefficients $a_{2,3}$ and $\bar{a}_{2,3}$ are given in \eqref{a_123456_sing}.

\paragraph{Third contribution.}

It remains for us to calculate the asymptotics of the most bulky contribution that comes from the term on the second line of \eqref{sing_int}. Introduce the notation
\begin{equation}\label{V_kl_V_klij}
\begin{split}
    V^{ab}_{kl}(s):=&\,\frac{-(2\e)^{-s}}{16\pi c\cos(\pi s/2)}\int_{-\pi/2}^{\pi/2}d\psi\int_0^\La d\vrho \vrho^3\cos^a\psi\sin^b\psi \partial_{kl}f(s),\\
    V^{ab}_{klij}(s):=&\,\frac{-(2\e)^{-s}}{16\pi cs(s+1)\cos(\pi s/2)}\int_{-\pi/2}^{\pi/2}d\psi\int_0^\La d\vrho \vrho^3\cos^a\psi\sin^b\psi \partial_{ijkl}f(s+2),
\end{split}
\end{equation}
where $(a,b)$ take values $(2,0)$, $(1,1)$, and $(0,2)$. Then we use formulas \eqref{partial_f}, \eqref{K_L_M} and obtain
\begin{align*}
    V^{ab}_{kl}(s):=&\,\frac{-(2\e)^{-s}\La^{s}}{16\pi cs\cos(\pi s/2)}\Big\{\de_{kl}N^{ab}(s) +\be_k\be_l K^{a+2,b}(s-2) -2 \be_{(k}\tau_{l)} K^{a+1,b+1}(s-2) +\tau_k\tau_l K^{a,b+2}(s-2)\Big\},\\
    V^{ab}_{klij}(s):=&\,\frac{-(2\e)^{-s}\La^{s}}{16\pi cs^2(s+1)\cos(\pi s/2)} \Big\{(\de_{ij}\de_{kl}+\cycle(i,j,k)) K^{ab}(s)+\ttg\label{V_kl_V_klij_expl}\\
    &+(\de_{ij}\be_k\be_l+\be_i\be_j\de_{kl} +\cycle(i,j,k)) L^{a+2,b}(s)-2(\de_{ij}\be_{(k}\tau_{l)}+\be_{(i}\tau_{j)}\de_{kl}+\cycle(i,j,k)) L^{a+1,b+1}(s)+\\
    &+(\de_{ij}\tau_k\tau_l +\tau_i\tau_j\de_{kl} +\cycle(i,j,k)) L^{a,b+2}(s) +\be_i\be_j\be_k\be_l M^{a+4,b}(s)-\\
    &-(\be_i\be_j\be_k\tau_l +\cycle(i,j,k,l)) M^{a+3,b+1}(s) +(\be_i\be_j\tau_k\tau_l +\tau_i\tau_j\be_k\be_l +\cycle(i,j,k)) M^{a+2,b+2}(s)-\\
    &-(\be_i\tau_j\tau_k\tau_l +\cycle(i,j,k,l))M^{a+1,b+3}(s) +\tau_i\tau_j\tau_k\tau_l M^{a,b+4}(s)\Big\},
\end{align*}
where the function $M^{kn}(s)$ is defined in \eqref{N_K_L_def}. The expansion of this function in the vicinity of the point $s=0$ is given in formula \eqref{K_L_expans}. In the limit $\e\rightarrow+0$, the nonvanishing contributions to the self-force stem from the residues of the expressions \eqref{V_kl_V_klij_expl} at the point $s=0$. Employing the expansions \eqref{N_expans}, \eqref{K_L_expans}, and \eqref{eps_La_exp_0}, and the notation \eqref{n_k_l}, we arrive at
\begin{align*}
    r^{ab}_{kl}:=\res_{s=0}V^{ab}_{kl}=&\,\frac{1}{16\pi c} \big[\de_{kl} n^{ab} +\be_k\be_l \tilde{k}^{a+2,b}
    -2\be_{(k}\tau_{l)} \tilde{k}^{a+1,b+1}
    +\tau_{k}\tau_{l} \tilde{k}^{a,b+2}  \big],\ttg\label{V_kl_V_klij_fin}\\
    r^{ab}_{klij}:=\res_{s=0}V^{ab}_{klij}=&\,\frac{1}{16\pi c} \big[(\de_{ij}\de_{kl}+\cycle(i,j,k)) k^{ab}
    +(\de_{ij}\be_k\be_l+\be_i\be_j\de_{kl} +\cycle(i,j,k)) l^{a+2,b}-\\
    &-2(\de_{ij}\be_{(k}\tau_{l)}+\be_{(i}\tau_{j)}\de_{kl}+\cycle(i,j,k)) l^{a+1,b+1}
    +(\de_{ij}\tau_k\tau_l +\tau_i\tau_j\de_{kl} +\cycle(i,j,k)) l^{a,b+2}+\\
    &+\be_i\be_j\be_k\be_l m^{a+4,b}
    -(\be_i\be_j\be_k\tau_l +\cycle(i,j,k,l)) m^{a+3,b+1}+\\
    &+(\be_i\be_j\tau_k\tau_l +\tau_i\tau_j\be_k\be_l +\cycle(i,j,k)) m^{a+2,b+2}
    -(\be_i\tau_j\tau_k\tau_l +\cycle(i,j,k,l)) m^{a+1,b+3} +\\
    &+\tau_i\tau_j\tau_k\tau_l m^{a,b+4}\big].
\end{align*}
These contributions contain logarithmic divergences as well.

The sum of the transversal and longitudinal contributions to \eqref{sing_int} has the form
\begin{equation}
\begin{split}
    \bar{J}_i^{(3)}=&\,c_t^{-1}\big[c^{-1}r^{20}_{kl}\dot{\be}_l -2r^{11}_{kl}\be'_l +r^{02}_{kl}\tau'_l\big] \vf_{ik}\Big|_{c\rightarrow c_t} -c_t^{-1}\big[c^{-1}r^{20}_{klij}\dot{\be}_l -2r^{11}_{klij}\be'_l +r^{02}_{klij}\tau'_l\big]\vf_{jk}\Big|_{c\rightarrow c_t}+\\
    &+c_l^{-1}\big[c^{-1}r^{20}_{klij}\dot{\be}_l -2r^{11}_{klij}\be'_l +r^{02}_{klij}\tau'_l\big]\vf_{jk} \Big|_{c\rightarrow c_l},
\end{split}
\end{equation}
where $\vf_{jk}\equiv\vf_{jk}(0,0)$. The minus sign before the factor $\bs{\be}'$ and the factors $c^{-1}$ originate from the change of variables \eqref{syst_coords}. The contraction with the Burgers vector is given by
\begin{align*}
    b_i\bar{J}_i^{(3)}=&\,\frac{\rho}{16\pi}\Big\{ \btbep \Big[ \frac{(\spb D\spbep)}{c^2} a_2 + \frac{(\spb \spt')}{c} a_3 +\frac{(\spbep\spbep')}{c}(\spb\spt) a_4 +\Big(\frac{(\spbep D\spbep)}{c^2} a_5+\frac{(\spbep \spt')}{c}a_6 \Big)(\spb\spbep) \Big]+\\
    &+2c_t^2\frac{(\spbep\spt D\spbep)}{c^2}\big[\spb^2(k^{20}-n^{20}/2) +(\spb\spt)^2 l^{22} +(\spb\spbep)^2 l^{40}\big]+\ttg\label{bJ3}\\
    &+2c_t^2\frac{(\spt\spt' \spbep)}{c}\big[\spb^2(k^{02}-n^{02}/2) +(\spb\spbep)^2 l^{22} +(\spb\spt)^2 l^{04}\big] +8c_t^2 (\spt\spbep'\spbep) \frac{(\spb\spbep)(\spb\spt)}{c}l^{22}\Big\}-\\
    &-\frac{\rho}{16\pi}\Big\{\cdots\Big\}_{c\rightarrow c_l,n^{kn}\rightarrow0,\tilde{k}^{kn}\rightarrow0},
\end{align*}
where we have introduced the coefficients $a_{4,5,6}$ and $\bar{a}_{4,5,6}$ (see \eqref{a16}) that depend only on $\spbep^2$ and the sound velocities $c_{t,l}$. As above, the ellipsis in the braces in \eqref{bJ3} describe the contribution of the longitudinal part of the Green's function and denote the same expression as in the curly brackets above but with the corresponding replacements. This contribution contains the coefficients $\bar{a}_{4,5,6}$. The explicit expressions for the divergent terms in the coefficients $a_{4,5,6}$ and $\bar{a}_{4,5,6}$ are given in \eqref{a_123456_sing}.

\paragraph{Sum.}

Eventually, the divergent and finite parts of the sum of the longitudinal and transversal contributions \eqref{sing_int} is written as
\begin{equation}\label{bJtot}
    b_i\bar{J}_i=b_i\bar{J}_i^{(1)}+b_i\bar{J}_i^{(2)}+b_i\bar{J}_i^{(3)},
\end{equation}
where the explicit expressions for the terms on the right-hand side are given in the formulas \eqref{bJ1}, \eqref{bJ2}, and \eqref{bJ3}. Then
\begin{equation}\label{sL_tot}
    \mathbf{f}^{\text{sL}}=c[\spt\spbep]b_i(\bar{J}_i+J^{\perp}_{i\bar{D}_{\perp}}+\tilde{J}^{\perp}_i + J^{\parallel}_{i\bar{D}_{\parallel}}+\tilde{J}^{\parallel}_i).
\end{equation}
The first contribution in this expression is local, i.e., it depends on the embedding map $\mathbf{z}(t,\la)$ and its derivatives up to the second order taken at the point where the self-force is evaluated. The local contribution has the form
\begin{align*}
    b_i\bar{J}_i=&\,\frac{\rho}{16\pi}\Big\{-\frac{2}{\La} \frac{(\spb\spbep)}{c}\btbep a_1 +\Big[\btbep(\spb D\spbep) +2D\btbep (\spb\spbep)\Big] \frac{a_2}{c^2}+\\
    &+ \Big[ \btbep(\spb \spt') +2\btbep' (\spb\spt)\Big] \frac{a_3}{c} +\btbep \Big[\frac{(\spbep\spbep')}{c}(\spb\spt) a_4+\\
    &+\Big(\frac{(\spbep D\spbep)}{c^2} a_5+\frac{(\spbep \spt')}{c}a_6 \Big)(\spb\spbep) \Big]-\ttg\label{bJtot1}\\
    &-2c_t^2\frac{(\spbep\spt D\spbep)}{c^2}\big[\spb^2(k^{20}-n^{20}/2) +(\spb\spt)^2 l^{22} +(\spb\spbep)^2 l^{40}\big]-\\
    &-2c_t^2\frac{(\spt\spt' \spbep)}{c}\big[\spb^2(k^{02}-n^{02}/2) +(\spb\spbep)^2 l^{22} +(\spb\spt)^2 l^{04}\big] -8c_t^2 (\spt\spbep'\spbep) \frac{(\spb\spbep)(\spb\spt)}{c}l^{22}\Big\}_{c\rightarrow c_t}-\\
    &-\frac{\rho}{16\pi}\Big\{\cdots\Big\}_{c\rightarrow c_l,\overset{(-1)}{N}{}^{10}(0)\rightarrow0,n^{kn}\rightarrow0,\tilde{k}^{kn}\rightarrow0}.
\end{align*}
The divergences contained in this expression give the leading order contribution to the Lorentzian force \eqref{sL_tot}. The explicit expressions for these divergences are given in \eqref{a_123456_sing}.

\subsection{Peach-K\"{o}hler force}\label{Peach_Kohler_f}

Now we consider the contribution to the dislocation self-force given by the Peach-K\"{o}hler force. This contribution can be represented in the form
\begin{equation}\label{sPK_1}
    f^{\text{sPK}}_i=\e_{ipn}\tau_p (P^1_n+P^2_n)=[\spt\mathbf{P}^1]_i+[\spt\mathbf{P}^2]_i,
\end{equation}
where
\begin{equation}
\begin{split}
    P^1_n&=b_i\int d t d\s \partial_\tau G^{2\e}_{ij}\big(-t,\spy_a(\tau,\la)-\spy_a(\tau+t,\la+\s)\big) \rho\chi_{nj}(t,\s),\\
    P^2_n&=-\int d t d\s \partial_k G^{2\e}_{ij}\big(-t,\spy_a(\tau,\la)-\spy_a(\tau+t,\la+\s)\big) \rho\chi_{nkij}(t,\s).
\end{split}
\end{equation}
The derivative $\partial_\tau$ denotes the partial derivative with respect to the first argument of the Green's function. We have introduced the notation
\begin{equation}\label{chi_expr}
\begin{split}
    \chi_{nj}(t,\s):=&\,cb_s\la_{nsmj}\e_{mkl}\tau_k\beta_l=c\rho\big[(c_l^2-2c_t^2)b_n[\spt\spbe]_j +c_t^2 b_j[\spt\spbe]_n +c_t^2\de_{nj}(\spb\spt\spbe) \big],\\
    \chi_{nkij}(t,\s):=&\,b_s\la_{nsmi}\la_{jklr}\e_{mlp}b_r\tau_p/\rho=\rho\big[(c_l^2-2c_t^2)^2 b_n\de_{jk}[\spb\spt]_i -2c_t^2(c_l^2-2c_t^2) b_n b_{(k}\e_{j)is}\tau_s+\\
    &+c_t^2(c_l^2-2c_t^2) b_i\de_{jk} [\spb\spt]_n -2c_t^4b_{i}b_{(k}\e_{j)ns}\tau_s -2c_t^4\de_{ni}b_{(k}[\spb\spt]_{j)} \big],\\
\end{split}
\end{equation}
and
\begin{equation}
    [\mathbf{a},\mathbf{b}]_i\equiv[\mathbf{a}\mathbf{b}]_i:=\e_{ijk}a_jb_k.
\end{equation}
Notice that $\chi_{nkij}$ is symmetric with respect to transposition of the indices $j$, $k$. Besides, the terms in $P^1_n$ and $P^2_n$  proportional to $\tau_n$ can be discarded as they vanish on substituting to \eqref{sPK_1}.

\subsubsection{Contribution $P^1_n$}

At first, we calculate the contribution $P^1_n$. The procedure for the calculation of this contribution is the same as in the previous section \ref{Lor_Force_Sec}. Therefore, we will use the similar notation as in Sec. \ref{Lor_Force_Sec}. The power counting shows that all the divergences of $P_n^1$ are contained in the local expression
\begin{equation}\label{sing_P1}
\begin{split}
    \bar{P}^{1\perp,\parallel}_n:=&\,\rho b_i\int_{D_{\perp,\parallel}} d t d\s \Big\{\partial_\tau G^{2\e\perp,\parallel}_{ij}\big(-t,-\overset{(1)}{\spy}_a) \big[\chi_{nj}(0,0) +\dot{\chi}_{nj}(0,0)t +\chi'_{nj}(0,0)\s\big]-\\
    &-\partial_\tau\partial_lG^{2\e\perp,\parallel}_{ij}\big(-t,-\overset{(1)}{\spy}_a\big) \chi_{nj}(0,0) \overset{(2)}{y}{}^l_a \Big\}.
\end{split}
\end{equation}
The contributions to $P_n^1$ that do not vanish in the limit $\e\rightarrow+0$ are written as (see the notation in Sec. \ref{Lor_Force_Sec})
\begin{equation}
    P^1_n=\bar{P}^{1\perp}_n+\bar{P}^{1\parallel}_n + P^{1\perp}_{n\bar{D}_{\perp}}+\tilde{P}^{1\perp}_n + P^{1\parallel}_{n\bar{D}_{\parallel}}+\tilde{P}^{1\parallel}_n.
\end{equation}
The last four terms are nonlocal and finite in the limit $\varepsilon\rightarrow+0$. Thus, we have to find the divergent and finite parts of the local contributions $\bar{P}^{1\perp,\parallel}_n$. As in the previous section, we perform the gauge transformation \eqref{gauge_trans}, \eqref{gauge_trans_1}. It is convenient to perform it on calculating the contractions presented in Appendix \ref{App_Conctract}. However, to shorten the calculations we will use the relations,
\begin{equation}
    \be_\parallel=0,\qquad\bs\tau^2=1,\qquad\bs\tau\bs\be'=0,\qquad\bs\tau\bs\tau'=0,\qquad \spt\dot{\spbe}=0,
\end{equation}
immediately and only in the final answer will we perform the replacement \eqref{gauge_trans_cons}.

\paragraph{First contribution.}

Substituting \eqref{Green_func_reg_split} into \eqref{sing_P1}, we see from the power counting that the leading order contribution to the self-force when $\e\rightarrow+0$ is determined by the integral
\begin{equation}\label{V_V_ij_0}
\begin{split}
    V(s):=&\,\frac{-(2\e)^{-s}}{8\pi c\cos(\pi s/2)} \int_{-\pi/2}^{\pi/2} d\psi\int_0^\La d\vrho \vrho \partial_t f(s)= \frac{-(2\e)^{-s}\La^{s-1}}{8\pi\cos(\pi s/2)} G^{00}_1(s-1),\\
    V_{ij}(s):=&\,\frac{-(2\e)^{-s}}{8\pi cs(s+1)\cos(\pi s/2)}\int_{-\pi/2}^{\pi/2}d\psi\int_0^\La d\vrho \vrho \partial_t\partial_{ij}f(s+2)=\\
    =&\,\frac{-(2\e)^{-s}\La^{s-1}}{8\pi (s-1)s(s+1)\cos(\pi s/2)}\big[\de_{ij}N^{00}_t(s+2)+\be_i\be_j K_t^{20}(s)+\tau_i\tau_j K_t^{02}(s)\big],
\end{split}
\end{equation}
where the integration contour over $s$ runs upwards along the line $\re s=1+0$ and we have introduced the notation \eqref{N_K_L_def}. When $|\im s|\rightarrow\infty$, the absolute value of the functions $P^{kn}_l(s)$ and $G^{kn}_l(s)$ grows not faster than some power of $s$. Therefore, the integration contour in the $s$ plane can be translated along the real axis taking into account the contributions of the singular points of $V(s)$ and $V_{ij}(s)$ at a finite $s$. We are interested in the contributions from the poles at the points $s=0$ and $s=1$.

The series of the integrand in the vicinities of these points are presented in \eqref{K_L_expans}, \eqref{eps_La_exp_0}. As expected, $N^{00}_t(s+2)$ and $K^{kn}_t(s)$ for $(k,n)$ equal to $(2,0)$ and $(0,2)$ possess a zero of the second order at the point $s=1$ (see \eqref{rels2}). Consequently, the residues of the functions $V(s)$ and $V_{ij}(s)$ at this point vanish. Using the expansions \eqref{K_L_expans}, \eqref{eps_La_exp_0}, we obtain
\begin{equation}
    r:=\res_{s=0}V=-\frac{1}{8\pi\La}\overset{(-1)}{G}{}^{00}_1(-1),\qquad r_{ij}:=\res_{s=0}V_{ij}=\frac{1}{8\pi\La}\big[\de_{ij}N^{00}_t(2) +\be_i\be_j K^{20}_t(0) +\tau_i\tau_j K^{02}_t(0) \big].
\end{equation}
We see that this contribution to the Peach-K\"{o}hler force is finite in the no-regularization limit.

The sum of the transversal and longitudinal contributions to \eqref{sing_P1} stemming from these terms is written as
\begin{equation}
    \bar{P}_n^{1(1)}=c_t^{-1}[rb_i\chi_{ni}-b_ir_{ij}\chi_{nj}]_{c\rightarrow c_t}+c_l^{-1}b_ir_{ij}\chi_{nj}|_{c\rightarrow c_l},
\end{equation}
where $\chi_{nj}\equiv\chi_{nj}(0,0)$.

\paragraph{Second contribution.}

Now we consider the contribution to \eqref{sing_P1} of the integrand terms that are linear with respect to $t$ and $\s$. In this case, the contribution to the self-force is determined by the integrals
\begin{align*}
    V^t(s):=&\,\frac{(2\e)^{-s}}{8\pi c^2\cos(\pi s/2)} \int_{-\pi/2}^{\pi/2} d\psi\int_0^\La d\vrho \vrho^2\cos\psi \partial_t f(s)= \frac{(2\e)^{-s}\La^{s}(s-1)}{8\pi cs\cos(\pi s/2)} G^{10}_1(s-1),\\
    V^\s(s):=&\,\frac{-(2\e)^{-s}}{8\pi c\cos(\pi s/2)} \int_{-\pi/2}^{\pi/2} d\psi\int_0^\La d\vrho \vrho^2\sin\psi \partial_t f(s)= 0,\\
    V^t_{ij}(s):=&\,\frac{(2\e)^{-s}}{8\pi c^2s(s+1)\cos(\pi s/2)}\int_{-\pi/2}^{\pi/2}d\psi\int_0^\La d\vrho \vrho^2\cos\psi \partial_t\partial_{ij}f(s+2)=\ttg\label{Vts_Vts_ij_0}\\
    =&\,\frac{(2\e)^{-s}\La^{s}}{8\pi c s^2(s+1)\cos(\pi s/2)} \big[\de_{ij}N^{10}_t(s+2)+\be_i\be_j K_t^{30}(s)+\tau_i\tau_j K_t^{12}(s)\big],\\
    V^\s_{ij}(s):=&\,\frac{-(2\e)^{-s}}{8\pi cs(s+1)\cos(\pi s/2)}\int_{-\pi/2}^{\pi/2}d\psi\int_0^\La d\vrho \vrho^2\sin\psi \partial_t\partial_{ij}f(s+2)=\\
    =&\,\frac{(2\e)^{-s}\La^{s}}{8\pi s^2(s+1)\cos(\pi s/2)} 2\be_{(i} \tau_{j)} K_t^{12}(s),
\end{align*}
where we have neglected the terms vanishing when $\be_\parallel=0$.

Employing the expansions \eqref{eps_La_exp_0} and the notation \eqref{n_k_l}, the residues at the point $s=0$ can be cast into the form
\begin{equation}
\begin{gathered}
    r^t:=\res_{s=0}V^t=\frac{g_t^{10}}{8\pi c},\qquad r^t_{ij}:=\res_{s=0}V^t_{ij}=-\frac{1}{8\pi c}\big[n^{10}_t\de_{ij} +\be_i\be_j k^{30}_t +\tau_i\tau_j k^{12}_t\big],\\
    r^\s_{ij}:=\res_{s=0}V^\s_{ij}=-\frac{1}{8\pi} 2\be_{(i} \tau_{j)} k^{12}_t.
\end{gathered}
\end{equation}
Then the sum of the transversal and longitudinal contributions to \eqref{sing_P1} coming from these terms becomes
\begin{equation}
    \bar{P}_n^{1(2)}=c_t^{-1} \big[r^t \dot{\chi}_{nj}b_j -r^t_{ij}b_i\dot{\chi}_{nj} -r^\s_{ij}b_i \chi'_{nj} \big]_{c\rightarrow c_t} +c_l^{-1} \big[r^t_{ij}b_i\dot{\chi}_{nj} +r^\s_{ij}b_i \chi'_{nj} \big]_{c\rightarrow c_l},
\end{equation}
where $\dot{\chi}_{nj}\equiv\dot{\chi}_{nj}(0,0)$ and $\chi'_{nj}\equiv \chi'_{nj}(0,0)$.

\paragraph{Third contribution.}

The contribution to \eqref{sing_P1} of the integrand terms quadratic with respect to $t$ and $\s$ is expressed through the integrals
\begin{align*}
    V_k^{ab}(s):=&\,\frac{(2\e)^{-s}\La^s}{16\pi c s\cos(\pi s/2)} \int_{-\pi/2}^{\pi/2} d\psi\cos^a\psi\sin^b\psi \partial_t\partial_k f(s)=\\
    =&\, \frac{(2\e)^{-s}\La^{s}}{16\pi s\cos(\pi s/2)} \big[\be_k N^{a+1,b}_t(s) -\tau_k N^{a,b+1}_t(s) \big],\\
    V^{ab}_{kij}(s):=&\,\frac{(2\e)^{-s}\La^s}{16\pi cs^2(s+1)\cos(\pi s/2)} \int_{-\pi/2}^{\pi/2}d\psi\cos^a\psi\sin^b\psi \partial_t\partial_{kij}f(s+2)=\\
    =&\,\frac{(2\e)^{-s}\La^s}{16\pi s^2(s+1)\cos(\pi s/2)} \big[(\de_{ij}\be_k+\cycle(i,j,k)) K_t^{a+1,b}(s)-\ttg\label{Vabk_Vab_ijk}\\
    &-(\de_{ij}\tau_k+\cycle(i,j,k))  K_t^{a,b+1}(s) +\be_i\be_j\be_k L_t^{a+3,b}(s)-\\
    &-(\be_i\be_j\tau_k+\cycle(i,j,k)) L_t^{a+2,b+1}(s)
    +(\be_i\tau_j\tau_k+\cycle(i,j,k))L_t^{a+1,b+2}(s)-\\
    &-\tau_i\tau_j\tau_k L_t^{a,b+3}(s)\big],
\end{align*}
where the function $L_t^{kn}(s)$ is defined in \eqref{N_K_L_def}. Its expansion in the neighbourhood of the point $s=0$ has the form \eqref{K_L_expans}. Using this expansion and formulas \eqref{eps_La_exp_0}, we deduce
\begin{equation}
\begin{split}
    r_k^{ab}:&=\res_{s=0}V^{ab}_k=\frac{-1}{16\pi}\big[\be_k\tilde{n}_t^{a+1,b} -\tau_k\tilde{n}_t^{a,b+1} \big],\\
    r_{kij}^{ab}:&=\res_{s=0}V^{ab}_k=\frac{-1}{16\pi}\big[(\de_{ij}\be_k+\cycle(i,j,k)) k_t^{a+1,b} -(\de_{ij}\tau_k+\cycle(i,j,k)) k_t^{a,b+1} +\be_i\be_j\be_k l_t^{a+3,b}-\\
    &-(\be_i\be_j\tau_k+\cycle(i,j,k))l_t^{a+2,b+1} +(\be_i\tau_j\tau_k+\cycle(i,j,k))l_t^{a+1,b+2} -\tau_i\tau_j\tau_k l_t^{a,b+3} \big],
\end{split}
\end{equation}
where we have used the notation introduced in formula \eqref{n_k_l}. As a result, the sum of the transversal and longitudinal contributions to \eqref{sing_P1} is written as
\begin{equation}
\begin{split}
    \bar{P}_n^{1(3)}=&\,c_t^{-1}\big[c^{-1}r^{20}_{k}\dot{\be}_k -2r^{11}_{k}\be'_k +r^{02}_{k}\tau'_k\big] \chi_{nj}b_j\Big|_{c\rightarrow c_t} -c_t^{-1}\big[c^{-1}r^{20}_{kij}\dot{\be}_k -2r^{11}_{kij}\be'_k +r^{02}_{kij}\tau'_k\big] b_i\chi_{nj}\Big|_{c\rightarrow c_t}+\\
    &+c_l^{-1}\big[c^{-1}r^{20}_{kij}\dot{\be}_k -2r^{11}_{kij}\be'_k +r^{02}_{kij}\tau'_k\big] b_i\chi_{nj}\Big|_{c\rightarrow c_l}.
\end{split}
\end{equation}

\subsubsection{Contribution $P^2_n$}

Consider the contribution $P^2_n$. By power counting we see that the divergences of the contribution $P_n^2$ are isolated in
\begin{equation}\label{sing_P2}
\begin{split}
    \bar{P}^{2\perp,\parallel}_n:=&\,-\rho\int_{D_{\perp,\parallel}} d t d\s \Big\{\partial_k G^{2\e\perp,\parallel}_{ij}\big(-t,-\overset{(1)}{\spy}_a) \big[\chi_{nkij}(0,0) +\dot{\chi}_{nkij}(0,0)t +\chi'_{nkij}(0,0)\s\big]-\\
    &-\partial_{kl} G^{2\e\perp,\parallel}_{ij}\big(-t,-\overset{(1)}{\spy}_a\big) \chi_{nkij}(0,0) \overset{(2)}{y}{}^l_a \Big\}.
\end{split}
\end{equation}
The contributions to $P_n^2$ nonvanishing in the limit $\e\rightarrow+0$ take the form (see the notation in Sec. \ref{Lor_Force_Sec})
\begin{equation}
    P^2_n=P^{2\perp}_{n\bar{D}_{\perp}}+\tilde{P}^{2\perp}_n+\bar{P}^{2\perp}_n + P^{2\parallel}_{n\bar{D}_{\parallel}}+\tilde{P}^{2\parallel}_n+\bar{P}^{2\parallel}_n.
\end{equation}
Thus we have to find the divergent and finite parts of the integrals $\bar{P}^{2\perp,\parallel}_n$. As in the previous section, we suppose that the gauge transformation \eqref{gauge_trans}, \eqref{gauge_trans_1} is performed.

The expression \eqref{sing_P2} coincides with \eqref{sing_int} up to a common sign and a change of the notation. Therefore, we can use the results of Sec. \ref{Lor_Force_Sec} and write the divergent and finite parts of the integral \eqref{sing_P2} at one stroke. Let
\begin{align*}
    \tilde{r}_k:=&\,\frac{1}{8\pi c\La}\be_k \overset{(-1)}{N}{}^{10}(0),\\
    \tilde{r}_{kij}:=&\,\frac{1}{8\pi c\La}\big[(\de_{ij}\be_k+\cycle(i,j,k)) K^{10}(0) +\be_i\be_j\be_k L^{30}(0)+(\be_i\tau_j\tau_k +\cycle(i,j,k)) L^{12}(0) \big],\\
    \tilde{r}_k^t:=&\,-\frac{1}{8\pi c^2} \be_k n^{20},\\
    \tilde{r}_k^\s:=&\,-\frac{1}{8\pi c} \tau_k n^{02},\\
    \tilde{r}_{kij}^t:=&\,-\frac{1}{8\pi c^2} \big[(\de_{ij}\be_k+\cycle(i,j,k)) k^{20}
     +\be_i\be_j\be_k l^{40}+
    (\be_i\tau_j\tau_k +\cycle(i,j,k)) l^{22}\big],\\
    \tilde{r}_{kij}^\s:=&\,-\frac{1}{8\pi c} \big[(\de_{ij}\tau_k+\cycle(i,j,k)) k^{02}+(\be_i\be_j\tau_k +\cycle(i,j,k)) l^{22} +\tau_i\tau_j\tau_k l^{04}\big],\ttg\\
    \tilde{r}^{ab}_{kl}:=&\,-\frac{1}{16\pi c} \big[\de_{kl} n^{ab} +\be_k\be_l \tilde{k}^{a+2,b}
    -2\be_{(k}\tau_{l)} \tilde{k}^{a+1,b+1}
    +\tau_{k}\tau_{l} \tilde{k}^{a,b+2}  \big],\\
    \tilde{r}^{ab}_{klij}:=&\,-\frac{1}{16\pi c} \big[(\de_{ij}\de_{kl}+\cycle(i,j,k)) k^{ab}
    +(\de_{ij}\be_k\be_l+\be_i\be_j\de_{kl} +\cycle(i,j,k)) l^{a+2,b}-\\
    &-2(\de_{ij}\be_{(k}\tau_{l)}+\be_{(i}\tau_{j)}\de_{kl}+\cycle(i,j,k)) l^{a+1,b+1}
    +(\de_{ij}\tau_k\tau_l +\tau_i\tau_j\de_{kl} +\cycle(i,j,k)) l^{a,b+2}+\\
    &+\be_i\be_j\be_k\be_l m^{a+4,b}
    -(\be_i\be_j\be_k\tau_l +\cycle(i,j,k,l)) m^{a+3,b+1}+\\
    &+(\be_i\be_j\tau_k\tau_l +\tau_i\tau_j\be_k\be_l +\cycle(i,j,k)) m^{a+2,b+2}
    -(\be_i\tau_j\tau_k\tau_l +\cycle(i,j,k,l)) m^{a+1,b+3} +\\
    &+\tau_i\tau_j\tau_k\tau_l m^{a,b+4}\big].
\end{align*}
The contributions to $\bar{P}^2_n$ nonvanishing in the limit $\e\rightarrow+0$ take the form
\begin{equation}
\begin{split}
    \bar{P}^2_n=&\,c_t^{-1}[\tilde{r}_k\chi_{nkii}-\tilde{r}_{kij}\chi_{nkij}]_{c\rightarrow c_t}
    +c_l^{-1}\tilde{r}_{kij}|_{c\rightarrow c_l}\chi_{nkij} +c_t^{-1}[\tilde{r}^t_k\dot{\chi}_{nkii}-\tilde{r}^t_{kij}\dot{\chi}_{nkij}+\\
    &+\tilde{r}^\s_k \chi'_{nkii}-\tilde{r}^\s_{kij} \chi'_{nkij}]_{c\rightarrow c_t} +c_l^{-1}[\tilde{r}^t_{kij}\dot{\chi}_{nkij} +\tilde{r}^\s_{kij} \chi'_{nkij}]_{c\rightarrow c_l}+\\
    &+c_t^{-1}\big[c^{-1}\tilde{r}^{20}_{kl}\dot{\be}_l -2\tilde{r}^{11}_{kl}\be'_l +\tilde{r}^{02}_{kl}\tau'_l\big]\chi_{nkii}\Big|_{c\rightarrow c_t} -c_t^{-1}\big[c^{-1}\tilde{r}^{20}_{klij}\dot{\be}_l -2\tilde{r}^{11}_{klij}\be'_l +\tilde{r}^{02}_{klij}\tau'_l\big]\chi_{nkij}\Big|_{c\rightarrow c_t}+\\
    &+c_l^{-1}\big[c^{-1}\tilde{r}^{20}_{klij}\dot{\be}_l -2\tilde{r}^{11}_{klij}\be'_l +\tilde{r}^{02}_{klij}\tau'_l\big]\chi_{nkij}\Big|_{c\rightarrow c_l}.
\end{split}
\end{equation}

\subsubsection{Sum}

The sum of the local contributions $\bar{P}^1_n$ and $\bar{P}^2_n$ consists of the following terms. The first term is
\begin{equation}\label{P112}
\begin{split}
    \bar{P}_n^{1(1)}+\bar{P}_n^{2(1)}=&\,\frac{\rho}{8\pi\La}\Big\{\frac{d_1}{c^2}\btbep b_n+ \frac{d_2}{c^2}(\spb\spbep) [\spb\spt]_n +\frac{c_t^2}{c^2}\big[d_3\spb^2 +d_4(\spb\spbep)^2 +d_5(\spb\spt)^2\big] [\spt\spbep]_n+\\
    &+\frac{c_t^2}{c^2}d_4(\spb\spbep)\btbep\be_{\perp n}  \Big\}_{c\rightarrow c_t} -\frac{\rho}{8\pi\La}\Big\{\cdots\Big\}_{c\rightarrow c_l,\overset{(-1)}{G}{}^{00}_1(-1)\rightarrow0,\overset{(-1)}{N}{}^{10}(0)\rightarrow0},
\end{split}
\end{equation}
where we have introduced the coefficients $d_k$ and $\bar{d}_k$ that depend only on $\spbep^2$ and $c_{t,l}$. Similar to the previous section, where we considered the contribution of the Lorentzian force to the dislocation self-interaction, the overbarred coefficients  $\bar{d}_{1-5}$ enter the contribution of the longitudinal part of the Green's function. The contribution to the self-force stemming from the term \eqref{P112} is written as
\begin{equation}\label{sPK_1loc}
\begin{split}
    \mathbf{f}^{\text{sPK}(1)}=&\,\frac{\rho}{8\pi\La}\Big\{\frac{d_1}{c^2}\btbep [\spt\spb]+ \frac{d_2}{c^2}(\spb\spbep) \spb_\perp -\frac{c_t^2}{c^2}\big[d_3\spb^2 +d_4(\spb\spbep)^2 +d_5(\spb\spt)^2\big] \spbep+\\
    &+\frac{c_t^2}{c^2}d_4(\spb\spbep)\btbep[\spt\spbep]  \Big\} -\frac{\rho}{8\pi\La}\Big\{\cdots\Big\}_{c\rightarrow c_l,\overset{(-1)}{G}{}^{00}_1(-1)\rightarrow0,\overset{(-1)}{N}{}^{10}(0)\rightarrow0}.
\end{split}
\end{equation}
The explicit expressions for the coefficients $d_k$ and $\bar{d}_k$ are given in \eqref{dk}, \eqref{d15}.

The sum of the second and third contributions,
\begin{equation}
    \mathbf{f}^{\text{sPK}(23)}:=[\spt\bar{\mathbf{P}}^{1(2)}]+[\spt\bar{\mathbf{P}}^{1(3)}] +[\spt\bar{\mathbf{P}}^{2(2)}]+[\spt\bar{\mathbf{P}}^{2(3)}],
\end{equation}
results in
\begin{align*}
    \mathbf{f}^{\text{sPK}(23)}=&\,-\frac{\rho}{16\pi}\Big\{ [\spt\spb]\Big[ 2d_6 \frac{D\btbep}{c} - (c_l^2-2c_t^2)(\spb\spbep) \big[ k^{30}_t\frac{(\spbep\spt D\spbep)}{c} +k^{12}_t(\spt\spt'\spbep)\big]+\\
    &+2\frac{d_{7}}{c^2}(\spb\spt)(\spbep\spt\spbep') +\frac{d_8}{c^3}(\spb\spt D\spbep) -\frac{d_9}{c^2}(\spt'\spb\spt) +\btbep\big[\frac{d_{10}}{c^3}(\spbep D\spbep) +\\
    &+\frac{d_{11}}{c^2}(\spbep\spt')\big] \Big]+ \spb_\perp \Big[\frac{d_{12}}{c^3}(\spb D\spbep)  +\frac{d_{13}}{c^2}(\spbep\spbep')(\spb\spt) +\frac{d_{14}}{c^2}(\spb\spt')+\\
    &+\frac{d_{15}}{c^3}(\spbep D\spbep)(\spb\spbep) +\frac{d_{16}}{c^2}(\spbep\spt')(\spb\spbep) \Big] -\frac{c_t^2}{c^3}(D\spbep)_\perp\Big[d_{17}\spb^2 +2d_{18}(\spb\spbep)^2+\\
    &+2d_{19}(\spb\spt)^2 \Big] -\frac{c_t^2}{c^2}\spt'\Big[d_{20}\spb^2 -2c_t^2l^{22}(\spb\spbep)^2 -2d_{21}(\spb\spt)^2 \Big] +\frac{c_t^2}{c^2}\spbep \Big[\frac{(\spbep D\spbep)}{c}\times\\
    &\times\big[d_{22}\spb^2-d_{23} (\spb\spbep)^2 -d_{24}(\spb\spt)^2 \big] +(\spbep\spt')\big[d_{25}\spb^2-d_{24}(\spb\spbep)^2 -d_{26}(\spb\spt)^2 \big]-\ttg\label{sPK_23loc}\\
    &-2\frac{d_{27}}{c}(\spb D\spbep)(\spb\spbep) -2d_{28}(\spbep\spbep')(\spb\spt)(\spb\spbep) -2d_{29} \big[(\spb\spbep)(\spb\spt') +2(\spb\spt)(\spb\spbep') \big]\Big]+\\
    &+\frac{c_t^2}{c^2}[\spt\spbep]\Big[ \btbep \big[\frac{d_{27}}{c}(\spb D\spbep) +d_{29}(\spb\spt') +d_{28}(\spb\spt)(\spbep\spbep') +\frac{d_{23}}{c}(\spb\spbep)(\spbep D\spbep)+\\
    &+d_{30}(\spb\spbep) (\spbep\spt') \big] +2c_t^2l^{40} (\spb\spbep) \Big(\frac{(\spb\spt D\spbep)}{c} +2(\spbep\spb\spbep') \Big) +2c_t^2l^{22}\big[(\spb\spbep)(\spt\spb\spt')+\\ &+2(\spb\spt)(\spb\spt'\spbep) \big] -2ck_t^{30} (\spb\spbep)D\btbep -2c^2k_t^{12}(\spb\spt)\btbep' \Big]+\\
    &+\frac{c_t^2}{c^3}d_{27}[\spt D\spbep]\btbep(\spb\spbep)
    +\frac{c_t^2}{c^2}d_{29}\btbep\big((\spb\spbep)[\spt\spt'] +2(\spb\spt)[\spt\spbep']\big)-\\ &-2\frac{d_{31}}{c^2}(\spb\spbep)(\spb\spt)(\spbep')_\perp\Big\}_{c\rightarrow c_t}
    +\frac{\rho}{16\pi}\Big\{\cdots\Big\}_{c\rightarrow c_l,g^{10}_t\rightarrow0,\tilde{n}_t^{kn}\rightarrow0,n^{kn}\rightarrow0,\tilde{k}^{kn}\rightarrow0},
\end{align*}
where we have used the notation introduced in formula \eqref{dk} and the relation
\begin{equation}
    (\spbep, \spb, \spbep' - (\spt\spbep')\spt) = (\spb\spt) (\spbep, \spt, \spbep' - (\spt \spbep')\spt) = (\spb \spt) (\spbep \spt \spbep').
\end{equation}
Thus, the contribution of the Peach-K\"{o}hler force to the self-interaction of a dislocation is given by
\begin{equation}
    \mathbf{f}^{\text{sPK}}=\mathbf{f}^{\text{sPK}(1)} +\mathbf{f}^{\text{sPK}(23)} +\sum_{r=1,2}[\spt,\mathbf{P}^{r\perp}_{\bar{D}_{\perp}} +\tilde{\mathbf{P}}^{r\perp} + \mathbf{P}^{r\parallel}_{\bar{D}_{\parallel}}+\tilde{\mathbf{P}}^{r\parallel}].
\end{equation}
The first two terms in this expression are local whereas the rest of the terms are nonlocal. The leading contribution to this expression comes from the divergences containing in the term \eqref{sPK_23loc}.

\subsection{Total self-force}\label{Total_Self_Force}

As a result, the expression for the total self-force \eqref{self_force} has the form
\begin{equation}\label{self_force_fin}
\begin{split}
    \mathbf{f}^{\text{self}}=&\,c[\spt\spbep]b_i(\bar{J}_i+J^{\perp}_{i\bar{D}_{\perp}}+\tilde{J}^{\perp}_i + J^{\parallel}_{i\bar{D}_{\parallel}}+\tilde{J}^{\parallel}_i)+\\ &+\mathbf{f}^{\text{sPK}(1)} +\mathbf{f}^{\text{sPK}(23)} +\sum_{r=1,2}[\spt,\mathbf{P}^{r\perp}_{\bar{D}_{\perp}} +\tilde{\mathbf{P}}^{r\perp} + \mathbf{P}^{r\parallel}_{\bar{D}_{\parallel}}+\tilde{\mathbf{P}}^{r\parallel}]=\\
    =&\,\mathbf{f}^{\text{sing}}\ln\frac{2\e}{\La}+\mathbf{f}^{\text{fin}}.
\end{split}
\end{equation}
Let us mention some properties of the expression \eqref{self_force_fin}. First, the contributions of the second and third terms in \eqref{self_force_fin}, which were denoted by the indices (2) and (3) in the previous sections, are invariant with respect to the inversion $\tau\rightarrow-\tau$. Such a replacement yields
\begin{equation}
    \partial_\tau\rightarrow-\partial_\tau,\qquad\spbep\rightarrow-\spbep,\qquad\be_\parallel\rightarrow-\be_\parallel.
\end{equation}
In particular, the divergences that are contained only in these contributions obey such a symmetry. The contributions of the first terms to the self-force, which were denoted by the index (1) in the previous sections, and the nonlocal terms are not invariant under the inversion $\tau\rightarrow-\tau$. These contributions describe, in particular, the irreversible processes in the dislocation dynamics related to the radiation of sound waves.

Second, the expression \eqref{self_force_fin} does not depend on the parameter $\La$. The only restrictions on this parameter are that $\La$ is smaller than the dislocation length and $\La\gg\e$. It is convenient to choose this parameter such that the contribution of the divergences to the self-force \eqref{self_force_fin} is maximized. For example, for $\e\simeq 5$ {\AA} and $\La=1$ cm, we have
\begin{equation}
    \ln\frac{2\e}{\La}\simeq-16.1.
\end{equation}
In this case, the main contribution to the self-force is determined by the divergences. If one neglects the impact of the reaction of radiation of sound waves on the dislocation dynamics, which is described by the contributions to \eqref{self_force_fin} not invariant under time reversion, then a rather good approximation for \eqref{self_force_fin} is obtained by taking into account only the divergent contributions. In virtue of the symmetry mentioned above, such approximate dynamics are reversible.

The self-force transforms under the conformal transformations,
\begin{equation}\label{conf_map}
    \spy\rightarrow\la\spy,\qquad t\rightarrow\la t\qquad\la>0,
\end{equation}
as
\begin{equation}
    \mathbf{f}^{\text{self}}(\e)\rightarrow \la^{-1}\mathbf{f}^{\text{self}}(\la^{-1}\e).
\end{equation}
In particular, in the absence of external forces, the self-consistent approximate equations of motion of a dislocation that include solely the divergent contributions are not only time reversible but also conformal invariant.

Third, the finite and divergent parts of the self-force \eqref{self_force_fin} do not depend on the choice of the representative in the class of spherically-symmetric regularizations \eqref{delta_reg_gen} of the dislocation core. All the dependence on the choice of the representative boils down to a change of the regularization parameter $\e$. Indeed, in order to obtain the self-force with an arbitrary spherically-symmetric regularization characterized by the form-factor $g_L(\s\spk)$, it is sufficient to replace $2\e\rightarrow\e+\e'$ in the self-force and to convolve \eqref{self_force_fin} with $g(\e/\s)g(\e'/\s)$ as in formula \eqref{regul_convol}. Then we derive
\begin{equation}\label{regul_differ}
    \int_0^\infty \frac{d\e d\e'}{\s^2}g\Big(\frac{\e}{\s}\Big)g\Big(\frac{\e'}{\s}\Big) \big(\mathbf{f}^{\text{sing}} \ln\frac{\e+\e'}{\La}+\mathbf{f}^{\text{fin}}\big)=\mathbf{f}^{\text{sing}} \ln\frac{2\tilde{\e}}{\La}+\mathbf{f}^{\text{fin}},
\end{equation}
where
\begin{equation}
    \tilde{\e}=\frac{\s}{2}\exp\Big[\int_0^\infty d\e d\e' g(\e)g(\e')\ln(\e+\e')\Big],
\end{equation}
and it is assumed that
\begin{equation}
    \Big|\int_0^\infty d\e d\e' g(\e)g(\e')\ln(\e+\e')\Big|<\infty.
\end{equation}
We have used the normalization condition \eqref{norm_cond} in obtaining \eqref{regul_differ}.

Now we account for the fact that dislocations do not virtually climb at low temperatures, i.e., we take into account the constraint \eqref{constrants} and the constraint reaction force \eqref{constr_reac_force}. The gauge transformation \eqref{gauge_trans_1} brings the constraint reaction force to
\begin{equation}\label{constr_reac_force1}
    \mathbf{f}^c=[\spb,\vk'c\spbep-\dot{\vk}\spt].
\end{equation}
The use of the constraint condition \eqref{constr_reac_force} considerably simplifies the expressions for the local contributions to  $f^{\text{self}}_i$. The explicit expressions for these contributions are presented in formulas \eqref{sL_loc}-\eqref{sPK_23loc2} of Appendix \ref{Useful_Forms_Local}. Then the complete system of equations of motion of a dislocation is written as
\begin{equation}\label{eff_eq_mot}
    \mathbf{f}^{\text{self}}+\mathbf{f}^c+\mathbf{f}^{\text{ext}}=0,\qquad \btbep=0.
\end{equation}
The Lagrange multiplier $\vk$ is found from the solution of the given system of equations. If we assume that $\ln(2\e/\La)$ is a constant parameter then Eqs. \eqref{eff_eq_mot} without external forces are also invariant under the conformal transformations \eqref{conf_map}. The existence of such a symmetry gives rise to the property of dislocation dynamics that an increase of the spatial size of a dislocation by a factor of $k$ leads to a deceleration of its free dynamics by a factor of $k$.

\section{Some physical implications}\label{Some_Phys_Cons}

Let us consider some particular cases of the general equations derived in the previous section. In the near-sonic limit, where the dislocation velocity $|\dot{\spy}|$ tends to the shear wave velocity $c_t<c_l$, we have
\begin{equation}
    v|_{c\rightarrow c_t}\rightarrow0.
\end{equation}
Then the leading contribution of the divergent part of the self-force comes from the terms in the expressions \eqref{sL_tot}, \eqref{bJtot1}, and \eqref{sPK_23loc} standing at the coefficients
\begin{equation}
    a_5\approx\frac{12 c_t^2}{v^5}\ln\frac{2\e}{\La},\qquad d_{23}\approx\frac{12 c_t^2}{v^5}\ln\frac{2\e}{\La}.
\end{equation}
In that case,
\begin{equation}\label{self_force_ultra}
    \mathbf{f}^{\text{self}}\approx\frac{3\rho c_t}{4\pi}\ln\frac{2\e}{\La} (\spb\spbep)^2\frac{(\spbep D\spbep)}{v^5}\spbep\Big|_{c\rightarrow c_t}.
\end{equation}
Notice that, in deriving the above expression, the contribution from the Lorentzian force is canceled by the corresponding contribution from the Peach-K\"{o}hler force. The gliding condition \eqref{constrants} has not been used. As long as the force \eqref{self_force_ultra} diverges in the limit $v\rightarrow0$, the dislocation cannot be accelerated up to the shear wave velocity by external forces, at least, for the model of dislocation we consider. In the degenerate case, $(\spb\spbep)=0$, the expression \eqref{self_force_ultra} is unapplicable and the terms of the order $v^{-3}$ in the self-force should be taken into account. Notice that the same power of $v$ in the self-force experienced by a straight near-sonic edge dislocation was found in \cite{HirZbLoth98}.

Now we turn to the case of a slowly moving dislocation, where
\begin{equation}
    |\dot{\spy}|\ll c_t.
\end{equation}
Expanding the expressions \eqref{sL_tot}, \eqref{bJtot1}, \eqref{sPK_1loc}, and \eqref{sPK_23loc} in powers of $1/c_t$, $1/c_l$, the leading contribution to the local part of the self-force becomes
\begin{equation}\label{self_force_nonrel}
    \mathbf{f}^{\text{self}}\approx\frac{\rho}{16\pi}\Big\{\frac{d_9}{c^2}(\spt'\spb\spt)[\spt\spb] -\frac{d_{14}}{c^2}(\spb\spt')\spb_\perp +\frac{c_t^2}{c^2}(d_{20}\spb^2-2 d_{21}(\spb\spt)^2)\spt' \Big\}_{c\rightarrow c_t}- \frac{\rho}{16\pi}\Big\{\cdots \Big\}_{c\rightarrow c_l}.
\end{equation}
Hereinafter, the omission marks denote the same contribution as the first expression in the braces but with the replacements $d_k\rightarrow\bar{d}_k$ and $c_t\rightarrow c_l$. The contributions kept in \eqref{self_force_nonrel} are of the order $c_{t,l}^2$. In fact, such an approximation for the force corresponds to a static limit. In particular, the Lorentzian force does not contribute to  \eqref{self_force_nonrel}. In this limit, the necessary coefficients $d_k$ and $\bar{d}_k$ are written as
\begin{align*}
    d_9&=4c_l^2c_t^2(1-2q^2)\ln\frac{2\e}{\La}+2c_l^2c_t^2\big[1+\ln2-q^2(1+5\ln2)\big],\\ \bar{d}_9&=4c_l^4(1-4q^2+3q^4)\ln\frac{2\e}{\La}+2c_l^4\big[(1+\ln2) (1-4q^2) +4q^4\big],\\
    d_{14}&=4c_t^4\ln\frac{2\e}{\La}+8c_t^4\ln2,\\
    \bar{d}_{14}&=4c_t^2c_l^2(1-q^2)\ln\frac{2\e}{\La} +2c_l^2c_t^2\big[1+\ln2 +2q^2(\ln2-1)\big],\\
    d_{20}&=2c_t^2(3\ln2-1),\ttg\label{dk_nonrel}\\
    \bar{d}_{20}&=4c_t^2\ln\frac{2\e}{\La}+8c_t^2\ln2,\\
    d_{21}&=2c_t^2\ln\frac{2\e}{\La}+2c_t^2(5\ln2-1),\\
    \bar{d}_{21}&=-2c_l^2(2-5q^2)\ln\frac{2\e}{\La}-2c_l^2\big[1+\ln2 -q^2(1+7\ln2)\big].
\end{align*}

Let us find the static configurations of a dislocation following from the approximate expression for the self-force \eqref{self_force_nonrel}, where only the divergent contributions are retained, i.e., the divergent part of the coefficients $d_k$ and $\bar{d}_k$ in \eqref{dk_nonrel} is taken. In that case,
\begin{equation}
    \mathbf{f}^{\text{self}}\approx-\frac{\rho c_t^2}{4\pi}\ln\frac{2\e}{\La}\Big\{(3q^2-2)(\spt'\spb\spt)[\spt\spb] +q^2(\spb\spt')\spb_\perp +[q^2\spb^2 +(3-5q^2)(\spb\spt)^2]\spt' \Big\}.
\end{equation}
Now it is not difficult to see that, in the absence of the external stresses and the constraint force, the equilibrium configuration of a dislocation is a straight line
\begin{equation}
    \spt'=0.
\end{equation}
However, if one takes into account the constraint \eqref{constrants} and its reaction \eqref{constr_reac_force1}, i.e., if one solves the equation
\begin{equation}
    \mathbf{f}^{\text{self}}+\mathbf{f}^c=0,\qquad \mathbf{f}^c=\dot{\vk}[\spt\spb],
\end{equation}
then the equilibrium configurations of dislocations will obey the equation
\begin{equation}\label{stat_conf}
    (\spb\spt')=0.
\end{equation}
In other words, the angle between the Burgers vector and the tangent vector to the dislocation curve must be constant along this curve. It follows from Eq. \eqref{stat_conf} that
\begin{equation}
    \spt'=\ga[\spb\spt],\qquad(\spt')^2=b^2_\perp \ga^2.
\end{equation}
Substituting such a representation for $[\spb\spt]$ into the initial equation, we obtain the derivative of the Lagrange multiplier with respect to time
\begin{equation}
    \dot{\vk}=-\ga\frac{\rho c_t^2}{4\pi}\ln\frac{2\e}{\La}\big[2(1-q^2)b_\perp^2 +(3-4q^2) b^2_\parallel\big].
\end{equation}
The expression on the right-hand side in the square brackets is a positive constant. Hence,
\begin{equation}
    \vk'=-\ga'\frac{\rho c_t^2}{4\pi}\ln\frac{2\e}{\La}\big[2(1-q^2)b_\perp^2 +(3-4q^2) b^2_\parallel\big]t+\vk_0'(\la),
\end{equation}
where $\vk_0(\la)$ is an arbitrary function.

If the dislocation is not static, then the total expression for the constraint force \eqref{constr_reac_force1} contains the term proportional to $\vk'$. Therefore, for the static configurations to be stable, it is necessary that the term growing with time, which is present in $\vk'$, vanishes. This is possible only if
\begin{equation}\label{stat_conf1}
    \ga'=0\;\;\Rightarrow\;\; (\spt'^2)'=0.
\end{equation}
Consequently, the stable static configurations of dislocations are the curves of a constant curvature. Bearing in mind the condition \eqref{stat_conf}, we conclude that the stable static configurations of dislocations are the circular helices with the axes directed along the Burgers vector. The degenerate cases of such helices are an edge dislocation forming a circle lying in the plane normal to the Burgers vector and a straight screw dislocation. Such configurations of dislocations are observed in experiments \cite{Weert57,GrilheThes,Friedel64}.

Consider the leading correction to the local part of the self-force stemming from the dislocation motion, viz., let us find the corrections to \eqref{self_force_nonrel} that are at most linear in the dislocation velocity. The Lorentzian force is at least quadratic in the velocity of dislocation and so it does not contribute to the expression we need. Using \eqref{sPK_1loc2} and \eqref{sPK_23loc2}, we obtain that the local part of the Peach-K\"{o}hler force gives
\begin{equation}\label{self_force_nonrel_plus}
\begin{split}
    \mathbf{f}^{\text{self}}\approx&\,\frac{\rho}{16\pi}\Big\{\frac{2}{\La}\big[\frac{d_2}{c^2}(\spb\spbep)\spb_\perp -\frac{c_t^2}{c^2}(d_3\spb^2+d_5b_\parallel^2)\spbep \big] +\frac{d_9}{c^2}(\spt'\spb\spt)[\spt\spb] -\big[\frac{d_{12}}{c^3}(\spb D\spbep)+\frac{d_{14}}{c^2}(\spb\spt')\big]\spb_\perp+\\
    &+\frac{c_t^2}{c^3}(d_{17}\spb^2+2 d_{19}b_\parallel^2)(D\spbep)_\perp +\frac{c_t^2}{c^2}(d_{20}\spb^2-2 d_{21}b_\parallel^2)\spt' \Big\}_{c\rightarrow c_t}- \frac{\rho}{16\pi}\Big\{\cdots \Big\}_{c\rightarrow c_l},
\end{split}
\end{equation}
where the constraint condition \eqref{constrants} is supposed to be fulfilled and the constraint force has the form \eqref{constr_reac_force1}. Since the coefficients $d_k$ and $\bar{d}_k$ are analytic functions of $\be_\perp^2$ in the vicinity of zero, the expression for the local part of the self-force \eqref{self_force_nonrel_plus} involves only their static values
\begin{equation}
    d_k|_{\spbep=0},\qquad \bar{d}_k|_{\spbep=0}.
\end{equation}
Several coefficients $d_k$ and $\bar{d}_k$ entering into expression \eqref{self_force_nonrel_plus} are presented in formulas \eqref{dk_nonrel} and \eqref{d15}. In the static limit, the remaining coefficients appear to be
\begin{align*}
    d_{12}&=-2c_t^4\ln\frac{2\e}{\La}-4c_t^4\ln2,\\
    \bar{d}_{12}&=-2c_t^2c_l^2(2-3q^2)\ln\frac{2\e}{\La} -2c_l^2c_t^2(2+\ln2 -4q^2),\\
    d_{17}&=2c_t^2\ln\frac{2\e}{\La}+2c_t^2(3\ln2-2),\ttg\\
    \bar{d}_{17}&=-2c_l^2(2-q^2)\ln\frac{2\e}{\La} -4c_l^2(1-q^2\ln2),\\
    d_{19}&=c_t^2\ln\frac{2\e}{\La}+2c_t^2(4-5\ln 2),\\
    \bar{d}_{19}&=c_l^2(2-q^2)\ln\frac{2\e}{\La}+2c_l^2\big[2-\ln2+2q^2(1-2\ln2)\big].
\end{align*}
Further, we will use only the divergent part of these coefficients in \eqref{self_force_nonrel_plus}. The terms standing at $D\spbep$ in formula \eqref{self_force_nonrel_plus} can be interpreted as the mass matrix $M_{ij}$ given in the approximation we employ. The expression for $M_{ij}$ following from \eqref{self_force_nonrel_plus} coincides with formula (8.16) of \cite{Kos65}, where the consequence of the constraint \eqref{constr_consq} should be used in the linear order in velocity.

Let us linearize the divergent part of the self-force near the static solution \eqref{stat_conf}, \eqref{stat_conf1} and obtain the equation for small oscillations of a dislocation. In fact, we can use the approximate expression for the local part of self-interaction force \eqref{self_force_nonrel_plus}, where one has to bear in mind that the unperturbed solution is a circular helix. On linearizing near such static solutions, we have
\begin{equation}
    \de D\spbep= D\de\spbep=\de\dot{\spbe}_\perp,\qquad  (\spt D\de\spbep)=0,\qquad (\spb\spt D\de\spbep)=0,\qquad \de d_k=0.
\end{equation}
Then the divergent part of the force \eqref{self_force_nonrel_plus} can be cast into the form
\begin{equation}
\begin{split}
    \de\mathbf{f}^{\text{sPK(23)}}=&\,\frac{\rho}{16 \pi}\Big\{\frac{d_9}{c^2}\big[(\spt'\spb\spt)[\de\spt\spb] +((\de\spt'\spb\spt)+(\spt'\spb\de\spt))[\spt\spb] \big] -\big[d_{12}(\spb D\de\spbep) +cd_{14}(\spb\de\spt')\big]\frac{\spb_\perp}{c^3} +\\
    &+(d_{17}\spb^2+2d_{19}b^2_\parallel )\frac{c_t^2}{c^3}(D\de\spbep)_\perp +(d_{20}\spb^2-2d_{21}b^2_\parallel )\frac{c_t^2}{c^2} \de\spt' -4\frac{c_t^2}{c^2}d_{21} b_\parallel(\spb\de\spt)\spt'  \Big\}_{c\rightarrow c_t}-\\
    &-\frac{\rho}{16 \pi}\Big\{\cdots\Big\}_{c\rightarrow c_l}.
\end{split}
\end{equation}
The constraint force \eqref{constr_reac_force1} is linearized as
\begin{equation}
    \de\mathbf{f}^c=[\spb,c\vk'\de\spbep-\de\dot{\vk}\spt-\dot{\vk}\de\spt].
\end{equation}
It is clear that the projection of this force onto the Burgers vector is zero, while
\begin{equation}\label{wave_eqn}
    (\spb\de\mathbf{f}^{\text{sPK(23)}})=\frac{\rho}{4\pi}\ln\frac{2\e}{\La}\Big\{\big[(1+q^4)b_\perp^2+b_\parallel^2\big](\spb\de\ddot{\spy}) -c_t^2\big[2q^2b_\perp^2 +(3-4q^2)b^2_\parallel\big](\spb\de\spy'') \Big\}.
\end{equation}
The coefficient at $(\spb\de\spy'')$ can be interpreted as the tension of a dislocation string. This expression coincides with formula (1) of \cite{Pegel66} and formula (6-82) of \cite{HirthLothe}. The coefficient standing at $(\spb\de\ddot{\spy})$ agrees with formula (2) of \cite{Pegel66}. A thorough discussion of the physical consequences of the string model for description of dislocation dynamics can be found in \cite{GranLuck66,HirthLothe,Nabar67,Friedel64,Mason66,AlshInden}.

In the lack of external forces, the small oscillations of a dislocation near its equilibrium configuration are described by the wave equation with the signal velocity squared
\begin{equation}\label{wave_veloc}
    s^2=c_t^2\frac{2q^2 b_\perp^2+(3-4q^2)b_\parallel^2}{(1+q^4) b_\perp^2 +b_\parallel^2}.
\end{equation}
Since $q^2\in(0,3/4)$, the expression on the right-hand side is positive for any Burgers vector. Formula \eqref{wave_veloc} coincides with the squared velocity of propagation of dislocation perturbations following from formulas of the paper \cite{Pegel66} and disagrees with formula (21) of \cite{LaubEsh66}. The quantity $s^2$ is a monotonic function of the angle between the tangent vector to the dislocation curve and the Burgers vector. It takes the values between
\begin{equation}
    s_\perp^2=\frac{2c_t^2q^2}{1+q^4}<\frac{24}{25}c_t^2,\qquad s_\parallel^2=(3-4q^2)c_t^2.
\end{equation}
Notice that both velocities are less then $c_l$, as expected. For $q^2\approx0.54$, the velocity \eqref{wave_veloc} does not depend on the direction of the Burgers vector and
\begin{equation}
    s^2=s_\perp^2=s_\parallel^2\approx0.84 c_t^2.
\end{equation}

The remaining components of the effective equations of motion of a dislocation,
\begin{equation}
    \de\mathbf{f}^{\text{sPK}(23)}+\de\mathbf{f}^{c}=0,
\end{equation}
allow one to obtain $\de\dot{\vk}$ and do not enforce additional restrictions on $(\spb\de\spy)$. The other components of $\de\spy$ are found from the constraint condition \eqref{constrants}. Indeed, in a general position, the vectors $\spb$, $\spt$, and $[\spb\spt]$ constitute a basis. Therefore, it is sufficient to find the projections of $\de\spy$ onto the vectors $\spt$ and $[\spb\spt]$. Linearizing the constraint condition, we obtain
\begin{equation}
    \partial_\tau(\spb\spt\de\spy)=0.
\end{equation}
Putting
\begin{equation}
    (\spb\spt\de\spy)|_{t=0}=0,
\end{equation}
we are left with
\begin{equation}
    (\spb\spt\de\spy)\equiv0.
\end{equation}
Furthermore,
\begin{equation}
    (\spt\de\spt)=0\;\;\Rightarrow\;\;(\spt\de\spy)'=(\spt'\de\spy)=\ga(\spb\spt\de\spy)=0.
\end{equation}
As a result,
\begin{equation}
    (\spt\de\spy)=\vf(\tau)\;\;\Rightarrow\;\;\de\be_\parallel=\dot{\vf}(\tau).
\end{equation}
We see that the projection of the velocity $\de\be_\parallel$ describes an unobservable flow of the dislocation along itself. This projection can be removed by means of the gauge transformation \eqref{gauge_trans}. Then
\begin{equation}
    (\spt\de\spy)=const,
\end{equation}
where the constant can be set to zero without loss of generality.

\section{Conclusion}

Let us sum up the results. We studied the effective dynamics of dislocations in crystals in the realm of linear elastodynamics. We elaborated the action principle \eqref{action}, \eqref{action1} for linear elastic media with dislocations, the finite size of the dislocation core being modeled via smearing of an ideally thin dislocation over its small neighborhood. The equations of motion following from this action principle reproduce the Peach-K\"{o}hler and Lorentzian forces acting on a dislocation. The self-force is contained in the the Peach-K\"{o}hler and Lorentzian forces and stems from the action of a stress and a medium motion produced by a dislocation onto itself. In the case of an isotropic elastic medium, we derived the explicit expressions for the divergent and finite parts of the self-force for a dislocation of a general form moving with an arbitrary velocity less than the shear wave velocity. The total expression \eqref{self_force_fin} is rather bulky. Schematically, it has the form
\begin{equation}
    f^{\text{self}}_i=A_{ij}(\spb,\dot{\spy},\spy')\ddot{y}_j+2B_{ij}(\spb,\dot{\spy},\spy')\dot{y}'_j+C_{ij}(\spb,\dot{\spy},\spy')y''_j +D_i(\spb,\dot{\spy},\spy')+\text{nonlocal terms},
\end{equation}
where the first three terms contain the logarithmic divergencies. Nevertheless, the expression \eqref{self_force_fin} can be used in analyzing certain particular cases when it is simplified. The general properties of the self-force \eqref{self_force_fin} were described in Sec. \ref{Total_Self_Force}. The immediate application of the obtained formulas for the self-force is related to numerical simulations. The explicit expression for the regularization removal asymptotics $\e\rightarrow0$ of the self-force allows one not to evaluate numerically rapidly varying integrals stemming from the self-stresses acting on dislocations. This will speed-up the numerical simulations and make them more stable.

In order to check the expression obtained, we considered its slow velocity limit and found coincidence with the known results. Namely, imposing the constraint that the dislocation is gliding, we obtained that the stable static configurations of a dislocation in an infinite elastic medium without external stresses are helices with the axes directed along the Burgers vector. In the degenerate cases, such helices are an edge dislocation in the form of a circle lying in the plane perpendicular to the Burgers vector and a straight screw dislocation. The equation of motion of small perturbations of such a helix is the wave equation (see \eqref{wave_eqn}). The expressions for the effective mass density and the tension of a dislocation coincide with the expressions known in the literature for straight dislocations.

We also investigated the near-sonic limit when the velocity of dislocation approaches the shear wave velocity from below. We found the explicit expression \eqref{self_force_ultra} for the leading contribution to the self-force in this limit. As expected, the self-force acting on a dislocation diverges in the near-sonic limit. Therefore, a dislocation cannot be accelerated up to the shear wave velocity by applying external stresses in the framework of the model we consider. Notice that, in deriving the expression for the self-force, we heavily relied on the assumption that the velocity of a dislocation is smaller than the shear wave velocity. If this is not the case, the expression for the self-force will change. It would be interesting to find the explicit expression for the self-force experienced by the dislocation created with the speed higher than the shear wave velocity \cite{GumGao99}. The corresponding calculations can be done along the lines of the present paper.

\paragraph{Acknowledgments.}

We are thankful to A.A. Sharapov for stimulating discussions and critical remarks. We are also indebted to the anonymous referees for useful comments. This study was supported by the Tomsk State University Development Programme (Priority-2030).

\appendix

\section{Derivatives}\label{Useful_Forms_Deriv}

In this appendix, we present the formulas for derivatives of the functions appearing in the Green's function \eqref{Green_func_reg_split} and arising in calculation of the self-force. Let $f=f(r)$. Then
\begin{equation}\label{partial_f}
\begin{split}
    \partial_k f=&\,n_k f',\\
    \partial_{jk} f=&\,\de_{jk} \frac{f'}{r} +n_jn_k \Big(f'' -\frac{f'}{r}\Big),\\
    \partial_{ijk}f=&\,(\de_{ij}n_k +\cycle(i,j,k))\Big(\frac{f''}{r} -\frac{f'}{r^2}\Big) +n_in_jn_k \Big(f''' -3\frac{f''}{r} +3\frac{f'}{r^2}\Big),\\
    \partial_{ijkl}f=&\,(\de_{ij}\de_{kl} +\cycle(i,j,k))\Big(\frac{f''}{r^2} -\frac{f'}{r^3}\Big) +(\de_{ij}n_kn_l+n_in_j\de_{kl} +\cycle(i,j,k)) \Big(\frac{f'''}{r} -3\frac{f''}{r^2} +3\frac{f'}{r^3}\Big)+\\
    &+n_in_jn_kn_l\Big(f^{(4)}-6\frac{f'''}{r} +15\frac{f''}{r^2} -15\frac{f'}{r^3} \Big),
\end{split}
\end{equation}
where $n_i:=x_i/r$. If
\begin{equation}
    f(r)=d(c;s)/r=(|r+ct|^{s-1}-|r-ct|^{s-1})/r,
\end{equation}
then
\begin{equation}
\begin{split}
    f'=&\,(s-1)\frac{h(s-1)}{r}-\frac{d(s)}{r^2},\\
    f''=&\,(s-1)(s-2)\frac{d(s-2)}{r}-2(s-1)\frac{h(s-1)}{r^2}+2\frac{d(s)}{r^3},\\
    f'''=&\,(s-1)(s-2)(s-3)\frac{h(s-3)}{r} -3(s-1)(s-2)\frac{d(s-2)}{r^2} +6(s-1)\frac{h(s-1)}{r^3} -6\frac{d(s)}{r^4},\\
    f^{(4)}=&\,(s-1)(s-2)(s-3)(s-4)\frac{d(s-4)}{r} -4(s-1)(s-2)(s-3)\frac{h(s-3)}{r^2}+\\
    &+12(s-1)(s-2)\frac{d(s-2)}{r^3} -24(s-1)\frac{h(s-1)}{r^4}+24\frac{d(s)}{r^5},
\end{split}
\end{equation}
where $d(s)\equiv d(c;s)$ and
\begin{equation}
    h(s)\equiv h(c;s)=|r+ct|^{s-1}_\s-|r-ct|^{s-1}_\s.
\end{equation}
The shorthand notation has been introduced
\begin{equation}
    |z|_\s^\nu:=\sgn(z)|z|^\nu,\quad z\in \mathbb{R}.
\end{equation}
Besides,
\begin{align*}
    \tilde{K}:=&\,\frac{f''}{r} -\frac{f'}{r^2}=(s-1)(s-2)\frac{d(s-2)}{r^2} -3(s-1)\frac{h(s-1)}{r^3} +3\frac{d(s)}{r^4},\\
    \tilde{L}:=&\,f''' -3\frac{f''}{r} +3\frac{f'}{r^2}=(s-1)(s-2)(s-3)\frac{h(s-3)}{r} -6(s-1)(s-2)\frac{d(s-2)}{r^2}+\\ &+15(s-1)\frac{h(s-1)}{r^3}-15\frac{d(s)}{r^4},\ttg\label{K_L_M}\\
    \tilde{M}:=&\,f^{(4)}-6\frac{f'''}{r} +15\frac{f''}{r^2} -15\frac{f'}{r^3}=(s-1)(s-2)(s-3)(s-4)\frac{d(s-4)}{r}-\\ &-10(s-1)(s-2)(s-3)\frac{h(s-3)}{r^2} +45(s-1)(s-2)\frac{d(s-2)}{r^3} -105(s-1)\frac{h(s-1)}{r^4} +105\frac{d(s)}{r^5}.
\end{align*}

Let us also introduce the notation
\begin{equation}
    g(c;s)\equiv g(s):=|r+ct|_\s^{s-1}+ |r-ct|_\s^{s-1},\qquad p(c;s)\equiv p(s):=|r+ct|^{s-1}+ |r-ct|^{s-1}.
\end{equation}
It is clear that
\begin{equation}
    \partial_t d(s)=c(s-1)g(s-1),\qquad \partial_t h(s)=c(s-1)p(s-1).
\end{equation}
Then
\begin{equation}
\begin{split}
    \partial_tf&=c(s-1)\frac{g(s-1)}{r},\\
    \partial_t\partial_kf&=cn_k r\tilde{N}_t(s),\\
    \partial_t\partial_{jk}f&=c\de_{jk}\tilde{N}_t(s) +cn_jn_k r\tilde{K}_t(s),
\end{split}
\end{equation}
where
\begin{equation}
\begin{split}
    \tilde{N}_t(s):=&\,(s-1)(s-2)\frac{p(s-2)}{r^2} -(s-1)\frac{g(s-1)}{r^3},\\
    \tilde{K}_t(s):=&\,(s-1)(s-2)(s-3)\frac{g(s-3)}{r^2} -3(s-1)(s-2)\frac{p(s-2)}{r^3} +3(s-1)\frac{g(s-1)}{r^4},\\
    \tilde{L}_t(s):=&\,(s-1)(s-2)(s-3)(s-4)\frac{p(s-4)}{r} -6(s-1)(s-2)(s-3)\frac{g(s-3)}{r^2}+\\
    &+15(s-1)(s-2)\frac{p(s-2)}{r^3} -15(s-1)\frac{g(s-1)}{r^4}.
\end{split}
\end{equation}

\section{Integrals}\label{Useful_Forms_Ints}

In evaluating the local part of the self-force, the following integrals arise
\begin{equation}\label{H_D_ints}
\begin{gathered}
    D^{kn}_l(s):=\int_{-\pi/2}^{\pi/2}d\psi\cos^k\psi \sin^n\psi \frac{\bar{d}(s)}{\bar{r}^l},\qquad H^{kn}_l(s):=\int_{-\pi/2}^{\pi/2}d\psi\cos^k\psi \sin^n\psi \frac{\bar{h}(s)}{\bar{r}^l},\\
    P^{kn}_l(s):=\int_{-\pi/2}^{\pi/2}d\psi\cos^k\psi \sin^n\psi \frac{\bar{p}(s)}{\bar{r}^l},\qquad G^{kn}_l(s):=\int_{-\pi/2}^{\pi/2}d\psi\cos^k\psi \sin^n\psi \frac{\bar{g}(s)}{\bar{r}^l},\\
\end{gathered}
\end{equation}
where
\begin{equation}\label{b_quants_not}
\begin{gathered}
    \bar{d}(s):=|\bar{r}-\cos\psi|^{s-1}-|\bar{r}+\cos\psi|^{s-1},\qquad \bar{h}(s):=|\bar{r}-\cos\psi|_\s^{s-1}-|\bar{r}+\cos\psi|_\s^{s-1},\\
    \bar{p}(s):=|\bar{r}-\cos\psi|^{s-1}+|\bar{r}+\cos\psi|^{s-1},\qquad \bar{g}(s):=|\bar{r}-\cos\psi|_\s^{s-1}+|\bar{r}+\cos\psi|_\s^{s-1},\\
    \bar{r}:=\sqrt{\be^2\cos^2\psi-2\be\cos\theta\sin\psi\cos\psi+\sin^2\psi}.
\end{gathered}
\end{equation}
The integrands of the integrals \eqref{H_D_ints} possess singularities at the points
\begin{equation}
    \tan\psi_\pm=\be\cos\theta\pm\sqrt{1-\be^2\sin^2\theta}=\be_\parallel\pm\sqrt{1-\be_\perp^2}.
\end{equation}
The condition $\be<1$ leads to $\pm\psi_\pm>0$. The parameters $\pm\tan\psi_\pm$ can be interpreted as the velocities of sound propagation along the dislocation curve at a given point divided by the corresponding sound speed. The following notation will be useful
\begin{equation}
    v:=\tan\psi_+.
\end{equation}

The integrals $D^{kn}_l(s)$, $H^{kn}_l(s)$, $P^{kn}_l(s)$, and $G^{kn}_l(s)$ are absolutely convergent provided $\re s>0$ and they are analytical functions of $s$ in this domain. The analytical continuation of these integrals to the domain $\re s\leqslant0$ is constructed by means of the standard procedure \cite{GSh,ParKam01,Wong,KalKaz3}. We expand the integrand near its singular points
\begin{equation}\label{H_D_expans}
\begin{split}
    \cos^k\psi \sin^n\psi \frac{|\bar{r}-\cos\psi|^{s-1}}{\bar{r}^l}=&\,\sum_{r=0}^\infty c_{knlr}^{-}(\be,s)(\psi-\psi_-(\be))^r|\psi-\psi_-(\be)|^{s-1}=\\
    =&\,\sum_{r=0}^\infty c_{knlr}^{+}(\be,s)(\psi-\psi_+(\be))^r|\psi-\psi_+(\be)|^{s-1},\\
    \cos^k\psi \sin^n\psi \frac{|\bar{r}-\cos\psi|^{s-1}_\s}{\bar{r}^l}=&\,\sum_{r=0}^\infty b_{knlr}^{-}(\be,s)(\psi-\psi_-(\be))^r|\psi-\psi_-(\be)|^{s-1}_\s=\\
    =&\,\sum_{r=0}^\infty b_{knlr}^{+}(\be,s)(\psi-\psi_+(\be))^r|\psi-\psi_+(\be)|^{s-1}_\s.
\end{split}
\end{equation}
Then the analytical continuation to the domain $\re s>-N-1$ is given by
\begin{align*}
    D^{kn}_l(s)=&\,\int_{-\pi/2}^0 d\psi\Big(\cos^k\psi\sin^n\psi\frac{\bar{d}(s)}{\bar{r}^l}-\sum_{r=0}^N c^-_{knlr} (\psi-\psi_-)^r|\psi-\psi_-|^{s-1} \Big)+\\
    &+\int_0^{\pi/2} d\psi\Big(\cos^k\psi\sin^n\psi\frac{\bar{d}(s)}{\bar{r}^l}-\sum_{r=0}^N c^+_{knlr} (\psi-\psi_+)^r|\psi-\psi_+|^{s-1} \Big)+\\
    &+\sum_{r=0}^N\frac{c^-_{knlr}\big[(-\psi_-)^{s+r}+(-1)^r(\pi/2+\psi_-)^{s+r}\big] +c^+_{knlr}\big[(\pi/2-\psi_+)^{s+r}+(-1)^r\psi_+^{s+r}\big]}{s+r},\\
    H^{kn}_l(s)=&\,\int_{-\pi/2}^0 d\psi\Big(\cos^k\psi\sin^n\psi\frac{\bar{h}(s)}{\bar{r}^l}-\sum_{r=0}^N b^-_{knlr} (\psi-\psi_-)^r|\psi-\psi_-|^{s-1}_\s \Big)+\ttg\label{DH_analyt}\\
    &+\int_0^{\pi/2} d\psi\Big(\cos^k\psi\sin^n\psi\frac{\bar{h}(s)}{\bar{r}^l}-\sum_{r=0}^N b^+_{knlr} (\psi-\psi_+)^r|\psi-\psi_+|^{s-1}_\s \Big)+\\
    &+\sum_{r=0}^N\frac{b^-_{knlr}\big[(-\psi_-)^{s+r}-(-1)^r(\pi/2+\psi_-)^{s+r}\big] +b^+_{knlr}\big[(\pi/2-\psi_+)^{s+r}-(-1)^r\psi_+^{s+r}\big]}{s+r}.
\end{align*}
The analytical continuations of the integrals $P^{kn}_l(s)$ and $G^{kn}_l(s)$ are constructed in the same way as for the functions $D^{kn}_l(s)$ and $H^{kn}_l(s)$ but with the replacements $\bar{d}(s)\rightarrow\bar{p}(s)$ and $\bar{h}(s)\rightarrow\bar{g}(s)$, respectively. We see that the functions $D^{kn}_l(s)$, $P^{kn}_l(s)$ possess singularities in the form of simple poles at the points $s=-2k$, $k=\overline{0,\infty}$, and the functions $H^{kn}_l(s)$, $G^{kn}_l(s)$ have singularities in the form of simple poles at the points $s=-2k-1$, $k=\overline{0,\infty}$. Let us denote the Laurent series coefficients as
\begin{equation}
    D^{kn}_l(s)=\sum_{r=-1}^\infty \overset{(r)}{D}{}^{kn}_l(0) s^r,\qquad H^{kn}_l(s)=\sum_{r=-1}^\infty \overset{(r)}{H}{}^{kn}_l(-1) (s+1)^r,
\end{equation}
and so forth. If the function possesses a finite limit at a given point, we will omit the superscript. For example,
\begin{equation}
    H^{kn}_l(0)\equiv \overset{(0)}{H}{}^{kn}_l(0).
\end{equation}
The residues at the poles of the functions $D^{kn}_l(s)$, $H^{kn}_l(s)$, $P^{kn}_l(s)$, and $G^{kn}_l(s)$ are determined by the coefficients of the expansions \eqref{H_D_expans} as
\begin{equation}\label{H_D_resid}
\begin{split}
    \res_{s=-2r}D^{kn}_l(s)&=\res_{s=-2r}P^{kn}_l(s)=2[c^+_{knl,2r}(\be,-2r)+c^-_{knl,2r}(\be,-2r)],\\
    \res_{s=-2r-1}H^{kn}_l(s)&=\res_{s=-2r-1}G^{kn}_l(s)=2[b^+_{knl,2r+1}(\be,-2r-1)+b^-_{knl,2r+1}(\be,-2r-1)].
\end{split}
\end{equation}
These coefficients have certain symmetry properties and they can be expressed in terms of the expansion coefficients of the function
\begin{equation}
\begin{split}
    \cos^k\psi \sin^n\psi \frac{(\bar{r}-\cos\psi)^{s-1}}{\bar{r}^l}=&\,\sum_{r=0}^\infty \tilde{c}_{knlr}(\be,s)(\psi-\psi_+(\be))^{r+s-1}.
\end{split}
\end{equation}
Namely,
\begin{equation}
\begin{split}
    c^+_{knlr}(\be,s)&=\tilde{c}_{knlr}(\be,s),\qquad c^-_{knlr}(\be,s)=(-1)^{r+n}\tilde{c}_{knlr}(-\be,s),\\
    b^+_{knlr}(\be,s)&=\tilde{c}_{knlr}(\be,s),\qquad b^-_{knlr}(\be,s)=(-1)^{r+n+1}\tilde{c}_{knlr}(-\be,s).
\end{split}
\end{equation}
Then the residues \eqref{H_D_resid} become
\begin{equation}
\begin{split}
    \res_{s=-2r}D^{kn}_l(s)&=\res_{s=-2r}P^{kn}_l(s)=2[\tilde{c}_{knl,2r}(\be,-2r)+(-1)^n\tilde{c}_{knl,2r}(-\be,-2r)],\\
    \res_{s=-2r-1}H^{kn}_l(s)&=\res_{s=-2r-1}G^{kn}_l(s)=2[\tilde{c}_{knl,2r+1}(\be,-2r-1)+(-1)^n\tilde{c}_{knl,2r+1}(-\be,-2r-1)].
\end{split}
\end{equation}
The several first coefficients $\tilde{c}_{knlr}$ are written as
\begin{align*}
    \tilde{c}_{knl0}(\be,s)=&\,(1-\be_\perp^2)^{(s-1)/2} v^n(1+v^2)^{(s+l-k-n-1)/2},\\
    \tilde{c}_{knl1}(\be,s)=&\,\tilde{c}_{knl0}(\be,s)\Big\{\frac{n}{v} -kv - l\big[v^3-\be_\parallel (1+v^2)\big] +(s-1)\frac{\be_\perp^2}{2}\frac{1+v^2}{(1-\be_\perp^2)^{1/2}} \Big\},\ttg\label{tilde_c}\\
    2\tilde{c}_{knl2}(\be,s)=&\,\frac{\tilde{c}^2_{knl1}(\be,s)}{\tilde{c}_{knl0}(\be,s)} +\tilde{c}_{knl0}(\be,s)\Big( (1+v^2) \Big\{ l\big[2\be_\parallel^2 (1+v^2) -2\be_\parallel (v+2v^3) -v^2 (1-2v^2) \big]-\\
    &-k-\frac{n}{v^2} \Big\} -(s-1)\frac{4+\be_\perp^2\big[9\be_\parallel^2(1+v^2)^2-6\be_\parallel(v+4v^3+3v^5) -1+3v^2+9v^4+9v^6 \big]}{12(1-\be_\perp^2)} \Big).
\end{align*}
These coefficients are sufficient to construct the analytical continuations of the functions $D^{kn}_l(s)$, $H^{kn}_l(s)$, $P^{kn}_l(s)$ and $G^{kn}_l(s)$ to the domain $\re s>-3$.

The integrals \eqref{H_D_ints} satisfy the recurrence relations
\begin{equation}\label{recurr_rels}
\begin{gathered}
    D^{k+2,n}_l(s)=D^{k,n}_l(s)-D^{k,n+2}_l(s),\quad H^{k+2,n}_l(s)=H^{kn}_l(s)-H^{k,n+2}_l(s),\\
    D^{k+2,n}_l(s)=D^{kn}_l(s+2)+D^{kn}_{l-2}(s) -2H^{kn}_{l-1}(s+1),\\
    H^{k+2,n}_l(s)=H^{kn}_l(s+2)+H^{kn}_{l-2}(s) -2D^{kn}_{l-1}(s+1),\\
    \be^2D^{k+2,n}_l(s)-2\be_\parallel D^{k+1,n+1}_l(s)+D^{k,n+2}_l(s)=D^{k,n}_{l-2}(s),\\
    \be^2H^{k+2,n}_l(s)-2\be_\parallel H^{k+1,n+1}_l(s)+H^{k,n+2}_l(s)=H^{k,n}_{l-2}(s).
\end{gathered}
\end{equation}
The same relations but with the simultaneous replacements $D^{kn}_l(s)\rightarrow P^{kn}_l(s)$, $H^{kn}_l(s)\rightarrow G^{kn}_l(s)$ are also valid. Moreover,
\begin{equation}\label{HGDP_rels}
\begin{gathered}
    H^{kn}_{l-1}(s)-G^{k+1,n}_l(s)=D^{kn}_l(s+1),\qquad D^{kn}_{l-1}(s)-P^{k+1,n}_l(s)=H^{kn}_l(s+1),\\
    P^{kn}_{l-1}(s)-G^{kn}_l(s+1)=D^{k+1,n}_l(s),\qquad G^{kn}_{l-1}(s)-P^{kn}_l(s+1)=H^{k+1,n}_l(s).
\end{gathered}
\end{equation}
It follows from the definition \eqref{H_D_ints} that
\begin{equation}\label{rels1}
    D^{kn}_l(1)=0.
\end{equation}
Setting $s=1$ on the second line of \eqref{recurr_rels} and \eqref{HGDP_rels}, we obtain
\begin{equation}\label{rels2}
    D^{kn}_l(3)=2H^{kn}_{l-1}(2),\qquad P^{kn}_{l-1}(1)=G_l^{kn}(2).
\end{equation}

In calculating the self-force, the following integrals arise
\begin{align*}
    N^{kn}(s):=\,&(s-1)H^{kn}_2(s-1)-D_3^{kn}(s),\\
    K^{kn}(s):=\,&(s+1)sD_3^{kn}(s)-3(s+1)H^{kn}_4(s+1)+3D^{kn}_5(s+2),\\
    L^{kn}(s):=\,&(s+1)s(s-1)H_4^{kn}(s-1)-6(s+1)sD^{kn}_5(s)+15(s+1)H^{kn}_6(s+1)-15D^{kn}_7(s+2),\\
    M^{kn}(s):=\,&(s+1)s(s-1)(s-2)D^{kn}_5(s-2) -10(s+1)s(s-1) H^{kn}_6(s-1)+\\
    &+45(s+1)sD^{kn}_7(s) -105(s+1)H^{kn}_8(s+1) +105D^{kn}_9(s+2),\ttg\label{N_K_L_def}\\
    N^{kn}_t(s+2):=&\,(s+1)s P^{kn}_2(s) -(s+1) G^{kn}_3(s+1),\\
    K^{kn}_t(s):=&\,(s+1)s(s-1) G^{kn}_3(s-1) -3(s+1) s P^{kn}_4(s) +3(s+1) G^{kn}_5(s+1),\\
    L_t^{kn}(s):=&\,(s+1)s(s-1)(s-2)P^{kn}_4(s-2) -6(s+1)s(s-1)G^{kn}_5(s-1)+\\
    &+15(s+1)sP^{kn}_6(s) -15(s+1)G^{kn}_7(s+1).
\end{align*}
The Laurent series in the punctured neighborhoods of $s=0$ and $s=1$ take the form
\begin{equation}\label{N_expans}
\begin{split}
    N^{kn}(s)&=-\frac{\overset{(-1)}{H}{}^{kn}_2(-1)+\overset{(-1)}{D}{}^{kn}_3(0)}{s} +\overset{(-1)}{H}{}^{kn}_2(-1) -\overset{(0)}{H}{}^{kn}_2(-1) -\overset{(0)}{D}{}^{kn}_3(0) +\cdots=\\
    &=(s-1)\big[H^{kn}_2(0) -\overset{(1)}{D}{}^{kn}_3(1) \big]+\cdots.
\end{split}
\end{equation}
In the last equality, we have employed the relation \eqref{rels1}. As for the remaining functions, we deduce
\begin{align*}
    K^{kn}(s)=&\,\frac{2\overset{(-1)}{D}{}^{kn}_3(-2) +3\overset{(-1)}{H}{}^{kn}_4(-1)+3\overset{(-1)}{D}{}^{kn}_5(0)}{s+2}-\\
    &-3\overset{(-1)}{D}{}^{kn}_3(-2) +2\overset{(0)}{D}{}^{kn}_3(-2) -3\overset{(-1)}{H}{}^{kn}_4(-1) +3\overset{(0)}{H}{}^{kn}_4(-1) +3\overset{(0)}{D}{}^{kn}_5(0)   + \cdots=\\
    =&\,\overset{(-1)}{D}{}^{kn}_3(0)-3H^{kn}_4(1)+3D^{kn}_5(2)+\\
    &+s\big[\overset{(-1)}{D}{}^{kn}_3(0) +\overset{(0)}{D}{}^{kn}_3(0) -3H^{kn}_4(1) -3\overset{(1)}{H}{}^{kn}_4(1) +3\overset{(1)}{D}{}^{kn}_5(2) \big]+ \cdots=\\
    =&\,2D^{kn}_3(1)-6H^{kn}_4(2)+3D^{kn}_5(3)+\\ &+(s-1)\big[3D_3^{kn}(1)+2\overset{(1)}{D}{}^{kn}_3(1)-3H_4^{kn}(2)-6\overset{(1)}{H}{}^{kn}_4(2) +3\overset{(1)}{D}{}^{kn}_5(3)\big]+\cdots,\\
    L^{kn}(s)=&\,-\overset{(-1)}{H}{}^{kn}_4(-1) -6\overset{(-1)}{D}{}^{kn}_5(0)+15 H_6^{kn}(1) -15D^{kn}_7(2)+\\
    &+s\big[-\overset{(0)}{H}{}^{kn}_4(-1) -6\overset{(-1)}{D}{}^{kn}_5(0) -6\overset{(0)}{D}{}^{kn}_5(0) +15H^{kn}_6(1) +15\overset{(1)}{H}{}^{kn}_6(1) -15\overset{(1)}{D}{}^{kn}_7(2)\big]+\cdots,\\
    =&\,-12D^{kn}_5(1)+30H^{kn}_6(2)-15D^{kn}_7(3)+\\
    &+(s-1)\big[2H^{kn}_4(0) -18D^{kn}_5(1) -12\overset{(1)}{D}{}^{kn}_5(1) +15 H^{kn}_6(2) +30\overset{(1)}{H}{}^{kn}_6(2) -15\overset{(1)}{D}{}^{kn}_7(3)\big]+\cdots,\\
    M^{kn}(s)=&\,2\overset{(-1)}{D}{}^{kn}_5(-2) +10 \overset{(-1)}{H}{}^{kn}_6(-1) +45 \overset{(-1)}{D}{}^{kn}_7(0) -105H^{kn}_8(1) +105D^{kn}_9(2)+\ttg\label{K_L_expans} \\
    &+s\big[-\overset{(-1)}{D}{}^{kn}_5(-2) +2\overset{(0)}{D}{}^{kn}_5(-2) +10\overset{(0)}{H}{}^{kn}_6(-1) +45\overset{(-1)}{D}{}^{kn}_7(0) +45\overset{(0)}{D}{}^{kn}_7(0)-\\
    &-105H^{kn}_8(1) -105\overset{(1)}{H}{}^{kn}_8(1) +105\overset{(1)}{D}{}^{kn}_9(2)\big]+\cdots,\\
    N^{kn}_t(s+2)=&\,\frac{2\overset{(-1)}{P}{}^{kn}_2(-2) +\overset{(-1)}{G}{}^{kn}_3(-1)}{s+2}-\\
    &-3\overset{(-1)}{P}{}^{kn}_2(-2) +2\overset{(0)}{P}{}^{kn}_2(-2) -\overset{(-1)}{G}{}^{kn}_3(-1) +\overset{(0)}{G}{}^{kn}_3(-1)+\cdots=\\
    =&\,\overset{(-1)}{P}{}^{kn}_2(0) -G^{kn}_3(1) +s\big[\overset{(-1)}{P}{}^{kn}_2(0) +\overset{(0)}{P}{}^{kn}_2(0) -G^{kn}_3(1) -\overset{(1)}{G}{}^{kn}_3(1)\big]+\cdots=\\
    =&\,2P^{kn}_2(1)-2G^{kn}_3(2)+(s-1)\big[3P^{kn}_2(1) +2\overset{(1)}{P}{}^{kn}_2(1) -G^{kn}_3(2) -2\overset{(1)}{G}{}^{kn}_3(2)\big]+\cdots,\\
    K^{kn}_t(s)=&\,-\overset{(-1)}{G}{}^{kn}_3(-1) -3\overset{(-1)}{P}{}^{kn}_4(0) +3G^{kn}_5(1)+\\
    &+s\big[-\overset{(0)}{G}{}^{kn}_3(-1) -3 \overset{(-1)}{P}{}^{kn}_4(0) -3\overset{(0)}{P}{}^{kn}_4(0) +3G^{kn}_5(1) +3\overset{(1)}{G}{}^{kn}_5(1)\big]+\cdots=\\
    =&\,-6P^{kn}_4(1)+6G^{kn}_5(2)+\\
    &+(s-1)\big[-9P^{kn}_4(1) -6\overset{(1)}{P}{}^{kn}_4(1) +2G^{kn}_3(0) +3G^{kn}_5(2) +6\overset{(1)}{G}{}^{kn}_5(2)\big]+\cdots,\\
    L_t^{kn}(s)=&\,2\overset{(-1)}{P}{}^{kn}_4(-2) +6\overset{(-1)}{G}{}^{kn}_5(-1) +15\overset{(-1)}{P}{}^{kn}_6(0) -15G^{kn}_7(1)
    + s\big[-\overset{(-1)}{P}{}^{kn}_4(-2)+\\
    &+2\overset{(0)}{P}{}^{kn}_4(-2) +6\overset{(0)}{G}{}^{kn}_5(-1) +15\overset{(-1)}{P}{}^{kn}_6(0) +15\overset{(0)}{P}{}^{kn}_6(0) -15G^{kn}_7(1) -15\overset{(1)}{G}{}^{kn}_7(1)\big]+\cdots.
\end{align*}
Also we have the expansions
\begin{align*}
    \frac{(2\e)^{-s}\La^{s-1}}{8\pi c(s-1)\cos(\pi s/2)}=&\,-\frac{1}{8\pi c\La}+O(s)=-\frac{1}{8\pi^2 c\e}\Big[\frac{1}{(s-1)^2} -\frac{\ln(2\e/\La)}{s-1}+O(1) \Big],\\
    \frac{(2\e)^{-s}\La^{s-1}}{8\pi c(s-1)s(s+1)\cos(\pi s/2)}=&\,-\frac{1}{8\pi c\La s}+O(1)=-\frac{1}{16\pi^2 c\e}\Big[\frac{1}{(s-1)^2} -\frac{\ln(2\e/\La)+3/2}{s-1} +O(1) \Big],\\
    \frac{(2\e)^{-s}\La^{s}}{8\pi c^2s\cos(\pi s/2)}=&\,\frac{1}{8\pi c^2}\Big[\frac{1}{s} -\ln(2\e/\La)+O(s) \Big],\ttg\label{eps_La_exp_0}\\
    \frac{(2\e)^{-s}\La^{s}}{8\pi c^2s^2(s+1)\cos(\pi s/2)}=&\,\frac{1}{8\pi c^2}\Big[\frac{1}{s^2} -\frac{\ln(2e\e/\La)}{s} +O(1) \Big],\\
    \frac{(2\e)^{-s}\La^{s-1}}{8\pi\cos(\pi s/2)}=&\,-\frac{1}{8\pi^2\e(s-1)}+O(1)=\frac{1}{8\pi\La}+O(s).
\end{align*}
In calculating the self-force, it is convenient to use the notation
\begin{align*}
    n^{kn}:&=\overset{(-1)}{N}{}^{kn}(0)\ln\frac{2\e}{\La}-\overset{(0)}{N}{}^{kn}(0),\\
    k^{kn}:&=K^{kn}(0)\ln\frac{2e\e}{\La}-\overset{(1)}{K}{}^{kn}(0),\\
    l^{kn}:&=L^{kn}(0)\ln\frac{2e\e}{\La}-\overset{(1)}{L}{}^{kn}(0),\\
    \tilde{k}^{kn}:&=\overset{(-1)}{K}{}^{kn}(-2)\ln\frac{2\e}{\La}-\overset{(0)}{K}{}^{kn}(-2),\\
    m^{kn}:&=M^{kn}(0)\ln\frac{2e\e}{\La}-\overset{(1)}{M}{}^{kn}(0),\ttg\label{n_k_l}\\
    g^{kn}_t:&=\overset{(-1)}{G}{}^{kn}_1(-1)\ln\frac{2\e}{\La} +\overset{(-1)}{G}{}^{kn}_1(-1)-\overset{(0)}{G}{}^{kn}_1(-1),\\
    n^{kn}_t:&=N^{kn}_t(2)\ln\frac{2e\e}{\La}-\overset{(1)}{N}{}^{kn}_t(2),\\
    k^{kn}_t:&=K^{kn}_t(0)\ln\frac{2e\e}{\La}-\overset{(1)}{K}{}^{kn}_t(0),\\
    \tilde{n}_t^{kn}:&=\overset{(-1)}{N}{}^{kn}_t(0)\ln\frac{2\e}{\La}-\overset{(0)}{N}{}^{kn}_t(0),\\
    l_t^{kn}:&=L_t^{kn}(0)\ln\frac{2e\e}{\La}-\overset{(1)}{L}{}^{kn}_t(0).
\end{align*}

If $\be_\parallel=0$, i.e., on performing the gauge transformation \eqref{gauge_trans}, the formulas listed above are drastically simplified. Owing to the fact that the integrand is an odd function in this case, we have
\begin{equation}
    D^{kn}_l(s)=H^{kn}_l(s)=P^{kn}_l(s)=G^{kn}_l(s)=0
\end{equation}
for $n$ odd. It is clear that in this case
\begin{equation}
    \tilde{c}_{knlr}(-\be,s)=\tilde{c}_{knlr}(\be,s),
\end{equation}
and, consequently,
\begin{equation}\label{res_DH}
\begin{split}
    \res_{s=-2r}D^{kn}_l(s)&=\res_{s=-2r}P^{kn}_l(s)=4\tilde{c}_{knl,2r}(\be,-2r),\\
    \res_{s=-2r-1}H^{kn}_l(s)&=\res_{s=-2r-1}G^{kn}_l(s)=4\tilde{c}_{knl,2r+1}(\be,-2r-1),
\end{split}
\end{equation}
for $n$ even. Furthermore,
\begin{equation}\label{tilde_c_sym}
\begin{split}
    \tilde{c}_{knl0}(\be,0)=&\,v^{n-1}(1+v^2)^{(l-k-n-1)/2},\\
    \tilde{c}_{knl1}(\be,-1)=&\,v^{n-3}(1+v^2)^{(l-k-n-2)/2}[n-1-kv^2-(l-1)v^4],\\
    2\tilde{c}_{knl2}(\be,-2)=&\,v^{n-5}(1+v^2)^{(l-k-n-3)/2}[n^2-4n+3-(2k+1)(n-1)v^2+\\
    &+(k(k-1)-(n-1)(2l-3))v^4+(l+k(2l-3))v^6+l(l-1)v^8].
\end{split}
\end{equation}
In evaluating the divergencies of the the self-force, the following integrals arise
\begin{equation}\label{HD_int}
    G^{k+1,n}_l(1)=H^{kn}_{l-1}(1)-D^{kn}_{l}(2)=4\int_{\psi_+}^{\pi/2}d\psi \frac{\cos^{k+1}\psi\sin^n\psi}{\bar{r}^l},
\end{equation}
where $n$ is even and $\be_\parallel=0$. These integrals are easily evaluated for even $k$ and $n$. In particular,
\begin{equation}\label{HD_int_expl}
\begin{gathered}
    G^{10}_3(1)=\frac{4}{1+v},\qquad G^{30}_5(1)=\frac43\frac{2+v}{(1+v)^2},\qquad G^{12}_5(1)=\frac43\frac{1+v+v^2}{1+v},\\
    G^{50}_7(1)=\frac4{15}\frac{8+9v+3v^2}{(1+v)^3},\qquad G^{32}_7(1)=\frac4{15}\frac{2+4v+6v^2+3v^3}{(1+v)^2},\qquad
    G^{14}_7(1)=\frac4{5}\frac{1-v^5}{1-v^2},\\
    G^{70}_9(1)=\frac4{35}\frac{16+29 v+20 v^2+ 5v^3}{(1+v)^4},\qquad G^{52}_9(1)=\frac4{105}\frac{8+ 24 v+ 48 v^2+ 45 v^3+ 15 v^4}{(1+v)^3},\\
    G^{34}_9(1)=\frac4{35}\frac{2 -7 v^5 +5 v^7}{(1-v^2)^2}.
\end{gathered}
\end{equation}

\section{Coefficients in the self-force}\label{Useful_Forms_Ints_Coeff}

In calculating the contributions of the Lorentzian force to the self-interaction of a dislocation, the following coefficients arise
\begin{align*}
    a_1:&=(c_l^2-c_t^2)\overset{(-1)}{N}{}^{10}(0)-(5c_l^2-6c_t^2)K^{10}(0)-(c_l^2-2c_t^2)L^{10}_{-2}(0),\\
    \bar{a}_1:&=-(5c_l^2-6c_t^2)K^{10}(0)-(c_l^2-2c_t^2)L^{10}_{-2}(0),\\
    a_2:&=(c_l^2-c_t^2)n^{20}-(5c_l^2-6c_t^2)k^{20}-(c_l^2-2c_t^2)l^{20}_{-2},\\
    \bar{a}_2:&=-(5c_l^2-6c_t^2)k^{20}-(c_l^2-2c_t^2)l^{20}_{-2},\\
    a_3:&=(c_l^2-c_t^2)n^{02}-(5c_l^2-6c_t^2)k^{02}-(c_l^2-2c_t^2)l^{02}_{-2},\\
    \bar{a}_3:&=-(5c_l^2-6c_t^2)k^{02}-(c_l^2-2c_t^2)l^{02}_{-2},\ttg\label{a16}\\
    a_4:&=(c_l^2-c_t^2)(n^{20}+2\tilde{k}^{22}) -(5c_l^2-6c_t^2)k^{20} -2(7c_l^2-10c_t^2)l^{22}-(c_l^2-2c_t^2)(l^{20}_{-2}+2m^{22}_{-2}),\\
    \bar{a}_4:&= -(5c_l^2-6c_t^2)k^{20} -2(7c_l^2-10c_t^2)l^{22}-(c_l^2-2c_t^2)(l^{20}_{-2}+2m^{22}_{-2}),\\
    a_5:&=(c_l^2-c_t^2)\tilde{k}^{40}-(7c_l^2-10c_t^2)l^{40}-(c_l^2-2c_t^2)m^{40}_{-2},\\
    \bar{a}_5:&=-(7c_l^2-10c_t^2)l^{40}-(c_l^2-2c_t^2)m^{40}_{-2},\\
    a_6:&=(c_l^2-c_t^2)\tilde{k}^{22}-(7c_l^2-10c_t^2)l^{22}-(c_l^2-2c_t^2)m^{22}_{-2},\\
    \bar{a}_6:&=-(7c_l^2-10c_t^2)l^{22}-(c_l^2-2c_t^2)m^{22}_{-2}.
\end{align*}
The coefficients entering into \eqref{bJ1} have the form
\begin{equation}\label{a1ba1}
\begin{split}
    a_1=&\,4c_t^2[\frac{1-4v^2-4v^4}{v^3\sqrt{1+v^2}} +3G^{20}_5(1)\Big],\\
    \bar{a}_1=&\,-4c_l^2\Big[\frac{1}{v^3\sqrt{1+v^2}} -q^2\Big(2\frac{1-2v^2-2v^4}{v^3\sqrt{1+v^2}}+3G^{20}_5(1)\Big)\Big],
\end{split}
\end{equation}
where $q^2:=c_t^2/c_l^2$ and
\begin{equation}
    G^{20}_5(1)=4\int_{\arctan v}^{\pi/2}\frac{d\psi\cos^2\psi}{(\be_\perp^2\cos^2\psi+\sin^2\psi)^{5/2}}.
\end{equation}
This integral can be expressed in terms of elliptic integrals.

In calculating the contribution of the Peach-K\"{o}hler force to the self-interaction of a dislocation, the corresponding coefficients are
\begin{align*}
    d_1:=&\,-c^2(c_l^2-c_t^2)\big[\overset{(-1)}{G}{}^{00}_1(-1)+N_t^{00}(2)\big] +(c_l^4-5c_l^2c_t^2+5c_t^4)\overset{(-1)}{N}{}^{10}(0)-\\
    &-(5c_l^4-20c_l^2c_t^2+18c_t^4)K^{10}(0) -(c_l^2-2c_t^2)^2 L^{10}_{-2}(0),\\
    d_2:=&\,c_t^2(c_l^2-2c_t^2)\big[\overset{(-1)}{N}{}^{10}(0)-L_{-2}^{10}(0)\big] -(5c_l^2c_t^2-8c_t^4)K^{10}(0),\\
    d_3:=&\,-c^2\big[\overset{(-1)}{G}{}^{00}_1(-1)+N_t^{00}(2)\big] -c_t^2\big[\overset{(-1)}{N}{}^{10}(0) -2K^{10}(0)\big],\\
    d_4:=&\,-c^2K_t^{20}(0) +2c_t^2L^{30}(0),\\
    d_5:=&\,-c^2 K_t^{02}(0) +2c_t^2L^{12}(0),\\
    d_6:=&\,-(c_l^2-c_t^2)(g^{10}_t+n^{10}_t),\\
    d_7:=&\,-c^2(c_l^2-2c_t^2)k_t^{12} +(c_l^4-5c_l^2c_t^2+5c_t^4)n^{20} -(5c_l^4-20c_l^2c_t^2+18c_t^4)k^{20} -(c_l^2-2c_t^2)^2 l^{20}_{-2},\\
    d_8:=&\,(c_l^4-5c_l^2c_t^2+5c_t^4)n^{20} -(5c_l^4-20c_l^2c_t^2+18c_t^4)k^{20} -(c_l^2-2c_t^2)^2 l^{20}_{-2},\\
    d_9:=&\,(c_l^4-5c_l^2c_t^2+5c_t^4)n^{02} -(5c_l^4-20c_l^2c_t^2+18c_t^4)k^{02} -(c_l^2-2c_t^2)^2 l^{02}_{-2},\\
    d_{10}:=&\,(c_l^4-5c_l^2c_t^2+5c_t^4)\tilde{k}^{40} -(7c_l^4-28c_l^2c_t^2+26c_t^4)l^{40} -(c_l^2-2c_t^2)^2 m^{40}_{-2} +c^2(c_l^2-c_t^2)(\tilde{n}^{30}_t-k^{30}_t),\\
    d_{11}:=&\,(c_l^4-5c_l^2c_t^2+5c_t^4)\tilde{k}^{22} -(7c_l^4-28c_l^2c_t^2+26c_t^4)l^{22} -(c_l^2-2c_t^2)^2 m^{22}_{-2} +c^2(c_l^2-c_t^2)(\tilde{n}^{12}_t-k^{12}_t),\\
    d_{12}:=&\,c_t^2(c_l^2-2c_t^2)(n^{20}-l^{20}_{-2}) -(5c_l^2c_t^2-8c_t^4)k^{20},\\
    d_{13}:=&\,c_t^2(c_l^2-2c_t^2)(n^{20}-l^{20}_{-2}+2\tilde{k}^{22}-2m^{22}_{-2}) -(5c_l^2c_t^2-8c_t^4)k^{20} -2(7c_l^2c_t^2-12c_t^4)l^{22},\\
    d_{14}:=&\,c_t^2(c_l^2-2c_t^2)(n^{02}-l^{02}_{-2}) -(5c_l^2c_t^2-8c_t^4)k^{02},\\
    d_{15}:=&\,c_t^2(c_l^2-2c_t^2)(\tilde{k}^{40}-m^{40}_{-2}) -(7c_l^2c_t^2-12c_t^4)l^{40},\ttg\label{dk}\\
    d_{16}:=&\,c_t^2(c_l^2-2c_t^2)(\tilde{k}^{22}-m^{22}_{-2}) -(7c_l^2c_t^2-12c_t^4)l^{22},\\
    d_{17}:=&\,-2c^2(g^{10}_t+n^{10}_t)-c_t^2(n^{20}-2k^{20}),\\
    d_{18}:=&\,c_t^2l^{40} -c^2 k^{30}_t,\\
    d_{19}:=&\,c_t^2l^{22} -c^2 k^{12}_t,\\
    d_{20}:=&\,c_t^2(n^{02}-2k^{02}),\\
    d_{21}:=&\,c_t^2l^{04} +(c_l^2-2c_t^2)(l^{02}_{-2}-n^{02}) +(5c_l^2-8c_t^2)k^{02},\\
    d_{22}:=&\,c_t^2(\tilde{k}^{40}- 2l^{40}) -c^2(\tilde{n}^{30}_t-k^{30}_t),\\
    d_{23}:=&\,2c_t^2m^{60} -c^2l_t^{50},\\
    d_{24}:=&\,2c_t^2m^{42} -c^2l_t^{32},\\
    d_{25}:=&\,c_t^2(\tilde{k}^{22}- 2l^{22}) -c^2(\tilde{n}^{12}_t-k^{12}_t),\\
    d_{26}:=&\,2c_t^2(m^{24}+4l^{22}) -c^2(l^{14}_t+4k^{12}_t),\\
    d_{27}:=&\,2c_t^2 l^{40} -c^2 k^{30}_t,\\
    d_{28}:=&\,2c_t^2(2m^{42}+l^{40}) -c^2(2l^{32}_t+k^{30}_t),\\
    d_{29}:=&\,2c_t^2 l^{22} -c^2 k^{12}_t,\\
    d_{30}:=&\,2c_t^2 (m^{42}+2l^{40}) -c^2 l^{32}_t,\\
    d_{31}:=&\,c_t^2(c_l^2-2c_t^2)(n^{20}-l^{20}_{-2}) -(5c_l^2c_t^2-8c_t^4)k^{20} -2c^2c_t^2k_t^{12}.
\end{align*}
The coefficients $\bar{d}_{k}$ are obtained from the coefficients $d_{k}$ by the replacements
\begin{equation}
    c\rightarrow c_l,\qquad\overset{(-1)}{G}{}^{00}_1(-1)\rightarrow0,\qquad\overset{(-1)}{N}{}^{10}(0)\rightarrow0,\qquad g^{10}_t\rightarrow0,\qquad\tilde{n}_t^{kn}\rightarrow0,\qquad n^{kn}\rightarrow0,\qquad\tilde{k}^{kn}\rightarrow0.
\end{equation}
The coefficients $\bar{d}_{k}$ are present in the contribution of the longitudinal part of the Green's function to the self-force. As seen from \eqref{dk}, there are obvious relations
\begin{equation}
\begin{gathered}
    d_{13}=d_{12}+2 d_{16},\qquad d_{28}=d_{27}+2d_{24},\qquad d_{30}=d_{24}+4c_t^2l^{40},\qquad d_{27}=d_{18}+c_t^2 l^{40},\\
    d_{31}=d_{12}-2c^2c_t^2k_t^{12},\qquad d_{29}=2d_{19}+c^2 k_t^{12},\qquad d_7=d_8-c^2(c_l^2-2c_t^2)k^{12}_t.
\end{gathered}
\end{equation}
The coefficients appearing in the first contribution to the Peach-K\"{o}hler force \eqref{sPK_1loc} have the form
\begin{align*}
    d_1&=(c_l^2c_t^2-c_t^4) G^{00}_3(1)-6c_t^4 G^{20}_5(1) -4c_l^4q^2(1-3q^2)\frac{\sqrt{1+v^2}}{v},\\
    \bar{d}_1&=(c_l^4-c_l^2c_t^2) G^{00}_3(1)-6c_t^4 G^{20}_5(1) -\frac{4c_l^4}{v^3\sqrt{1+v^2}} \big[1+v^2+v^4-q^2 (4+v^2+v^4)+2q^4(2-v^2-v^4)\big],\\
    d_2&=c_t^4\Big[6G^{20}_5(1) -8\frac{\sqrt{1+v^2}}{v} \Big],\\
    \bar{d}_2&=c_l^4\Big[6q^4G^{20}_5(1) -\frac{4q^2}{v^3\sqrt{1+v^2}}\big[1-2q^2(1-v^2-v^4)\big] \Big],\ttg\label{d15}\\
    d_3&=c_t^2\Big[G^{00}_3(1)-6G^{20}_5(1) +4\frac{\sqrt{1+v^2}}{v} \Big],\\
    \bar{d}_3&=c_l^2\Big[G^{00}_3(1)-6q^2G^{20}_5(1) -4(1-2q^2)\frac{\sqrt{1+v^2}}{v} \Big],\\
    d_4&=-c_t^2\Big[3G^{20}_5(1) -30G^{40}_7(1) -4\frac{1-5v^2-5v^4}{v^3\sqrt{1+v^2}} \Big],\\
    \bar{d}_4&=-c_l^2\Big[3G^{20}_5(1)-30q^2G^{40}_7(1) +\frac{4}{v^3\sqrt{1+v^2}}\big[1-v^2-v^4 -2q^2(1-3v^2-3v^4)\big] \Big],\\
    d_5&=-c_t^2\Big[3G^{02}_5(1) -30G^{22}_7(1) +4\frac{1+7v^2+5v^4}{v\sqrt{1+v^2}} \Big],\\
    \bar{d}_5&=-c_l^2\Big[3G^{02}_5(1)-30q^2G^{22}_7(1) -\frac{4}{v\sqrt{1+v^2}}\big[1+3v^2+v^4 -2q^2(1+5v^2+3v^4)\big] \Big].
\end{align*}
The integrals $G^{nk}_l(1)$ entering these expressions can be expressed in terms of elliptic integrals.

Using the relations \eqref{res_DH}, \eqref{tilde_c_sym}, \eqref{HD_int}, and \eqref{HD_int_expl}, the divergent parts of the coefficients arising in the contribution of the Lorentzian force to the self-force \eqref{bJtot1} become
\begin{align*}
    a_2&=4c_t^2\frac{(1-v)(1+3v)}{v^3(1+v)^2}\ln\frac{2\e}{\La}+\cdots,\\
    \bar{a}_2&=-4c_l^2\Big(\frac{1}{v^3}-2q^2\frac{1+2v-v^2}{v^3(1+v)^2}\Big)\ln\frac{2\e}{\La}+\cdots,\\
    a_3&=-4c_t^2\frac{1-3v}{v(1+v)}\ln\frac{2\e}{\La}+\cdots,\\
    \bar{a}_3&=4c_l^2\Big(\frac{1}{v}-2q^2\frac{1-v}{v(1+v)}\Big)\ln\frac{2\e}{\La}+\cdots,\\
    k^{20}-\frac12n^{20}&=-2\frac{1+2v-v^2}{v^3(1+v)^2}\ln\frac{2\e}{\La}+\cdots,\ttg\label{a_123456_sing}\\
    k^{02}-\frac12n^{02}&=2\frac{1-v}{v(1+v)}\ln\frac{2\e}{\La}+\cdots,\\
    l^{40}&=4\frac{1+3v}{v^3(1+v)^3}\ln\frac{2\e}{\La}+\cdots,\\
    l^{22}&=-\frac{4}{v(1+v)^2}\ln\frac{2\e}{\La}+\cdots,\\
    l^{04}&=\frac{12}{1+v}\ln\frac{2\e}{\La}+\cdots,\\
    a_4&=-4c_t^2\frac{(1-v)(1+3v)}{v^3(1+v)^2}\ln\frac{2\e}{\La}+\cdots,\\
    \bar{a}_4&=4c_l^2\Big(\frac{1}{v^3}-2q^2\frac{1+2v-v^2}{v^3(1+v)^2} \Big)\ln\frac{2\e}{\La}+\cdots,\\
    a_5&=4c_t^2\frac{3+9v+5v^2-9v^3}{v^5(1+v)^3}\ln\frac{2\e}{\La}+\cdots,\\
    \bar{a}_5&=-4c_l^2\Big(\frac{3}{v^5} -2q^2\frac{3+9v+7v^2-3v^3}{v^5(1+v)^3}\Big)\ln\frac{2\e}{\La}+\cdots,\\
    a_6&=-4c_t^2\frac{(1-v)(1+3v)}{v^3(1+v)^2}\ln\frac{2\e}{\La}+\cdots,\\
    \bar{a}_6&=4c_l^2\Big(\frac{1}{v^3}-2q^2\frac{1+2v-v^2}{v^3(1+v)^2}\Big)\ln\frac{2\e}{\La}+\cdots,
\end{align*}
where the ellipsis denote the contributions that are regular in the limit $\e\rightarrow+0$. We see that for the divergent parts of the coefficients
\begin{equation}
    \sing a_2=- \sing a_4=-\sing a_6,\qquad \sing \bar{a}_2=- \sing \bar{a}_4=-\sing \bar{a}_6.
\end{equation}

Similarly, the divergent parts of the coefficients appearing in the contribution of the Peach-K\"{o}hler force to the self-force \eqref{sPK_1loc}, \eqref{sPK_23loc} take the form
\begin{align*}
    d_6&=4c_t^2(q^{-2}-1)\frac{1+v-v^2}{v^3(1+v)}\ln\frac{2\e}{\La}+\cdots,\\
    \bar{d}_6&=-4c_l^2\frac{1-q^2}{v(1+v)}\ln\frac{2\e}{\La}+\cdots,\\
    k_t^{30}&=4\frac{1+2v}{v^3(1+v)^2}\ln\frac{2\e}{\La}+\cdots,\\
    k_t^{12}&=-\frac{4}{v(1+v)}\ln\frac{2\e}{\La}+\cdots,\\
    d_7&=4c_t^4 \Big(\frac{1}{v^3} -\frac{2}{(1+v)^2} -q^{-2}\frac{1+v-v^2}{v^3(1+v)} \Big)\ln\frac{2\e}{\La}+\cdots,\\
    \bar{d}_7&=-4c_l^4\Big(\frac{1+v-v^2}{v^3(1+v)}-2q^{2}\frac{2+2v-v^2}{v^3(1+v)} +2q^{4}\frac{2+4v+v^2}{v^3(1+v)^2}\Big) \ln\frac{2\e}{\La}+\cdots,\\
    d_8&=4c_t^4 \Big(\frac{1+2v+3v^2}{v^3(1+v)^2} - \frac{q^{-2}}{v^3}\Big)\ln\frac{2\e}{\La}+\cdots,\\
    \bar{d}_8&=-4c_l^4\Big(\frac{1-4q^{2}}{v^3} +2q^{4}\frac{2+4v+v^2}{v^3(1+v)^2}\Big)\ln\frac{2\e}{\La}+\cdots,\\
    d_9&=-4c_t^4 \Big(\frac{1+3v}{v(1+v)} -\frac{q^{-2}}{v} \Big)\ln\frac{2\e}{\La}+\cdots,\\
    \bar{d}_9&=4c_l^4\Big(\frac{1-4q^2}{v} +2q^{4}\frac{2+v}{v(1+v)}\Big)\ln\frac{2\e}{\La}+\cdots,\\
    d_{10}&=4c_t^4 \Big(\frac{3+9v+2v^2}{v^3(1+v)^3} -q^{-2}\frac{1+2v}{v^3(1+v)^2} \Big)\ln\frac{2\e}{\La}+\cdots,\\
    \bar{d}_{10}&=-4c_l^4 \Big(\frac{3+6v+4v^2+2v^3}{v^5(1+v)^2} -q^{2}\frac{(2+v)(6+9v+2v^2)}{v^5(1+v)^2} +2q^4\frac{6+18v+17v^2+3v^3}{v^5(1+v)^3} \Big)\ln\frac{2\e}{\La}+\cdots,\\
    d_{11}&=-4c_t^4 \Big(\frac{3+v}{v(1+v)^2} -\frac{q^{-2}}{v(1+v)} \Big)\ln\frac{2\e}{\La}+\cdots,\\
    \bar{d}_{11}&=4c_l^4 \Big(\frac{1+v+v^2}{v^3(1+v)} -q^{2}\frac{(2+v)^2}{v^3(1+v)} +2q^4\frac{2+4v+v^2}{v^3(1+v)^2} \Big)\ln\frac{2\e}{\La}+\cdots,\\
    d_{12}&=-\frac{8c_t^4}{v(1+v)^2}\ln\frac{2\e}{\La}+\cdots,\\
    \bar{d}_{12}&=-4c_l^4 q^2\Big(\frac{1}{v^3} -2q^{2}\frac{1+2v}{v^3(1+v)^2} \Big)\ln\frac{2\e}{\La}+\cdots,\\
    d_{13}&=\frac{8c_t^4}{v(1+v)^2} \ln\frac{2\e}{\La}+\cdots,\\
    \bar{d}_{13}&=4c_l^4q^2 \Big(\frac{1}{v^3} -2q^2\frac{1+2v}{v^3(1+v)^2} \Big)\ln\frac{2\e}{\La}+\cdots,\ttg\label{d_631_sing}\\
    d_{14}&=\frac{8c_t^4}{1+v}\ln\frac{2\e}{\La}+\cdots,\\
    \bar{d}_{14}&=4c_l^4 q^2\Big(\frac{1}{v} -\frac{2q^{2}}{v(1+v)} \Big)\ln\frac{2\e}{\La}+\cdots,\\
    d_{15}&=-8c_t^4\frac{1+3v}{v^3(1+v)^3}\ln\frac{2\e}{\La}+\cdots,\\
    \bar{d}_{15}&=-4c_l^4 q^2\Big(\frac{3}{v^5} -2q^{2}\frac{3+9v+8v^2}{v^5(1+v)^3} \Big)\ln\frac{2\e}{\La}+\cdots,\\
    d_{16}&=\frac{8c_t^4}{v(1+v)^2}\ln\frac{2\e}{\La}+\cdots,\\
    \bar{d}_{16}&=4c_l^4 q^2\Big(\frac{1}{v^3} -2q^{2}\frac{1+2v}{v^3(1+v)^2} \Big)\ln\frac{2\e}{\La}+\cdots,\\
    d_{17}&=4c_t^2\Big(\frac{1}{v^3} -\frac{2}{(1+v)^2} \Big)\ln\frac{2\e}{\La}+\cdots,\\
    \bar{d}_{17}&=-8c_l^2\Big(\frac{1}{v(1+v)} -\frac{q^{2}}{v(1+v)^2} \Big)\ln\frac{2\e}{\La}+\cdots,\\
    d_{18}&=-\frac{8c_t^2}{v(1+v)^3}\ln\frac{2\e}{\La}+\cdots,\\
    \bar{d}_{18}&=-4c_l^2\Big(\frac{1+2v}{v^3(1+v)^2} -q^2\frac{1+3v}{v^3(1+v)^3} \Big)\ln\frac{2\e}{\La}+\cdots,\\
    d_{19}&=\frac{4c_t^2}{(1+v)^2}\ln\frac{2\e}{\La}+\cdots,\\
    \bar{d}_{19}&=4c_l^2\Big(\frac{1}{v(1+v)} -\frac{q^2}{v(1+v)^2} \Big)\ln\frac{2\e}{\La}+\cdots,\\
    d_{20}&=-4c_t^2\frac{1-v}{v(1+v)}\ln\frac{2\e}{\La}+\cdots,\\
    \bar{d}_{20}&=\frac{8c_t^2}{1+v}\ln\frac{2\e}{\La}+\cdots,\\
    d_{21}&=\frac{4c_t^2}{1+v}\ln\frac{2\e}{\La}+\cdots,\\
    \bar{d}_{21}&=-4c_l^2\Big(\frac{1}{v} -q^2\frac{2+3v}{v(1+v)} \Big)\ln\frac{2\e}{\La}+\cdots,\\
    d_{22}&=-4c_t^2\frac{1+3v-2v^2}{v^3(1+v)^3}\ln\frac{2\e}{\La}+\cdots,\\
    \bar{d}_{22}&=4c_l^2\Big(\frac{1+2v}{v^3(1+v)^2} -2q^2\frac{1+3v}{v^3(1+v)^3} \Big)\ln\frac{2\e}{\La}+\cdots,\\
    d_{23}&=4c_t^2\frac{3+12v+13v^2-8v^3}{v^5(1+v)^4}\ln\frac{2\e}{\La}+\cdots,\\
    \bar{d}_{23}&=-4c_l^2\Big(\frac{3+9v+8v^2}{v^5(1+v)^3} -6q^2\frac{1+4v+5v^2}{v^5(1+v)^4} \Big)\ln\frac{2\e}{\La}+\cdots,\\
    d_{24}&=-4c_t^2\frac{1+3v-2v^2}{v^3(1+v)^3}\ln\frac{2\e}{\La}+\cdots,\\
    \bar{d}_{24}&=4c_l^2\Big(\frac{1+2v}{v^3(1+v)^2} -2q^2\frac{1+3v}{v^3(1+v)^3} \Big)\ln\frac{2\e}{\La}+\cdots,\\
    d_{25}&=4c_t^2\frac{1-v}{v(1+v)^2}\ln\frac{2\e}{\La}+\cdots,\\
    \bar{d}_{25}&=-4c_l^2\Big(\frac{1}{v(1+v)} -\frac{2q^2}{v(1+v)^2} \Big)\ln\frac{2\e}{\La}+\cdots,\\
    d_{26}&=-4c_t^2\frac{1-v}{v(1+v)^2}\ln\frac{2\e}{\La}+\cdots,\\
    \bar{d}_{26}&=4c_l^2\Big(\frac{1}{v(1+v)} -\frac{2q^2}{v(1+v)^2} \Big)\ln\frac{2\e}{\La}+\cdots,\\
    d_{27}&=4c_t^2\frac{1+3v-2v^2}{v^3(1+v)^3}\ln\frac{2\e}{\La}+\cdots,\\
    \bar{d}_{27}&=-4c_l^2\Big(\frac{1+2v}{v^3(1+v)^2} -2q^2\frac{1+3v}{v^3(1+v)^3} \Big)\ln\frac{2\e}{\La}+\cdots,\\
    d_{28}&=-4c_t^2\frac{1+3v-2v^2}{v^3(1+v)^3}\ln\frac{2\e}{\La}+\cdots,\\
    \bar{d}_{28}&=4c_l^2\Big(\frac{1+2v}{v^3(1+v)^2} -2q^2\frac{1+3v}{v^3(1+v)^3} \Big)\ln\frac{2\e}{\La}+\cdots,\\
    d_{29}&=-4c_t^2\frac{1-v}{v(1+v)^2}\ln\frac{2\e}{\La}+\cdots,\\
    \bar{d}_{29}&=4c_l^2\Big(\frac{1}{v(1+v)} -\frac{2q^2}{v(1+v)^2} \Big)\ln\frac{2\e}{\La}+\cdots,\\
    d_{30}&=4c_t^2\frac{3+9v+2v^2}{v^3(1+v)^3}\ln\frac{2\e}{\La}+\cdots,\\
    \bar{d}_{30}&=4c_l^2\Big(\frac{1+2v}{v^3(1+v)^2} +2q^2\frac{1+3v}{v^3(1+v)^3} \Big)\ln\frac{2\e}{\La}+\cdots,\\
    d_{31}&=\frac{8c_t^4}{(1+v)^2}\ln\frac{2\e}{\La}+\cdots,\\
    \bar{d}_{31}&=-4c_l^4q^2\Big(\frac{(1-v)(1+2v)}{v^3(1+v)} -2q^2\frac{1+2v}{v^3(1+v)^2} \Big)\ln\frac{2\e}{\La}+\cdots,
\end{align*}
where the ellipsis denote the contributions that are regular in the limit $\e\rightarrow+0$. There are the relations
\begin{equation}
\begin{gathered}
    \sing d_{22}= \sing d_{24} =-\sing d_{27} =\sing d_{28},\qquad \sing \bar{d}_{22}= \sing \bar{d}_{24} =-\sing \bar{d}_{27} =\sing \bar{d}_{28},\\
    \sing d_{25}= -\sing d_{26}= -\sing d_{29},\qquad \sing \bar{d}_{25}= -\sing \bar{d}_{26}= -\sing d_{29}.
\end{gathered}
\end{equation}

\section{Contractions}\label{App_Conctract}

In evaluating the contribution of $b_i\bar{J}_i^{(1)}$ to the self-force, the following contractions arise
\begin{equation}
\begin{split}
    b_i\bar{\vf}_{ik}\be_k=&\,c(c_l^2-c_t^2)(\spb\bs\be)\btbe,\\
    b_i(\de_{ij}\be_k+\cycle(i,j,k))\bar{\vf}_{jk}=&\,c(5c_l^2-6c_t^2)(\spb\bs\be)\btbe,\\
    b_i\be_i \be_j \be_k\bar{\vf}_{jk}=&\,c(c_l^2-2c_t^2)\be^2(\spb\bs\be)\btbe,\\
    b_i(\be_i \tau_j \tau_k+\cycle(i,j,k))\bar{\vf}_{jk}=&\,c(c_l^2-2c_t^2)[(\spb\bs\be) +2\be_\parallel (\spb\bs\tau)]\btbe,\\
\end{split}
\end{equation}
where $\bar{\vf}_{jk}:=\vf_{jk}(0,0)/\rho$. The contribution $b_i\bar{J}_i^{(2)}$ to the self-force involves the contractions
\begin{align*}
    b_i\dot{\bar{\vf}}_{ik}\be_k=&\,c(c_l^2-c_t^2)(\spb\bs\be)\btbe^\cdot+cc_t^2\spb^2(\bs\be\bs\tau\dot{\bs\be}),\\
    b_i\bar{\vf}'_{ik}\tau_k=&\,c(c_l^2-c_t^2)(\spb\bs\tau)\btbe'+cc_t^2\spb^2(\bs\tau\bs{\tau}'\bs\be),\\
    b_i(\de_{ij}\be_k+\cycle(i,j,k))\dot{\bar{\vf}}_{jk}=&\,c(5c_l^2-6c_t^2)(\spb\bs\be)\btbe^\cdot+2cc_t^2\spb^2(\bs\be\bs\tau\dot{\bs\be}),\\
    b_i(\de_{ij}\tau_k+\cycle(i,j,k))\bar{\vf}'_{jk}=&\,c(5c_l^2-6c_t^2)(\spb\bs\tau)\btbe'+2cc_t^2\spb^2(\bs\tau\bs{\tau}'\bs\be),\\
    b_i\be_i \be_j \be_k\dot{\bar{\vf}}_{jk}=&\,c(c_l^2-2c_t^2)\be^2(\spb\bs\be)\btbe^\cdot+2cc_t^2(\spb\bs\be)^2(\bs\be\bs\tau\dot{\bs\be}),\ttg\\
    b_i(\be_i \be_j \tau_k+\cycle(i,j,k))\bar{\vf}'_{jk}=&\,c(c_l^2-2c_t^2)[\be^2(\spb\bs\tau) +2\be_\parallel(\spb\bs\be)]\btbe'+\\
    &+4cc_t^2(\spb\bs\be)(\spb\bs\tau)(\bs\be\bs\tau\bs{\be}') +2cc_t^2(\spb\bs\be)^2(\bs\tau\bs{\tau}'\bs\be),\\
    b_i(\be_i \tau_j \tau_k+\cycle(i,j,k))\dot{\bar{\vf}}_{jk}=&\,c(c_l^2-2c_t^2)[(\spb\bs\be) +2\be_\parallel(\spb\bs\tau)]\btbe^\cdot+\\
    &+4c^2c_t^2(\spb\bs\be)(\spb\bs\tau)(\bs\tau\bs{\be}'\bs\be) +2cc_t^2(\spb\bs\tau)^2(\bs\be\bs\tau\dot{\bs\be}),\\
    b_i\tau_i \tau_j \tau_k\bar{\vf}'_{jk}=&\,c(c_l^2-2c_t^2)(\spb\bs\tau)\btbe' +2cc_t^2(\spb\bs\tau)^2(\bs\tau\bs{\tau}'\bs\be),
\end{align*}
where $\dot{\bar{\vf}}_{jk}:=\dot{\vf}_{jk}(0,0)/\rho$ and $\bar{\vf}'_{jk}:=\vf'_{jk}(0,0)/\rho$. In calculating the contribution of  $b_i\bar{J}_i^{(3)}$ to the self-force, we use the contractions
\begin{align*}
    b_i\de_{kl}\bar{\vf}_{ik}\dot{\be}_l=&\,c(c_l^2-c_t^2)(\spb\dot{\bs\be})\btbe +cc_t^2\spb^2(\dot{\bs\be}\bs\tau\bs\be),\\
    b_i\de_{kl}\bar{\vf}_{ik}\tau'_l=&\,c(c_l^2-c_t^2)(\spb\bs{\tau}')\btbe +cc_t^2\spb^2(\bs{\tau}'\bs\tau\bs\be),\\
    b_i\be_k\be_l\bar{\vf}_{ik}\dot{\be}_l=&\,c(c_l^2-c_t^2)(\spb\bs\be)(\bs\be\dot{\bs\be})\btbe,\\
    b_i\be_k\be_l\bar{\vf}_{ik}\tau'_l=&\,c(c_l^2-c_t^2)(\spb\bs\be)(\bs\be\bs{\tau}')\btbe,\\
    2b_i\be_{(k}\tau_{l)}\bar{\vf}_{ik}\be'_l=&\,c(c_l^2-c_t^2)(\bs\be\bs{\be}')(\spb\bs\tau)\btbe,\\
    b_i\tau_k\tau_l\bar{\vf}_{ik}\dot{\be}_l=&\,c(c_l^2-c_t^2)(\spb\bs\tau)(\bs\tau\dot{\bs\be})\btbe,\\
    b_i\tau_k\tau_l\bar{\vf}_{ik}\tau'_l=&\,0,\ttg\\
    b_i(\de_{ij}\de_{kl}+\cycle(i,j,k))\bar{\vf}_{jk}\dot{\be}_l=&\,c(5c_l^2-6c_t^2)(\spb\dot{\bs\be})\btbe +2cc_t^2\spb^2(\dot{\bs\be}\bs\tau\bs\be),\\
    b_i(\de_{ij}\de_{kl}+\cycle(i,j,k))\bar{\vf}_{jk}\tau'_l=&\,c(5c_l^2-6c_t^2)(\spb\bs{\tau}')\btbe +2cc_t^2\spb^2(\bs{\tau}'\bs\tau\bs\be),\\
    b_i(\de_{ij}\be_k\be_l+\be_i\be_j\de_{kl}+\cycle(i,j,k))\bar{\vf}_{jk}\dot{\be}_l=&\,c[(7c_l^2-10c_t^2)(\spb\bs\be)(\bs\be\dot{\bs\be}) +(c_l^2-2c_t^2)\be^2(\spb\dot{\bs\be})]\btbe+\\ &+2cc_t^2(\spb\bs\be)^2(\dot{\bs{\be}}\bs\tau\bs\be),\\
    2b_i(\de_{ij}\be_{(k}\tau_{l)}+\be_{(i}\tau_{j)}\de_{kl}+\cycle(i,j,k))\bar{\vf}_{jk}\be'_l=&\,c\big\{(7c_l^2-10c_t^2)
    (\spb\bs\tau)(\bs\be\bs{\be}')+\\
    &+2(c_l^2-2c_t^2)\be_\parallel(\spb\bs{\be}')\big\}\btbe
    +4cc_t^2(\spb\bs\be)(\spb\bs\tau)(\bs{\be}'\bs\tau\bs\be),\\
    b_i(\de_{ij}\tau_{k}\tau_{l}+\tau_{i}\tau_{j}\de_{kl}+\cycle(i,j,k))\bar{\vf}_{jk}\dot{\be}_l=&\,c[(7c_l^2-10c_t^2)(\spb\bs\tau)(\bs\tau\dot{\bs\be}) +(c_l^2-2c_t^2)(\spb\dot{\bs\be})]\btbe+\\ &+2cc_t^2(\spb\bs\tau)^2(\dot{\bs{\be}}\bs\tau\bs\be),\\
    b_i(\de_{ij}\tau_{k}\tau_{l}+\tau_{i}\tau_{j}\de_{kl}+\cycle(i,j,k))\bar{\vf}_{jk}\tau'_l=&\,c(c_l^2-2c_t^2)(\spb\bs{\tau}')\btbe
    +2cc_t^2(\spb\bs\tau)^2(\bs{\tau}'\bs\tau\bs\be),\\
    b_i\be_i \be_j \be_k \be_l\bar{\vf}_{jk}\dot{\be}_l=&\,c(c_l^2-2c_t^2)\be^2(\bs\be\dot{\bs\be})(\spb\bs\be)\btbe,\\
    b_i\be_i \be_j \be_k \be_l\bar{\vf}_{jk}\tau'_l=&\,c(c_l^2-2c_t^2)\be^2(\bs\be\bs{\tau}')(\spb\bs\be)\btbe,\\
    b_i(\be_i \be_j \be_k \tau_l+\cycle(i,j,k,l))\bar{\vf}_{jk}\be'_l=&\,c(c_l^2-2c_t^2)[\be^2(\bs\be\bs{\be}')(\spb\bs\tau) +2\be_\parallel (\spb\bs\be)(\bs\be \bs{\be}')]\btbe,\\
\end{align*}
and
\begin{align*}
    b_i(\be_i \be_j \tau_k \tau_l +\tau_i \tau_j \be_k \be_l +\cycle(i,j,k))\bar{\vf}_{jk}\dot{\be}_l=&\,c(c_l^2-2c_t^2)\big\{\be^2(\bs\tau\dot{\bs{\be}})(\spb\bs\tau) + (\bs\be\dot{\bs{\be}})(\spb\bs\be)+\\
    &+2\be_\parallel [(\bs\tau\dot{\bs\be})(\spb\bs\be)+(\bs\be\dot{\bs\be})(\spb\bs\tau)] \big\}\btbe,\\
    b_i(\be_i \be_j \tau_k \tau_l +\tau_i \tau_j \be_k \be_l +\cycle(i,j,k))\bar{\vf}_{jk}\tau'_l=&\,c(c_l^2-2c_t^2)(\bs\be\bs{\tau}')[(\spb\bs\be) +2\be_\parallel (\spb\bs\tau) ]\btbe,\\
    b_i(\be_i \tau_j \tau_k \tau_l+\cycle(i,j,k,l))\bar{\vf}_{jk}\be'_l=&\,c(c_l^2-2c_t^2)(\bs\be\bs{\be}')(\spb\bs\tau)\btbe,\ttg\\
    b_i\tau_i \tau_j \tau_k \tau_l\bar{\vf}_{jk}\dot{\be}_l=&\,c(c_l^2-2c_t^2)(\bs\tau\dot{\bs{\be}})(\spb\bs\tau)\btbe,\\
    b_i\tau_i \tau_j \tau_k \tau_l\bar{\vf}_{jk}\tau'_l=&\,0.
\end{align*}

The contribution $P^{1(1)}_n$ to the self-force contains the contractions
\begin{equation}
\begin{split}
    b_j\bar{\chi}_{nj}=&\,c(c_l^2-c_t^2)\btbe b_n+cc_t^2\spb^2[\spt\spbe]_n,\\
    \be_i\be_j b_i\bar{\chi}_{nj}=&\,cc_t^2 (\spb\spbe)[(\spb\spbe)[\spt\spbe]_n+\btbe\be_n],\\
    \tau_i\tau_j b_i\bar{\chi}_{nj}=&\,cc_t^2 (\spb\spt)^2[\spt\spbe]_n,
\end{split}
\end{equation}
where $\bar{\chi}_{nj}:=\chi_{nj}(0,0)/\rho$ and we have cast out the terms proportional to $\tau_n$.

The contractions appearing in the contribution $P^{1(2)}_n$ are
\begin{equation}
\begin{split}
    b_j\dot{\bar{\chi}}_{nj}=&\,c(c_l^2-c_t^2)\btbe^\cdot b_n+cc_t^2\spb^2[\spt\spbe]^\cdot_n,\\
    \be_i\be_j b_i\dot{\bar{\chi}}_{nj}=&\,c(\spb\spbe)[(c_l^2-2c_t^2)(\spbe\spt\dot{\spbe})b_n +c_t^2(\spb\spbe)[\spt\spbe]^\cdot_n+c_t^2\btbe^\cdot\be_n],\\
    \tau_i\tau_j b_i\dot{\bar\chi}_{nj}=&\,c(\spb\spt)[c(c_l^2-2c_t^2)(\spt \spbe'\spbe)b_n +c_t^2(\spb\spt)[\spt\spbe]^\cdot_n],\\
    2\be_{(i}\tau_{j)} b_i\bar{\chi}'_{nj}=&\,c[(c_l^2-2c_t^2)((\spb\spbe)(\spt\spt'\spbe)+(\spb\spt)(\spbe\spt\spbe'))b_n +2c_t^2(\spb\spbe)(\spb\spt)[\spt\spbe]'_n+c_t^2(\spb\spt) \btbe'\be_n],
\end{split}
\end{equation}
where $\dot{\bar{\chi}}_{nj}:=\dot{\chi}_{nj}(0,0)/\rho$, $\bar{\chi}'_{nj}:=\chi'_{nj}(0,0)/\rho$, and we have omitted the terms proportional to $\tau_n$.

The contractions appearing in the contribution $P^{1(3)}_n$ are written as
\begin{align*}
    \be_k\dot{\be}_k\bar{\chi}_{nj}b_j=&\,c(\spbe\dot{\spbe})\big[(c_l^2-c_t^2)\btbe b_n+c_t^2\spb^2[\spt\spbe]_n\big],\\
    \tau_k\be'_k\bar{\chi}_{nj}b_j=&\,0,\\
    \be_k\tau'_k\bar{\chi}_{nj}b_j=&\,c(\spbe\spt')\big[(c_l^2-c_t^2)\btbe b_n+c_t^2\spb^2[\spt\spbe]_n\big],\\
    (\de_{ij}\be_k+\cycle(i,j,k))b_i\dot{\be}_k\bar{\chi}_{nj}=&c\Big\{\big[ (c_l^2-c_t^2)(\spbe\dot{\spbe})\btbe +(c_l^2-2c_t^2)(\spb\spbe) (\dot{\spbe}\spt\spbe) \big] b_n +\\
    &+c_t^2\big[2(\spb\spbe)(\spb\dot{\spbe})+\spb^2(\spbe\dot{\spbe}) \big][\spt\spbe]_n  +c_t^2 \btbe((\spb\spbe)\be_n)^\cdot \Big\},\\
    \be_i\be_j\be_kb_i\dot{\be}_k\bar{\chi}_{nj}=&\,cc_t^2(\spb\spbe)(\spbe\dot{\spbe}) \big[(\spb\spbe)[\spt\spbe]_n+\btbe \be_n\big],\ttg\\
    (\be_i\tau_j\tau_k+\cycle(i,j,k))b_i\dot{\be}_k\bar{\chi}_{nj}=&\,cc_t^2(\spb\spt)^2(\spbe\dot{\spbe}) [\spt\spbe]_n,\\
    (\de_{ij}\tau_k+\cycle(i,j,k))b_i\be'_k\bar{\chi}_{nj}=&\,c\big[2c_t^2(\spb\spt)(\spb\spbe')[\spt\spbe]_n +(c_l^2-2c_t^2)(\spb\spt)(\spbe'\spt\spbe)b_n +c_t^2(\spb\spt)\btbe\be'_n \big],\\
    (\be_i\be_j\tau_k+\cycle(i,j,k))b_i\be'_k\bar{\chi}_{nj}=&\,cc_t^2(\spb\spt)(\spbe\spbe')\big[2(\spb\spbe)[\spt\spbe]_n +\btbe\be_n \big],\\
    \tau_i\tau_j\tau_kb_i\be'_k\bar{\chi}_{nj}=&\,0,\\
    (\de_{ij}\be_k+\cycle(i,j,k))b_i \tau'_k\bar{\chi}_{nj}=&c\Big\{\big[ (c_l^2-c_t^2)(\spbe\spt')\btbe +(c_l^2-2c_t^2)(\spb\spbe) (\spt'\spt\spbe) \big] b_n +\\
    &+c_t^2\big[2(\spb\spbe)(\spb\spt')+\spb^2(\spbe\spt') \big][\spt\spbe]_n  + c_t^2\btbe((\spb\spt')\be_n +(\spb\spbe)\tau'_n) \Big\},\\
    \be_i\be_j\be_kb_i\tau'_k\bar{\chi}_{nj}=&\,cc_t^2(\spb\spbe)(\spbe\spt') \big[(\spb\spbe)[\spt\spbe]_n+\btbe \be_n\big],\\
    (\be_i\tau_j\tau_k+\cycle(i,j,k))b_i\tau'_k\bar{\chi}_{nj}=&\,cc_t^2(\spb\spt)^2(\spbe\spt') [\spt\spbe]_n,
\end{align*}
where we have discarded the terms proportional to $\tau_n$, $\be_\parallel$, and $(\spt\dot{\spbe})$.

The contribution $P^2_n$ involves the contractions
\begin{equation}\label{chi_expr_1}
\begin{split}
    \bar{\chi}_{nkii}&= (c_l^4-5c_l^2c_t^2+5c_t^4)b_n[\spb\spt]_k +c_t^2(c_l^2-2c_t^2) [\spb\spt]_n b_k +c_t^4\spb^2\e_{nks}\tau_s,\\
    \bar{\chi}_{njij}&= (3c_l^4-10c_l^2c_t^2+8c_t^4)b_n[\spb\spt]_i +(3c_l^2c_t^2-4c_t^4) [\spb\spt]_n b_i,
\end{split}
\end{equation}
where $\bar{\chi}_{nkij}:=\chi_{nkij}/\rho$. The expressions for $\dot{\bar{\chi}}_{nkij}$, $\bar{\chi}'_{nkij}$, and their contractions are obtained from \eqref{chi_expr} and \eqref{chi_expr_1} by differentiating $\tau_k$ in these expressions.

Before going into calculations of the contractions appearing in $P^2_n$, we observe that the cross products
\begin{equation}
    [\spbe\dot{\spbe}]_n,\qquad [\spbe\dot{\spt}]_n=c[\spbe \spbe']_n,\qquad [\spbe \spt']_n
\end{equation}
pass into the vectors proportional to $\tau_n$ on performing the gauge transformation \eqref{gauge_trans}, \eqref{gauge_trans_1}. Therefore, the contributions proportional to these cross products can be omitted.

The contribution $P^2_n$ contains the contractions
\begin{align*}
    \bar{\chi}_{nkii}\be_k=&\, (c_l^4-5c_l^2c_t^2+5c_t^4)\btbe b_n +c_t^2(c_l^2-2c_t^2) (\spb\spbe) [\spb\spt]_n +c_t^4\spb^2[\spbe\spt]_n,\\
    (\de_{ij}\be_k+\cycle(i,j,k))\bar{\chi}_{nkij}=&\, (5c_l^4-20c_l^2c_t^2+18c_t^4)\btbe b_n +(5c_l^2c_t^2-8c_t^4) (\spb\spbe) [\spb\spt]_n+\\
    &+2c_t^4\spb^2[\spbe\spt]_n,\\
    \be_i\be_j\be_k\bar{\chi}_{nkij}=&\, (c_l^2-2c_t^2)^2\be^2\btbe b_n +c_t^2(c_l^2-2c_t^2)\be^2 (\spb\spbe) [\spb\spt]_n+\\
    &+2c_t^4 (\spb\spbe)^2 [\spbe\spt]_n -2c_t^4(\spb\spbe)\btbe\be_n,\\
    (\be_i\tau_j\tau_k+\cycle(i,j,k))\bar{\chi}_{nkij}=&\, (c_l^2-2c_t^2)^2\btbe b_n +c_t^2(c_l^2-2c_t^2) (\spb\spbe) [\spb\spt]_n +2c_t^4 (\spb\spt)^2 [\spbe\spt]_n,\\
    \dot{\bar{\chi}}_{nkii}\be_k=&\, c(c_l^4-5c_l^2c_t^2+5c_t^4)(\spbe\spb\spbe') b_n +cc_t^2(c_l^2-2c_t^2) (\spb\spbe) [\spb\spbe']_n,\\
    \bar{\chi}'_{nkii}\tau_k=&\, (c_l^4-5c_l^2c_t^2+5c_t^4)(\spt\spb\spt') b_n +c_t^2(c_l^2-2c_t^2) (\spb\spt) [\spb\spt']_n +c_t^4\spb^2[\spt\spt']_n,\\
    (\de_{ij}\be_k+\cycle(i,j,k))\dot{\bar{\chi}}_{nkij}=&\, c(5c_l^4-20c_l^2c_t^2+18c_t^4)(\spbe\spb\spbe') b_n +c(5c_l^2c_t^2-8c_t^4) (\spb\spbe) [\spb\spbe']_n,\\
    \be_i\be_j\be_k\dot{\bar{\chi}}_{nkij}=&\, c(c_l^2-2c_t^2)^2 \be^2 (\spbe\spb\spbe') b_n +cc_t^2(c_l^2-2c_t^2)\be^2 (\spb\spbe) [\spb\spbe']_n-\ttg\\
    &-2cc_t^4(\spb\spbe)(\spbe\spb\spbe')\be_n,\\
    (\be_i\tau_j\tau_k+\cycle(i,j,k))\dot{\bar{\chi}}_{nkij}=&\, c(c_l^2-2c_t^2)^2(\spbe\spb\spbe') b_n +cc_t^2(c_l^2-2c_t^2) (\spb\spbe) [\spb\spbe']_n+\\
    &+4cc_t^4(\spb\spbe)(\spb\spt)[\spt\spbe']_n -2cc_t^4 (\spb\spt) (\spt\spb\spbe') \spbe_n,\\
    (\de_{ij}\tau_k+\cycle(i,j,k))\bar{\chi}'_{nkij}=&\, (5c_l^4-20c_l^2c_t^2+18c_t^4)(\spt\spb\spt') b_n +(5c_l^2c_t^2-8c_t^4) (\spb\spt) [\spb\spt']_n+\\
    &+2c_t^4\spb^2[\spt\spt']_n,\\
    (\be_i\be_j\tau_k+\cycle(i,j,k))\bar{\chi}'_{nkij}=&\, (c_l^2-2c_t^2)^2\be^2(\spt\spb\spt') b_n +c_t^2(c_l^2-2c_t^2) \be^2(\spb\spt) [\spb\spt']_n+\\
    &+2c_t^4(\spb\spbe)^2[\spt\spt']_n -2c_t^4 \big((\spb\spt) (\spbe\spb\spt') +(\spb\spbe)(\spt\spb\spt') \big) \spbe_n,\\
    \tau_i\tau_j\tau_k\bar{\chi}'_{nkij}=&\, (c_l^2-2c_t^2)^2 (\spt\spb\spt') b_n +c_t^2(c_l^2-2c_t^2) (\spb\spt) [\spb\spt']_n+2c_t^4(\spb\spt)^2[\spt\spt']_n.
\end{align*}
The quadratic with respect to $t$ and $\s$ integrals in $P^2_n$ involve the contractions
\begin{align*}
    \bar{\chi}_{nlii}\dot{\be}_l=&\,(c_l^4-5c_l^2c_t^2+5c_t^4)(\dot{\spbe}\spb\spt)b_n +c_t^2(c_l^2-2c_t^2)(\spb\dot{\spbe})[\spb\spt]_n +c_t^4 \spb^2[\dot{\spbe}\spt]_n,\\
    \bar{\chi}_{nlii}\tau'_l=&\,(c_l^4-5c_l^2c_t^2+5c_t^4)(\spt'\spb\spt)b_n +c_t^2(c_l^2-2c_t^2)(\spb\spt')[\spb\spt]_n +c_t^4 \spb^2[\spt'\spt]_n,\\
    \be_k\be_l\bar{\chi}_{nkii}\dot{\be}_l=&\,(\spbe\dot{\spbe})\big[(c_l^4-5c_l^2c_t^2+5c_t^4)\btbe b_n +c_t^2(c_l^2-2c_t^2)(\spb\spbe)[\spb\spt]_n +c_t^4\spb^2[\spbe\spt]_n \big],\\
    \be_k\be_l\bar{\chi}_{nkii}\tau'_l=&\,(\spbe\spt')\big[(c_l^4-5c_l^2c_t^2+5c_t^4)\btbe b_n +c_t^2(c_l^2-2c_t^2)(\spb\spbe)[\spb\spt]_n +c_t^4\spb^2[\spbe\spt]_n \big],\ttg\\
    2\be_{(k}\tau_{l)}\bar{\chi}_{nkii}\be'_l=&\,c_t^2(c_l^2-2c_t^2)(\spbe\spbe')(\spb\spt)[\spb\spt]_n,\\
    \tau_k\tau_l\bar{\chi}_{nkii}\dot{\be}_l=&\,\tau_k\tau_l\bar{\chi}_{nkii}\tau'_l=0,
\end{align*}
and also
\begin{align*}
    (\de_{ij}\de_{kl}+\cycle(i,j,k))\bar{\chi}_{nkij}\dot{\be}_l=&\,(5c_l^4-20c_l^2c_t^2+18c_t^4)(\spb\spt\dot{\spbe}) +(5c_l^2c_t^2-8c_t^4)(\spb\dot{\spbe})[\spb\spt]_n+\\
    &+2c_t^4\spb^2[\dot{\spbe}\spt]_n,\\
    (\de_{ij}\de_{kl}+\cycle(i,j,k))\bar{\chi}_{nkij}\tau'_l=&\,(5c_l^4-20c_l^2c_t^2+18c_t^4)(\spb\spt\spt') +(5c_l^2c_t^2-8c_t^4)(\spb\spt')[\spb\spt]_n+\\
    &+2c_t^4\spb^2[\spt'\spt]_n,\\
    (\de_{ij}\be_k\be_l +\be_i\be_j\de_{kl}+\cycle(i,j,k))\bar{\chi}_{nkij}\dot{\be}_l=&\,\big[(7c_l^4-28c_l^2c_t^2+26c_t^4) (\spbe\dot{\spbe})\btbe+\\
    &+(c_l^2-2c_t^2)^2\be^2(\dot{\spbe}\spb\spt) \big]b_n +\big[(7c_l^2c_t^2 -12c_t^4)(\spbe\dot{\spbe})(\spb\spbe)+\\
    &+c_t^2(c_l^2-2c_t^2)\be^2(\spb\dot{\spbe}) \big][\spb\spt]_n +2 c_t^4\big[(\spbe\dot{\spbe})\spb^2+\\
    &+2(\spb\spbe)(\spb\dot{\spbe})\big][\spbe\spt]_n +2c_t^4(\spb\spbe)^2[\dot{\spbe}\spt]_n -2c_t^4\big[(\spb\dot{\spbe})\btbe+\\
    &+(\spb\spbe)(\dot{\spbe}\spb\spt)\big]\be_n -2c_t^4(\spb\spbe)\btbe\dot{\be}_n,\\
    (\de_{ij}\be_k\be_l +\be_i\be_j\de_{kl}+\cycle(i,j,k))\bar{\chi}_{nkij} \tau'_l=&\,\big[(7c_l^4-28c_l^2c_t^2+26c_t^4) (\spbe \spt')\btbe+\ttg\\
    &+(c_l^2-2c_t^2)^2\be^2(\spt'\spb\spt) \big]b_n +\big[(7c_l^2c_t^2 -12c_t^4)(\spbe\spt')(\spb\spbe)+\\
    &+c_t^2(c_l^2-2c_t^2)\be^2(\spb\spt') \big][\spb\spt]_n +2 c_t^4\big[(\spbe\spt')\spb^2+\\
    &+2(\spb\spbe)(\spb\spt')\big][\spbe\spt]_n +2c_t^4(\spb\spbe)^2[\spt'\spt]_n -2c_t^4\big[(\spb\spt')\btbe+\\
    &+(\spb\spbe)(\spt'\spb\spt)\big]\be_n -2c_t^4(\spb\spbe)\btbe\tau'_n,\\
    2(\de_{ij}\be_{(k}\tau_{l)}+\be_{(i}\tau_{j)}\de_{kl}+\cycle(i,j,k))\bar{\chi}_{nkij} \be'_l=&\,(7c_l^2c_t^2-12c_t^4)(\spbe\spbe')(\spb\spt)[\spb\spt]_n +4c_t^4(\spb\spbe')(\spb\spt)[\spbe\spt]_n+\\
    &+4c_t^4(\spb\spbe)(\spb\spt)[\spbe'\spt]_n -2c_t^4(\spb\spt)(\spbe'\spb\spt)\be_n-\\
    &-2c_t^4(\spb\spt)\btbe\be'_n,\\
    (\de_{ij}\tau_k\tau_l+\tau_i\tau_j\de_{kl}+\cycle(i,j,k))\bar{\chi}_{nkij}\dot{\be}_l=&\, (c_l^2-2c_t^2)^2(\dot{\spbe}\spb\spt)b_n +c_t^2(c_l^2-2c_t^2)(\spb\dot{\spbe})[\spb\spt]_n+\\
    &+2c_t^4(\spb\spt)^2 [\dot{\spbe}\spt]_n,\\
    (\de_{ij}\tau_k\tau_l+\tau_i\tau_j\de_{kl}+\cycle(i,j,k))\bar{\chi}_{nkij}\tau'_l=&\, (c_l^2-2c_t^2)^2(\spt'\spb\spt)b_n +c_t^2(c_l^2-2c_t^2)(\spb\spt')[\spb\spt]_n+\\
    &+2c_t^4(\spb\spt)^2 [\spt'\spt]_n,\\
    \be_i\be_j\be_k\be_l\bar{\chi}_{nkij}\dot{\be}_l=&\,(\spbe\dot{\spbe})\big[(c_l^2-2c_t^2)^2\be^2\btbe b_n +c_t^2(c_l^2-2c_t^2)\be^2 (\spb\spbe) [\spb\spt]_n+\\
    &+2c_t^4 (\spb\spbe)^2 [\spbe\spt]_n -2c_t^4(\spb\spbe)\btbe\be_n \big],\\
    \be_i\be_j\be_k\be_l\bar{\chi}_{nkij}\tau'_l=&\,(\spbe\spt')\big[(c_l^2-2c_t^2)^2\be^2\btbe b_n +c_t^2(c_l^2-2c_t^2)\be^2 (\spb\spbe) [\spb\spt]_n+\\
    &+2c_t^4 (\spb\spbe)^2 [\spbe\spt]_n -2c_t^4(\spb\spbe)\btbe\be_n \big],
\end{align*}
and more
\begin{align*}
    (\be_i\be_j\be_k\tau_l+\cycle(i,j,k,l))\bar{\chi}_{nkij}\be'_l=&\, c_t^2(c_l^2-2c_t^2)\be^2(\spbe\spbe')(\spb\spt)[\spb\spt]_n +4c_t^4(\spbe\spbe')(\spb\spt)(\spb\spbe)[\spbe\spt]_n-\\ &-2c_t^4(\spbe\spbe')(\spb\spt)\btbe\be_n,\\
    (\be_i\be_j\tau_k\tau_l+\tau_i\tau_j\be_k\be_l +\cycle(i,j,k))\bar{\chi}_{nkij}\dot{\be}_l=&\, (c_l^2-2c_t^2)^2(\spbe\dot{\spbe})\btbe b_n +c_t^2(c_l^2-2c_t^2)(\spbe\dot{\spbe})(\spb\spbe)[\spb\spt]_n+\\
    &+2c_t^4(\spbe\dot{\spbe})(\spb\spt)^2[\spbe\spt]_n,\\
    (\be_i\be_j\tau_k\tau_l+\tau_i\tau_j\be_k\be_l +\cycle(i,j,k))\bar{\chi}_{nkij}\tau'_l=&\, (c_l^2-2c_t^2)^2(\spbe\spt')\btbe b_n +c_t^2(c_l^2-2c_t^2)(\spbe\spt')(\spb\spbe)[\spb\spt]_n+\\
    &+2c_t^4(\spbe\spt')(\spb\spt)^2[\spbe\spt]_n,\\
    (\be_i\tau_j\tau_k\tau_l +\cycle(i,j,k,l))\bar{\chi}_{nkij}\be'_l=&\, c_t^2(c_l^2-2c_t^2) (\spbe\spbe') (\spb\spt)[\spb\spt]_n,\ttg\\
    \tau_i\tau_j\tau_k\tau_l\bar{\chi}_{nkij}\dot{\be}_l=&\, \tau_i\tau_j\tau_k\tau_l\bar{\chi}_{nkij}\tau'_l=0.
\end{align*}

\section{Local contributions to the self-force}\label{Useful_Forms_Local}

In this appendix, we give the explicit expressions for the local contributions \eqref{bJtot1}, \eqref{sPK_1loc}, and \eqref{sPK_23loc} to the self-force in the case of a strongly suppressed dislocation climb, i.e., the constraint \eqref{constrants} will be employed. For the Lorentzian force, we find
\begin{equation}\label{sL_loc}
    \mathbf{f}^{\text{sL}}_{\text{loc}}=c[\spt\spbep]b_i\bar{J}_i,
\end{equation}
where
\begin{align*}
    b_i\bar{J}_i=&\,-\frac{\rho c_t^2}{8\pi c}\Big\{ \frac{(\spbep\spt D\spbep)}{c} \big[\spb^2(k^{20}-n^{20}/2) +(\spb\spt)^2 l^{22} +(\spb\spbep)^2 l^{40}\big]+\\
    &+ (\spt\spt' \spbep) \big[\spb^2(k^{02}-n^{02}/2) +(\spb\spbep)^2 l^{22} +(\spb\spt)^2 l^{04}\big] +4 (\spt\spbep'\spbep) (\spb\spbep)(\spb\spt) l^{22} \Big\}_{c\rightarrow c_t}+\ttg\label{bJtot2}\\
    &+\frac{\rho c_t^2}{8\pi c}\Big\{\cdots\Big\}_{c\rightarrow c_l,\overset{(-1)}{N}{}^{10}(0)\rightarrow0,n^{kn}\rightarrow0,\tilde{k}^{kn}\rightarrow0}.
\end{align*}
As for the local contributions of the Peach-K\"{o}hler force, we have
\begin{equation}\label{sPK_1loc2}
\begin{split}
    \mathbf{f}^{\text{sPK}(1)}=&\, \frac{\rho}{8\pi\La}\Big\{ \frac{d_2}{c^2}(\spb\spbep) \spb_\perp -\frac{c_t^2}{c^2}\big[d_3\spb^2 +d_4(\spb\spbep)^2 +d_5(\spb\spt)^2\big] \spbep\Big\}_{c\rightarrow c_t}-\\
    &-\frac{\rho}{8\pi\La}\Big\{\cdots\Big\}_{c\rightarrow c_l,\overset{(-1)}{G}{}^{00}_1(-1)\rightarrow0,\overset{(-1)}{N}{}^{10}(0)\rightarrow0},
\end{split}
\end{equation}
and
\begin{align*}
    \mathbf{f}^{\text{sPK}(23)}=&\,-\frac{\rho}{16\pi}\Big\{ [\spt\spb]\Big[ (2c_t^2-c_l^2)(\spb\spbep) \big[ k^{30}_t\frac{(\spbep\spt D\spbep)}{c} +k^{12}_t(\spt\spt'\spbep)\big]+\frac{2d_{7}-d_8}{c^2}(\spb\spt)(\spbep\spt\spbep')-\\ &-\frac{d_9}{c^2}(\spt'\spb\spt)\Big]  + \spb_\perp \Big[\frac{d_{12}}{c^3}(\spb D\spbep)  +\frac{d_{13}}{c^2}(\spbep\spbep')(\spb\spt) +\frac{d_{14}}{c^2}(\spb\spt')+\\
    &+\frac{d_{15}}{c^3}(\spbep D\spbep)(\spb\spbep) +\frac{d_{16}}{c^2}(\spbep\spt')(\spb\spbep) \Big] -\frac{c_t^2}{c^3}(D\spbep)_\perp\Big[d_{17}\spb^2 +2d_{18}(\spb\spbep)^2+\\
    &+2d_{19}(\spb\spt)^2 \Big] -\frac{c_t^2}{c^2}\spt'\Big[d_{20}\spb^2 -2c_t^2l^{22}(\spb\spbep)^2 -2d_{21}(\spb\spt)^2 \Big] +\frac{c_t^2}{c^2}\spbep \Big[\frac{(\spbep D\spbep)}{c}\times\\
    &\times\big[d_{22}\spb^2-d_{23} (\spb\spbep)^2 -d_{24}(\spb\spt)^2 \big] +(\spbep\spt')\big[d_{25}\spb^2-d_{24}(\spb\spbep)^2 -d_{26}(\spb\spt)^2 \big]-\ttg\label{sPK_23loc2}\\
    &-2\frac{d_{27}}{c}(\spb D\spbep)(\spb\spbep) -2d_{28}(\spbep\spbep')(\spb\spt)(\spb\spbep) -2d_{29} \big[(\spb\spbep)(\spb\spt') +2(\spb\spt)(\spb\spbep') \big]\Big]+\\
    &+\frac{2c_t^4}{c^2}[\spt\spbep]\Big[ l^{40} (\spb\spbep) (\spbep,2\spb-(\spb\spt)\spt,\spbep') +l^{22}\big[(\spb\spbep)(\spb\spt'\spt)+2(\spb\spt)(\spb\spt'\spbep) \big]  \Big]-\\
    &-2\frac{d_{31}}{c^2}(\spb\spbep)(\spb\spt)(\spbep')_\perp\Big\}_{c\rightarrow c_t}
    +\frac{\rho}{16\pi}\Big\{\cdots\Big\}_{c\rightarrow c_l,g^{10}_t\rightarrow0,\tilde{n}_t^{kn}\rightarrow0,n^{kn}\rightarrow0,\tilde{k}^{kn}\rightarrow0},
\end{align*}
where we have used the differential consequence of the constraint \eqref{constrants}:
\begin{equation}\label{constr_consq}
    (\spb\spt D\spbep)=c(\spb\spt)(\spbep'\spt\spbep).
\end{equation}


\end{document}